%% file: pristine_hr.tex
\documentclass[fleqn,usenatbib]{mnras}

\usepackage{newtxtext,newtxmath}
\usepackage[T1]{fontenc}
\usepackage{ae,aecompl}
\usepackage{booktabs}

\usepackage{pdflscape}

\usepackage{graphicx}	% Including figure files
\usepackage{amsmath}	% Advanced maths commands
\newcommand{\eg}{$e.g.,$ }
\newcommand{\kms}{km\,s$^{-1}$}

\newcommand{\teff}{T$_{\rm eff}$ }
\newcommand{\logg}{log\,$g$ }
\newcommand{\Pristine}{{\it Pristine}}
\newcommand{\msun}{M$_{\odot}$}

\title[Pristine Survey XII - GRACES spectra of EMP stars]{The Pristine survey - XII: Gemini-GRACES chemo-dynamical study of newly discovered extremely metal-poor stars in the Galaxy}

\author[C. L. Kielty et al.]
{Collin L. Kielty,$^{1}$\thanks{E-mail: clkielty@uvic.ca.}
Kim A. Venn$^{1}$,
Federico Sestito$^{1,2,3}$,
Else Starkenburg$^{3,4}$,
\and
Nicolas F. Martin$^{2,5}$,
David S. Aguado$^{6}$,
Anke Arentsen$^{2,3}$,
S\'{e}bastien Fabbro$^{7}$,
\and
Jonay I. Gonz\'{a}lez Hern\'{a}ndez$^{8,9}$,
Vanessa Hill$^{10}$,
Pascale Jablonka$^{10,11}$,
\and
Carmela Lardo$^{11}$,
Lyudmila I. Mashonkina$^{12}$,
Julio F. Navarro$^{1}$,
Chris Sneden$^{13}$,
\and
Guillaume F. Thomas$^{7}$,
Kris Youakim$^{14}$,
Spencer Bialek$^{1}$,
Rub\'en S\'anchez-Janssen$^{15}$.
\\$\;$
\\
$^{1}$Department of Physics and Astronomy, University of Victoria, Victoria, BC, V8W 3P2, Canada \\
$^{2}$Universit\'e de Strasbourg, CNRS, Observatoire astronomique de Strasbourg, UMR 7550, F-67000 Strasbourg, France \\
$^{3}$Leibniz-Institut f\"ur Astrophysik Potsdam (AIP), An der Sternwarte 16, D-14482, Potsdam, Germany \\
$^{4}$Kapteyn Astronomical Institute, University of Groningen, Landleven 12, 9747 AD Groningen, The Netherlands\\
$^{5}$Max-Planck-Institut f\"ur Astronomie, K\"onigstuhl 17, D-69117, Heidelberg, Germany \\
$^{6}$Institute of Astronomy, University of Cambridge, Madingley Road, Cambridge CB3 0HA, UK \\
$^{7}$National Research Council of Canada, Herzberg Astronomy \& Astrophysics Program, 5071 West Saanich Road, \\
Victoria, BC, V9E 2E7, Canada \\
$^{8}$Instituto de Astrof\'{i}sica de Canarias, V\'{i}a Lactea, E-38205 La Laguna, Tenerife, Spain \\
$^{9}$Departamento de Astrof\'{i}sica, Universidad de La Laguna, E-38206 La Laguna, Tenerife, Spain \\
$^{10}$GEPI, Observatoire de Paris, Universit\'e PSL, CNRS, Place Jules Janssen, F-92190 Meudon, France \\
$^{11}$Dipartimento di Fisica e Astronomia, Universit\'a degli Studi di Bologna, Via Gobetti 93/2, I-40129 Bologna, Italy\\
$^{12}$Institute of Astronomy, Russian Academy of Sciences, RU-119017 Moscow, Russia \\
$^{13}$Department of Astronomy, University of Texas at Austin, Austin, TX, 78712, USA \\
$^{14}$Department of Astronomy, Stockholm University, SE-106 91 Stockholm, Sweden \\
$^{15}$UK Astronomy Technology Centre, Royal Observatory Edinburgh, Blackford Hill, Edinburgh EH9 3HJ, UK
}

\date{Accepted XXX. Received Dec, 2020}

\pubyear{2021}

\begin{document}
\label{firstpage}
\pagerange{\pageref{firstpage}--\pageref{lastpage}}
\maketitle

\begin{abstract}
High-resolution optical spectra of 30 metal-poor stars selected from the \Pristine\ survey are presented, based on observations taken with the Gemini Observatory GRACES spectrograph.  
Stellar parameters \teff and \logg are determined using a Gaia DR2 colour-temperature calibration and surface gravity from the Stefan-Boltzmann equation. 
GRACES spectra are used to determine chemical abundances (or upper-limits) for 20 elements  (Li, O, Na, Mg, K, Ca, Ti, Sc, Cr, Mn, Fe, Ni, Cu, Zn, Y, Zr, Ba, La, Nd, Eu). 
These stars are confirmed to be metal-poor ([Fe/H]$<-2.5$), with higher precision than from earlier medium-resolution analyses. 
The chemistry for most targets is similar to other extremely metal-poor stars in the Galactic halo.  
Three stars near [Fe/H]$=-3.0$ have unusually low Ca and high Mg, suggestive of contributions from few SN~II where alpha-element formation through hydrostatic nucleosynthesis was more efficient.  
Three new carbon-enhanced metal-poor stars are also identified (two CEMP-s and one potential CEMP-no star) when our chemical abundances are combined with carbon from previous medium-resolution analyses. 
The GRACES spectra also provide precision radial velocities ($\sigma_{\rm RV}\le0.2$km\,s$^{-1}$) for dynamical orbit calculations with the Gaia DR2 proper motions.
Most of our targets are dynamically associated with the Galactic halo;  however, five stars with [Fe/H]$<-3$ have planar-like orbits, including one retrograde star.
Another five stars are dynamically consistent with the Gaia-Sequoia accretion event; three have typical halo [$\alpha$/Fe] ratios for their metallicities, whereas two are [Mg/Fe]-deficient, and one is a new CEMP-s candidate.  
These results are discussed in terms of the formation and early chemical evolution of the Galaxy.
\end{abstract}


%%%%%%%%%%%%%%%%%%%%%%%%%%%%%%%%%%%%%%%%%%%%%%%%%%

%%%%%%%%%%%%%%%%% BODY OF PAPER %%%%%%%%%%%%%%%%%%

\section{Introduction}

The oldest and most metal-poor stars contain a fossil record of the star formation processes in the early Universe. 
These first stars would have been composed solely of hydrogen, helium, and trace amounts of lithium \citep{Steigman07, Cyburt16}; without metals to efficiently cool the gas, large Jeans masses, and thereby very massive stars ($M_* \gtrsim 100 M_{\odot}$) are predicted to have formed  \citep{Silk83, Tegmark97, Bromm1999, Abel02, Yoshida06}. 
In addition, improved gas fragmentation models \citep[e.g.,][]{Clark11, Schneider12} and the discovery of very low mass ultra metal-poor stars ([Fe/H]$<-4$; \citealt{Keller14, Starkenburg17a, Schlaufman18, Nordlander19}) have
suggested that lower mass, long lived stars may have also formed in these pristine environments \citep[e.g.,][]{Nakamura2001, Wise12}.
Notably, if $\leq0.8$ \msun\ stars were to form, they could still exist today on the main sequence, and are expected to have undergone very little surface chemical evolution over their lifetimes.
Detailed studies could provide invaluable constraints on (1) the nucleosynthetic yields from the first stars and the earliest supernovae
\citep[e.g.,][]{HegerWoosley10, Ishigaki14, Tominaga14, Clarkson18, Nordlander19}, (2) the physical conditions in the high redshift universe \citep[where stars are too faint to be observed individually;][]{Tegmark97, Freeman02, Cooke14, Hartwig18, Salvadori19}, (3) the primordial initial mass function \citep[e.g.,][]{Greif12, Susa14}, and (4)  early Galactic chemical evolution processes \citep[see][and references therein]{Sneden08, Tolstoy09, Roederer14, Yoon16, Wanajo18, Kobayashi2020}. 

Dedicated searches for old, metal-poor stars in the Milky Way and in its dwarf galaxy satellites were initiated over two decades ago \citep[e.g.,][]{bond1980, carney1981, Beers85, Beers99}.
Chemical abundances have now been measured for hundreds of extremely metal-poor stars (EMP), where [Fe/H]\footnote{We adopt the square bracket notation for chemical abundances relative to the Sun, such that [X/Y] = log(X/Y)$_\odot$ $-$ log(X/Y)$_*$, where X and Y are column densities for two given elements.} $\le -3$ 
\citep[e.g.,][]{Aoki13, Yong13, Cohen13, FN15, Matsuno17, Aguado19Pristine} and over a dozen stars with [Fe/H]$< -4$ \citep[see][and references therein]{Aguado18UMPa, Aguado18UMPb,  Bonifacio18, Starkenburg17a, Aguado19Pristine, Frebel19, Nordlander19}.
Most of the Galactic EMP stars show enhanced [$\alpha$/Fe] abundances and a diversity of neutron capture element ratios, interpreted as the yields from core-collapse supernovae with different progenitor masses and explosion prescriptions. 
More recently, the importance of compact binary mergers to the r-process abundance ratios has also been revealed \citep{Ji16, RoedererReticulum16, hansen2017}.
In dwarf galaxies, the abundance ratios of hydrostatic elements (\eg O, Na, Mg), explosive elements (\eg Si, K, Mn), complex elements (like Zn), and heavy elements (as Ba, Y) tend to be lower than those of their Galactic counterparts of similar metallicity. 
This reflects a range of differences in their star formation histories, including interstellar mixing efficiencies, star formation efficiencies, star formation rates, and effective mass functions of the dwarf galaxy systems \citep{Venn04, Tolstoy09, Nissen10, McWilliam13, FN15, Hasselquist17, Hayes18, Lucchesi20}.
The diversity of chemical abundance profiles seen in this sparse population of objects makes for challenging statistical studies. 
Targeted programs are needed to uncover larger samples of these (nearly) pristine stars.

The present day locations of metal-poor stars also a diagnostic of early galaxy formation.
Based on cosmological simulations of the Local Group, it is thought that the Galactic halo was formed through the accretion and disruption of dwarf galaxies at early epochs. 
Consequently, properties of the old, metal-poor stars seen in the halo manifest in the properties of their dwarf galaxy hosts \citep[e.g.,][]{Ibata94, Helmi99, Johnston2008, Wise12, Starkenburg17b, ElBadry18EMP}.
The arrival of precision proper motions from \textit{Gaia} DR2 \citep{GaiaDR2}, and increasingly large datasets of stars with precision radial velocities from high-resolution spectroscopy, has enabled the determination of orbits for halo stars.
The majority of the EMP halo stars have been shown to have high-velocities and eccentric orbits, consistent with accretion from a dwarf galaxy \citep{Sestito19UMP, Sestito19Pristine, dimatteo2020, Cordoni2020}.
A large population of halo stars has also been found with retrograde orbits and kinematics consistent with a halo merger remnant, e.g., \textit{Gaia-Enceladus/Sausage} \citep{Meza2005, Belokurov18, Helmi18, Myeong18} and \textit{Gaia-Sequoia} \citep{Barba19, Myeong19, Matsuno17, Monty2020, Cordoni2020}.

The Galactic halo is not the only place to look for old stars.  Curiously, some EMP stars have been found on nearly circular, planar orbits \citep{Caffau12, Sestito19UMP, Sestito19Pristine, Schlaufman18, Venn2020}.
Since the Galactic plane is thought to have formed $\sim 10$ Gyr ago \citep{Casagrande16}, these stars challenge the idea that the metal-poor stars are amongst the oldest stars accreted from the mergers of dwarf galaxies.
The Galactic bulge is another environment to look for relics of first stars \citep{White00, guo2010, Starkenburg17b, ElBadry18EMP}.
To date, $\sim 2000$ very metal-poor stars ([Fe/H] $< -2.0$) associated with the bulge have been found \citep{GarciaPerez15, Howes15, Howes16, Lamb17, Arentsen20b}.
Detailed chemical abundance analyses for these objects are limited, especially at the lowest metallicities, but they indicate that many of the metal-poor bulge candidates are chemically similar to halo stars \citep{Lucey2019}. 
In fact, estimates of their orbital properties using the exquisite astrometry from Gaia DR2 \citep{GaiaDR2} suggests that up to 50\% of these stars may be normal halo stars with highly elliptical and plunging orbits, a fraction which increases with decreasing metallicity \citep{Lucey2020}.
However, while \textit{Gaia} DR2 proper motions are extremely valuable in eliminating foreground metal-poor non-bulge members, the uncertainties on the parallax measurements for most stars in the bulge are currently too large for reliable distance estimates. 

Regardless of where metal-poor stars are found, 
the union of stellar dynamics and chemical cartography provides a foundation for studies of \textit{Galactic Archaeology} \citep[e.g.,][]{Carney03, Freeman02, Venn04, Hayden15, Bovy16b, Hasselquist17}.
Dedicated surveys have been successful in finding most of the metal-poor stars; these include the HK and Hamburg-ESO surveys \citep{Beers92, Christlieb02, BC05}, the SDSS SEGUE, BOSS, and APOGEE surveys \citep{Yanny09a, Eisenstein11, Majewski2015}, LAMOST \citep{Cui12}, and more recently the narrow band imaging surveys \textit{SkyMapper} \citep{Keller07, DaCosta2019, Onken2020} and \textit{Pristine} \citep{Starkenburg17a}.  
The discovery of more EMP stars is necessary to build a statistically significant sample to study the metal-poor Galaxy, and also to overcome cosmic variance in the chemo-dynamic analysis of stellar populations within it.

In this paper, 30 new candidate EMP stars have been selected from the \textit{Pristine} survey for follow-up high-resolution spectroscopy.  \textit{Pristine} is based on a MegaPrime/MegaCam imaging survey at the 3.6-metre Canada France Hawaii Telescope, using a unique narrow-band filter centered on the \ion{Ca}{II} H \& K spectral lines (\textit{CaHK}).  When \textit{CaHK} is combined with the broad-band SDSS \textit{gri} photometry \citep{York00} or Gaia DR2 photometry, this \textit{CaHK} filter can been calibrated to find metal-poor candidates with $4000 <$ T$_{\rm eff}$ $< 7000$~K \citep{Youakim17, Starkenburg17a, Arentsen20}.
The \textit{Pristine} survey has been shown to successfully predict stellar metallicities. $\sim56\%$ of stars with [Fe/H]$_{Phot.}<-2.5$ are confirmed to have [Fe/H]$<-2.5$ ($\sim23\%$ when [Fe/H]$_{Phot.}<-3.0$) based on follow-up spectroscopic studies \citep{Youakim17, Caffau2017, Bonifacio2019, Aguado19Pristine, Venn2020}. The target stars for this paper were selected from the medium-resolution spectroscopic follow-up studies of EMP candidates found within the initial $\sim$2500 sq. degree footprint\footnote{We note that the \textit{Pristine} survey has more than doubled its survey area since its initial footprint.} of the \textit{Pristine} survey (Martin et al., in prep.). In Section 4, we improve upon the the previous stellar parameter determinations; in Section 5, we calculate the chemical abundances or upper-limits for $\sim$15 elements; and in Section 6, we estimate the orbits using the Gaia DR2 database for a chemo-dynamical analysis.
Together, these properties permit a study of the origins of these metal-poor stars themselves, ultimately to unravel the formation history of the Milky Way Galaxy.

\section{Target Selection}

\begin{figure}
	\includegraphics[width=\columnwidth]{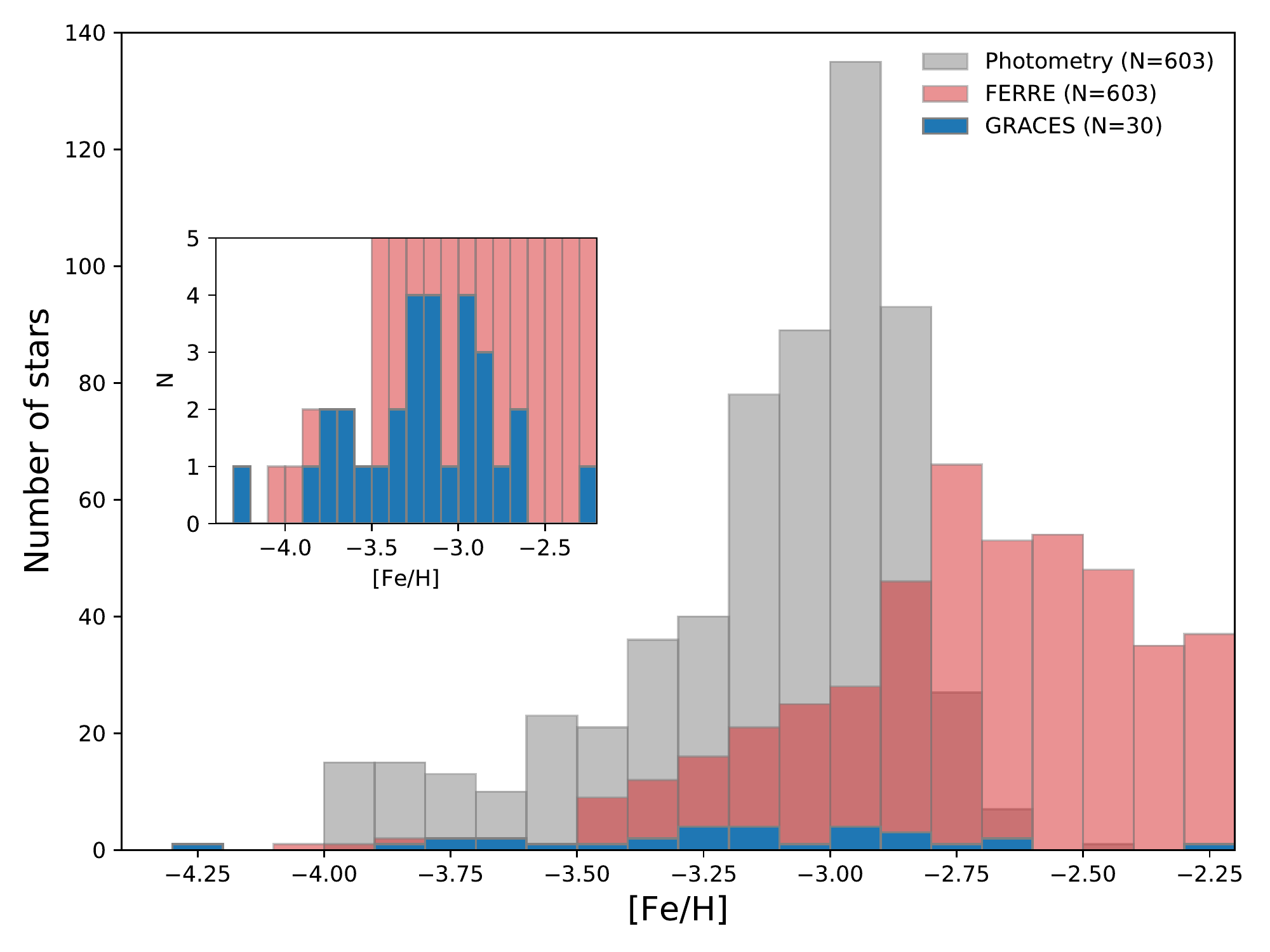}
    \caption{The metallicity distribution for the full sample of metal-poor stars found in the original \textit{Pristine} footprint.  The grey and pink distributions are for the same sample of stars \citep[V$<$18,][]{Aguado19Pristine}, showing that medium-resolution spectral follow-up found slightly higher metallicities for this sample.
    %All stars shown pass the selection criteria outlined in \citet{Youakim17} and \citet{Aguado2019} and have high probabilities ($>$ 90\%) of having a Pristine metallicity  [Fe/H]$_{\rm phot} <-2.5$ in both the ($g-i$) and ($g-r$) calibrations. \textit{Pristine} photometric metallicities are an average of \textit{g - i} and \textit{g - r}).
    GRACES metallicities from this paper are shown in blue (inset included for higher detail).  Clearly, there are many more EMP candidates available for high-resolution spectroscopic analyses.\label{fig:histogram}}
\end{figure}

Targets in this paper have been selected from the original 2500 sq. degree \textit{Pristine} photometric survey \citep{Starkenburg17a, Youakim17} and medium-resolution spectroscopic follow-up survey \citep{Youakim17, Aguado19Pristine}. All of the medium-resolution spectra (hereafter MRS) were observed at the Isaac Newton Telescope (INT), and were analysed using the FERRE data analysis pipeline \citep{AllendePrieto2006} and the ASSET grid of synthetic stellar spectra as described and published by \citealt{Aguado2017WHT}. 
FERRE searches for the best fit to the observed spectrum by simultaneously deriving the main stellar atmospheric parameters (effective temperature, surface gravity, and metallicity [Fe/H]), and it also determines the carbonicity, [C/Fe].  Uncertainties are found via Markov Chain Monte Carlo (MCMC) sampling around the best fit stellar parameters.
Stars with a very high probability ($>$ 90\%) for a \textit{Pristine} metallicity  [Fe/H]$_{\rm phot} <-2.75$ in both the ($g-i$) and ($g-r$) calibrations are shown in Fig.~\ref{fig:histogram} (in grey).  The metallicities from the INT medium-resolution spectral analyses by FERRE for those {\it same stars} are also shown in Fig.~\ref{fig:histogram} (in red).  Clearly some of our initial \textit{Pristine} metallicities were too low; however, significant improvements in the \textit{Pristine} selection function have been made since our original MRS spectroscopic follow-up programs, in anticipation of providing targets for the WEAVE survey \citep{Dalton14, Dalton2018}.  

\begin{figure*}
	\includegraphics[width=\textwidth]{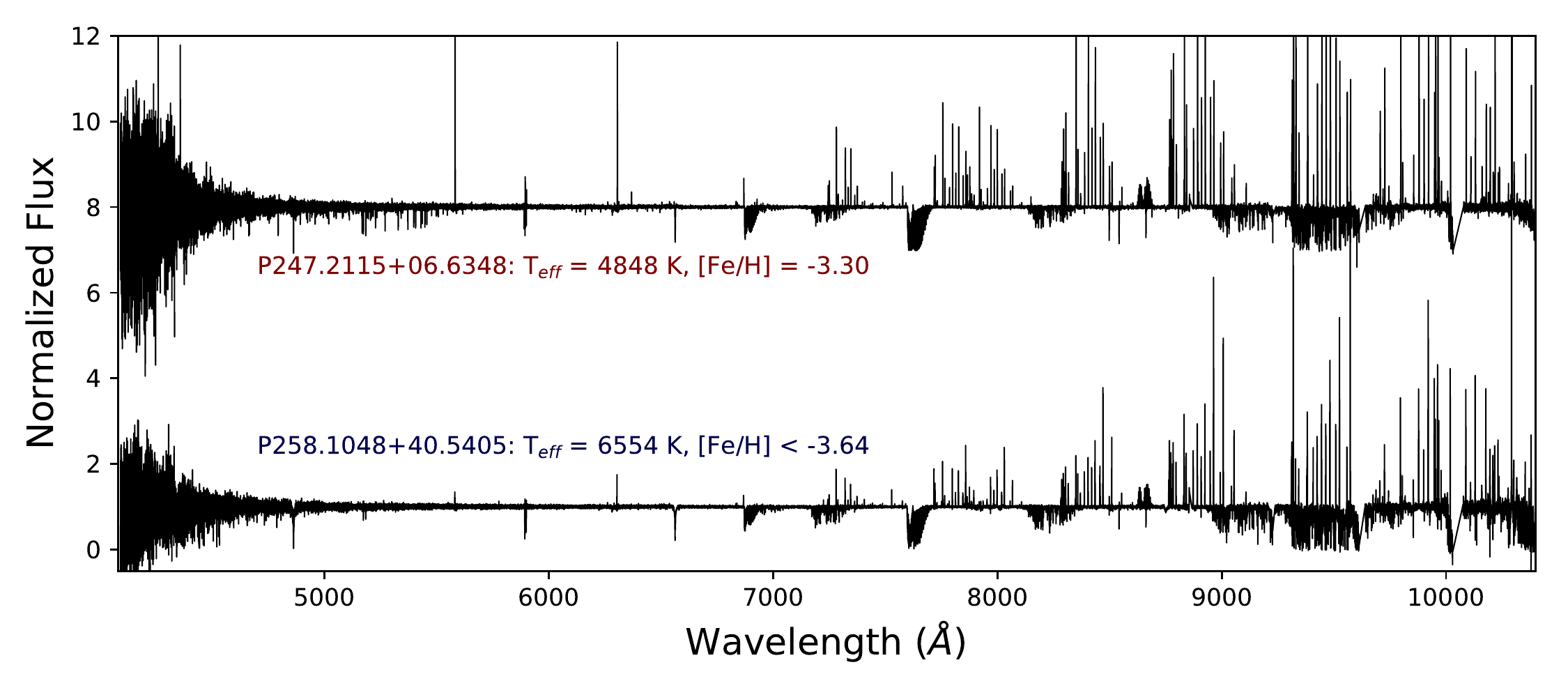}
    \caption{Full Gemini/GRACES spectra for the hottest and coolest stars in our sample.  Even though the star+sky mode was used, the sky subtraction is clearly imperfect due to the presence of emission lines at longer wavelengths.  The poor throughput below $\sim$4800\AA\ of the long 270-metre optical fibre can be seen as an decrease in the SNR at blue wavelengths.}
    \label{fig:all_spectra}
\end{figure*}

Targets in this paper with high-resolution spectroscopy have been selected
with [Fe/H]$_{\rm MRS} <-3$ and low temperatures (T$_{\rm eff}$ $<$ 6500\,K) in the magnitude range $14.8<$ V $<16.4$.  These are noted in blue in Fig.~\ref{fig:histogram}.  We have observed $<$5\% of the full sample from the original footprint area, showing that a spectroscopic survey like WEAVE is necessary to reach all of them.
Some stars with enhanced carbon were included ([C/Fe]$_{\rm FERRE} >1$); however, stars with non-normal carbon abundances were not given a higher priority in the target selection because the carbon abundances are typically unreliable when low SNR $(<25)$ and low-resolution data are analysed with the FERRE pipeline \citep[see below, also][Arentsen et al. in prep.]{Aguado19Pristine}.

The targets analysed in this paper are listed in Table~\ref{tab:photometry}. The stellar identifications (IDs) are a combination of their SDSS RA and DEC coordinates and $V$ are in observer magnitudes calculated from SDSS $u,g,r,i,z$ given the calibration by \citep{Lupton2005}. The other SDSS and \textit{Pristine} filter magnitudes have been dereddened using E(B-V) from \citep{Schlegel1998}. 
Table~\ref{tab:photometry} also lists the information on the exposures and the SNR of the final combined spectrum for each target.
Stellar parameters from the \textit{Pristine} survey are listed in Table~\ref{tab:ferre_params}.
This includes the photometric \textit{Pristine} metallicities ([Fe/H]$_{\rm phot}$) and the colour temperatures (T$_{\rm phot}$), which are an average of the dwarf and giant solutions from SDSS \textit{gri} and \textit{Pristine} \textit{CaHK} photometry \citep[see ][]{Starkenburg17a}.
Table~\ref{tab:ferre_params} also includes the stellar parameters from the FERRE analysis of the INT medium-resolution spectroscopic survey by \cite{Aguado19Pristine} for our sample.  
The average SNR $=28$ for the MRS of our targets, with a range $7 <$ SNR $<79$.  As mentioned above, the FERRE pipeline simultaneously determines the effective temperature (T$_{\rm eff}$), surface gravity (log g), metallicity [Fe/H]$_{\rm MRS}$, and carbonicity [C/Fe]$_{\rm MRS}$ for each star. Radial velocities are also derived from MRS ($\Delta$RV $\sim 1$ \kms).   
\citet{Aguado19Pristine} investigated the FERRE carbonicity measurements as functions of both surface gravity and SNR.
They find that carbon abundances are more reliably determined (systematic uncertainties of $\sim 0.2$ dex) for cool, lower gravity giants, primarily due to the increased strength of spectral lines at lower temperatures. Likewise, increased line strength means lower SNR is needed for a detection of the G band.
Conversely, the uncertainties in the carbon abundances increases up to $\sim$0.6 dex for the hotter, higher gravity stars, where weaker spectral features and less reliable surface gravities are expected.
Since the carbon measurement is so heavily dependent on both SNR and gravity, we do not report individual values for [C/Fe]$_{\rm MRS}$ in Table \ref{tab:ferre_params}, and instead flag stars as C-rich candidates if [C/Fe]$_{\rm MRS} > 1.0$.
The stars are flagged further to note whether the carbon abundances are \textit{reliable}, based on the SNR of the medium-resolution INT spectra analysed by FERRE.

\section{Gemini-GRACES Observations }

High-resolution spectra have been collected for 30 extremely metal-poor candidates with the Gemini Remote Access to CFHT ESPaDOnS Spectrograph
\citep[GRACES;][]{Chene2014, Pazder2014}.
A 270-metre optical fibre links the Cassegrain focus at the Gemini-North telescope to the Canada-France-Hawaii Telescope
ESPaDOnS spectrograph \citep[a cross-dispersed high-resolution 
{\'e}chelle spectrograph that covers the whole visible spectrum;][]{Donati2006}.
We note that the exposure times we used were in good agreement with the GRACES ITC provided by Gemini Observatory. 
In the 2-fibre (object+sky) mode, spectra are obtained with resolution R$\sim$65,000; however, light below $\sim$4800 \AA\ is severely limited by poor transmission through the optical fibre link.  

The GRACES spectra have been reduced using the Gemini ``Open-source Pipeline for ESPaDOnS Reduction and Analysis" tool \citep[OPERA,][]{OPERA}.  This includes standard calibrations (images are bias subtracted, flat fielded, extracted, wavelength calibrated, and heliocentric corrected).  Starting from the individually extracted and normalized {\'e}chelle orders (the *i.fits files), one continuous spectrum is stitched together by weighting the overlapping wavelength regions by their error spectrum, and co-adding as a weighted average.
All visits for a given star are examined for radial velocity (RV) variations, potentially indicating binarity; 
no evidence for binary systems was found based on the criteria $\Delta$RV$\le1$\,\kms, although we note radial velocity variations are unlikely to be measured due to the short cadence of our observations.  
All visits for a given star have then been co-added via a weighted mean using the error spectrum.  The coadded spectrum has then been radial velocity corrected through cross-correlation with a high SNR comparison spectrum of the metal-poor standard star HD~122563.  Other radial velocity standards were examined (e.g., Arcturus and a synthetic spectrum for a typical metal-poor RGB), however the solutions had the smallest uncertainties ($\sigma$RV $\le0.2$ \kms) when compared to HD~122563.
As a final step, the RV corrected spectra were  re-normalized using k-sigma clipping with a nonlinear filter (a combination of a median and a boxcar). The effective scale length of the filter ranged from 8 to 15 \AA, dependent on the crowding of the spectral lines, which was sufficient to follow the continuum when
used in conjunction with iterative clipping.
The full wavelength coverage of the final spectra for two sample stars is shown in Fig.~\ref{fig:all_spectra}, including the imperfect sky subtraction, as seen by the night sky emission lines above $\sim$8000 \AA. 

\begin{figure}
	\includegraphics[width=\columnwidth]{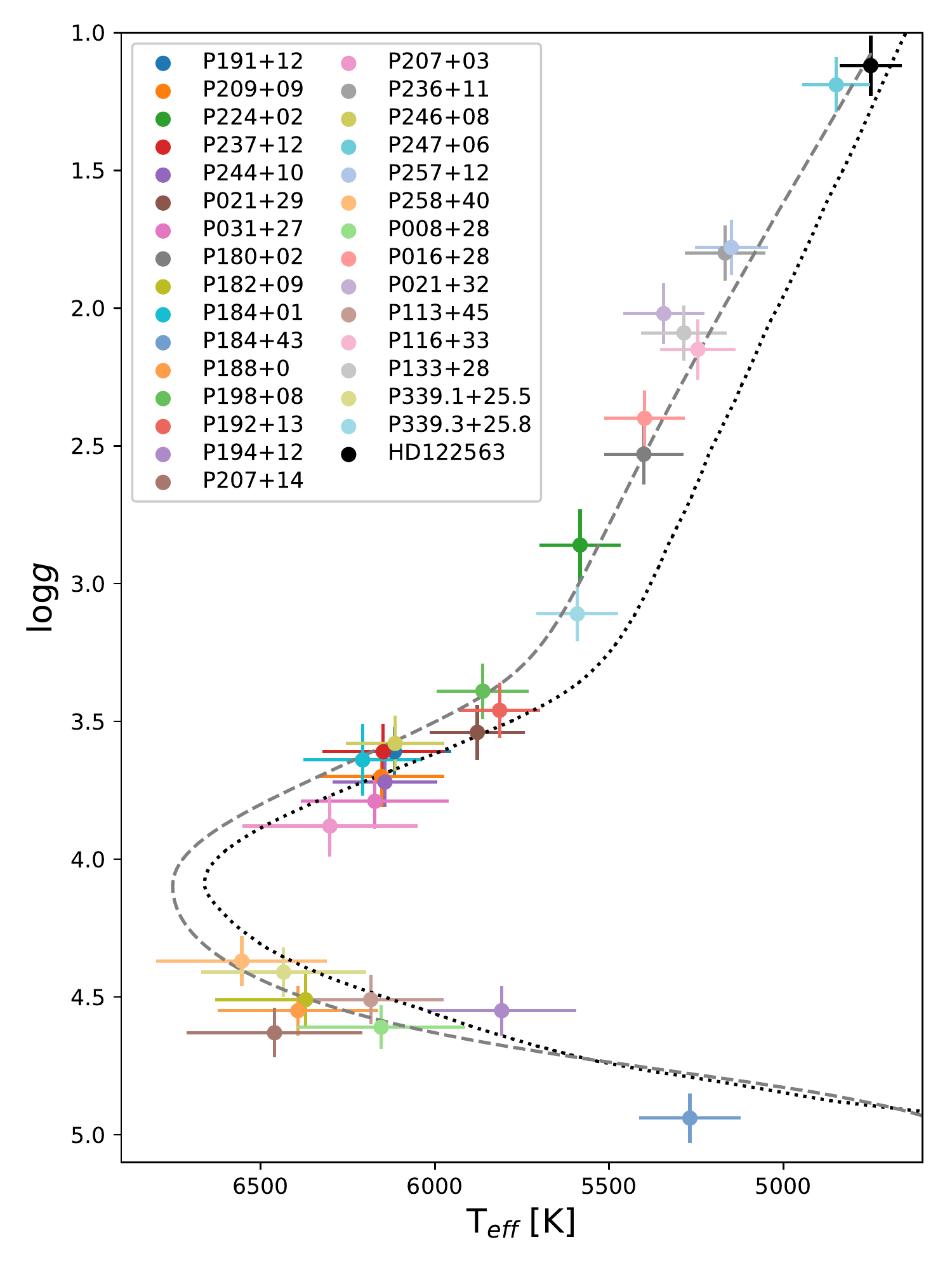}
    \caption{T$_{\rm eff}$ and log~$g$ derived from \citet{MB2020} and the Stefan-Boltzmann law. MIST/MESA (dashed grey line) and Yale-Yonsei (dotted black line) isochrones with [Fe/H] = -3.5 and age = 13 Gyr are shown for visual reference.}
    \label{fig:gaia_isochrone}
\end{figure} 

\section{Stellar Parameters and Radial Velocities}

The stellar temperatures (T$_{\rm eff}$) have been determined using the colour-temperature relation for Gaia photometry from \citet{MB2020} (MB2020). This calibration was selected based on their inclusion of very metal-poor stars  \citep[from][]{GHB2009}.  When calculating temperatures from MB2020, it is necessary to know if the star is a dwarf or giant (two sets of calibrations) and to have a metallicity estimate {\it a priori}.  Using the Gaia parallaxes, we were able to estimate the likelihood of an individual star as a dwarf or giant (see below), and we adopted the initial metallicities determined from the {\sl Pristine} survey photometry and INT medium-resolution spectroscopic analysis \citep{Aguado19Pristine}.
Colours were dereddened using \citet{SF2011}, adapted for Gaia DR2 photometry\footnote{Gaia DR2 extinction values (v 3.4) at http://stev.oapd.inaf.it/cgi-bin/cmd.}.
A comparison of our temperatures from the MB2020 calibration to those from the  \citet{Casagrande2020} calibration showed very good agreement, even though the latter had very few standard stars at very low metallicities.

Surface gravities (log $g$) were determined using the Stefan-Bolzmann equation (e.g., \citealt{Venn17, KI2003}).  This method requires (i) a distance, which we calculate from the Gaia DR2 parallax assuming the parallax zero point offset from \citet{Lindegren2018}, 
(ii) the solar bolometric magnitude\footnote{IAU 2015 Resolution B2 on the bolometric magnitude scale available at
https://www.iau.org/static/resolutions/IAU2015\_English.pdf.} of M$_{\rm bol}$ = 4.74, and 
(iii) bolometric corrections for Gaia DR2 photometry \citep{Andrae2018}. 

Uncertainties in T$_{\rm eff}$ and log~$g$ were determined from a Monte Carlo analysis on the uncertainties from the input parameters, as well as assuming a flat prior on the stellar mass, spanning 0.5 to 1.0 M$_\odot$.  
The final stellar parameter, microturbulence $\xi$, was determined from the surface gravity values, using the calibrations by \cite{Sitnova2015} for dwarf stars and \cite{Mashonkina17} for giants.

This method was selected over the use of isochrones due to significant differences between the MIST/MESA \citep{Paxton2011, Dotter2016, choi2016mesa} and Yale-Yonsei isochrones \citep{YY1, YYL} for old, metal-poor stars shown in Fig.~\ref{fig:gaia_isochrone}. 
The quality of the spectral parameters using the MIST/MESA isochrones for metal-poor stars has been questioned. For example, \citet{Monty2020} found good agreement between spectroscopic and isochrone-mapping temperature determinations for stars with [Fe/H]$>-2$. However, for lower metallicities, the temperatures determined from isochrone-mapping tended to be hotter, by up to $\Delta$\teff=+500\,K when [Fe/H]=$-3.5$. 
A similar result has been seen by \citet{Joyce2015, Joyce2018} due to a range of (optimized) values for the convective mixing length parameter. 
Nevertheless, a comparison of our temperatures from the MB2020 calibrations to those from the Bayesian inference method using MIST/MESA isochrones developed by \citet{Sestito19UMP} showed good agreement for dwarfs and red giants. Where we found significant deviations were for our EMP stars on the sub-giant branch and those predicted to be carbon-rich.  

\begin{figure}
	\includegraphics[width=\columnwidth]{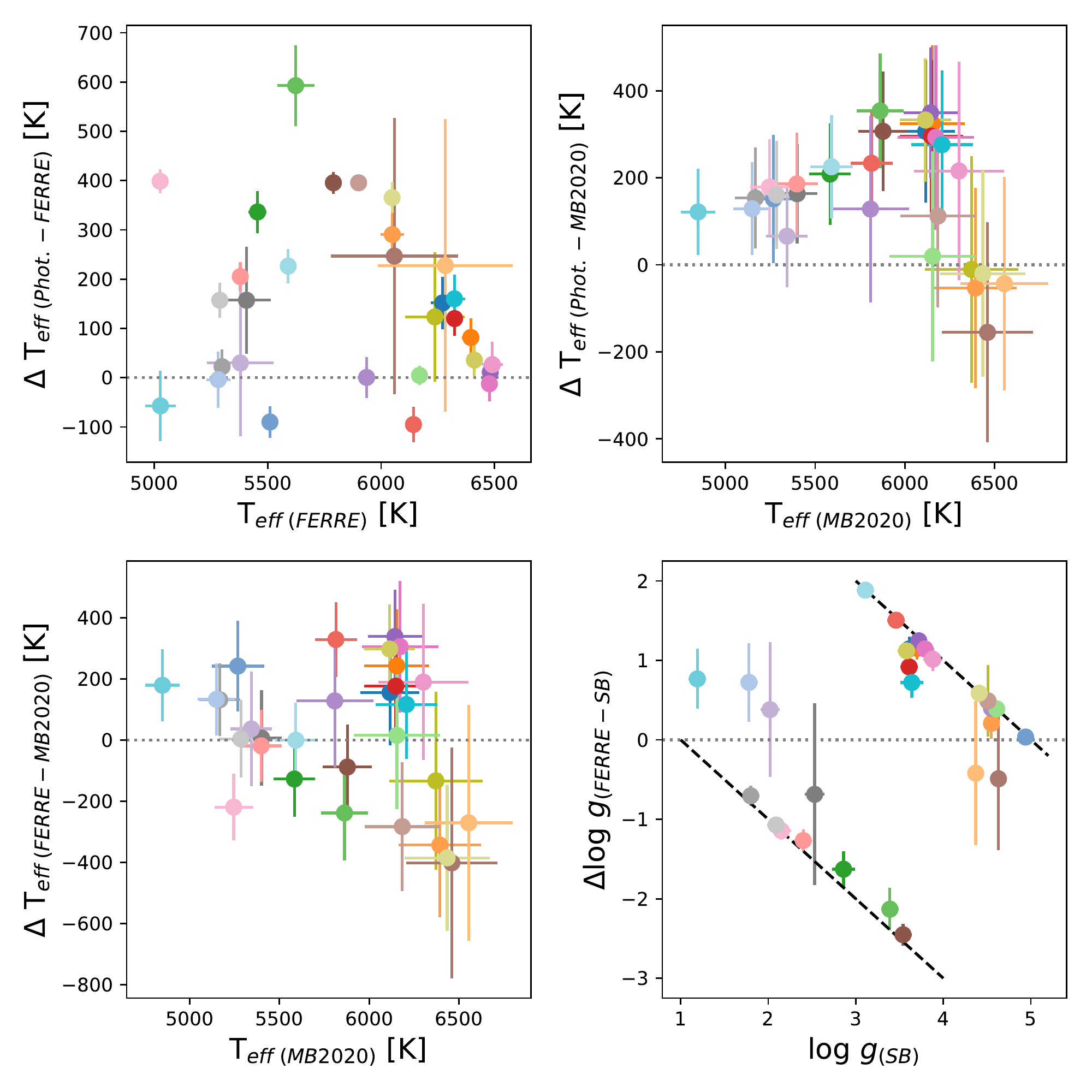}
    \caption{Comparison of \teff and \logg derived from SDSS photometry, medium-resolution spectroscopy, and our method of using the \citet{MB2020} calibration with the Stefan-Boltzmann law (see text). Dashed black lines of fixed log\,$g_{\rm FERRE} = 1.0$ and 5.0 (the bounds of the FERRE grid) are shown in the bottom right panel.  See Fig.~\ref{fig:gaia_isochrone} for star labels.
    \label{fig:stellar_params_compx}
    }
\end{figure}

%The targets in this paper are shown in Fig.~\ref{fig:gaia_isochrone}.  
A comparison of our stellar parameters  (Table~\ref{tab:graces_params_fe}) to those determined from our medium-resolution FERRE analysis (Table~\ref{tab:ferre_params}) are shown in the bottom row of Fig.~\ref{fig:stellar_params_compx}.
The FERRE stellar parameter errors are the FERRE MCMC fitting errors to the spectra. 
The SDSS photometric colour estimates for temperature (Table~\ref{tab:photometry}) are also shown for comparisons in the top row of Fig.~\ref{fig:stellar_params_compx}, where separate dwarf and giant solutions have been averaged together. %
We note that the MB2020 calibrations appear to be well constrained for EMP stars on the red giant branch and the main sequence (see their Fig. 2).   Nevertheless, for hotter stars, which are typically the main sequence or main sequence turn off stars, there are larger temperature uncertainties as seen in Fig.~4.  From tests in our MCMC analysis of temperature, we find this is simply due to a nearly constant uncertainty in the colours that scales to a slightly larger temperature uncertainty as the temperature increases.

Overall, the agreement between our temperatures from MB2020 and the FERRE MRS analysis are in good agreement ($\sigma$T$_{\rm eff} \sim 200$ K).  The SDSS photometric temperatures tend to be hotter than our temperatures or those from the FERRE MRS analysis, by $\Delta$\teff $\sim$200\,K.  Only the hottest stars show photometric and FERRE MRS temperatures that are cooler than ours. 
On the other hand, our values for \logg are more precise then those from the FERRE MRS analysis.
Due to the low SNR of the MRS, FERRE is forced to make a dwarf/giant distinction, forcing the gravities to values to be near the edges of their synthetic grid (\logg = 1.0 or 5.0).
Similar trends in gravity have been seen when comparing the FERRE MRS results to other high-resolution spectral analyses of EMP stars in our group \citep{Bonifacio19, Venn2020, Arentsen20b}. 

\begin{figure}
	\includegraphics[width=\columnwidth]{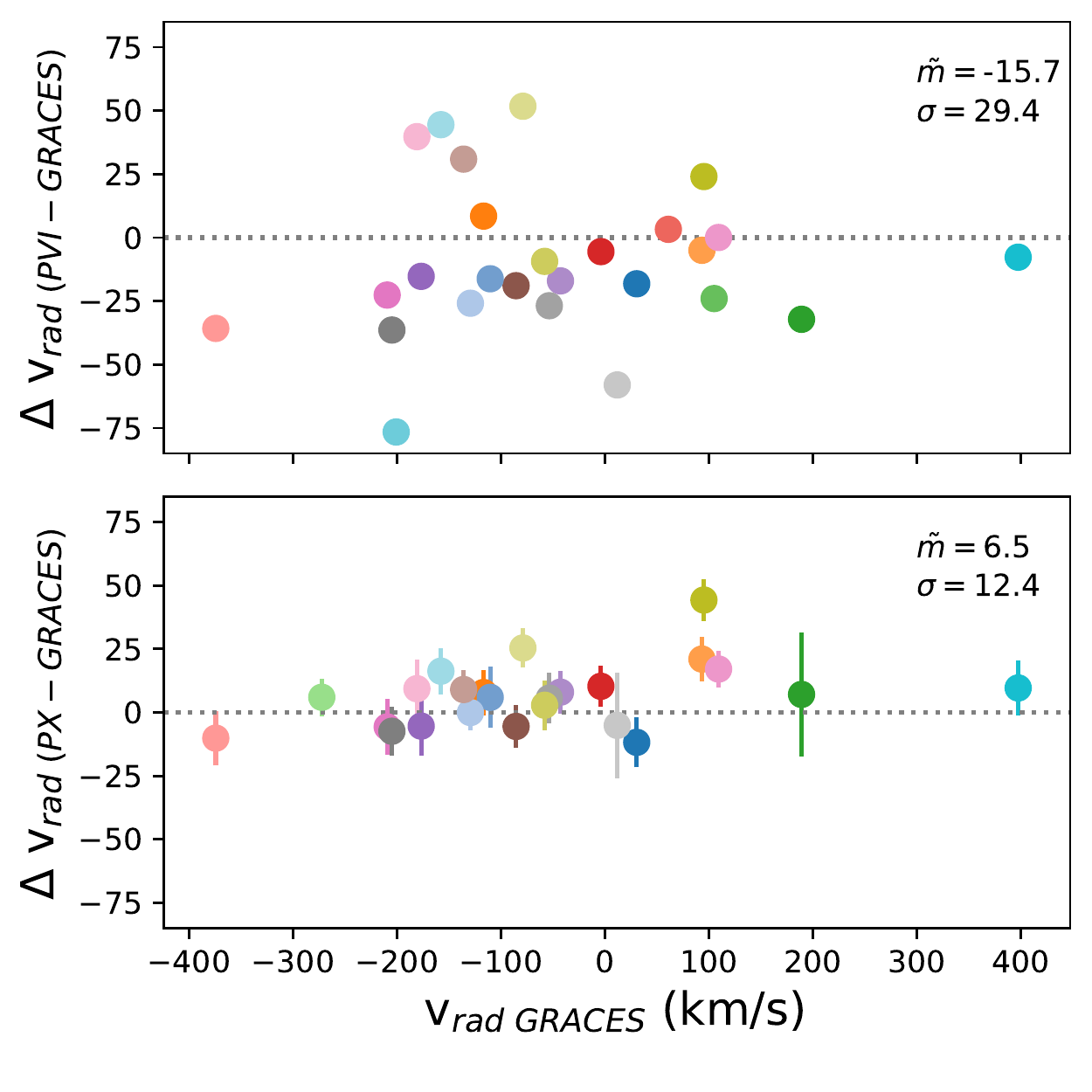}
    \caption{Comparison of the radial velocities (RVs) derived from our GRACES spectra and those from medium-resolution INT spectra.  The MRS radial velocities in the top panel are from \textit{Pristine} VI \citep[PVI][]{Aguado19Pristine}, and improved measurements with uncertainties are shown in the lower panel from  \textit{Pristine} X \citep[PX][]{Sestito19Pristine}.  The GRACES RV errorbars are smaller than the point sizes. The median offset ($\tilde m$) and standard deviation ($\sigma$), both in km/s, are shown in each panel. See Fig.~\ref{fig:gaia_isochrone} for star labels.}
    \label{fig:rv_comp}
\end{figure}

A comparison of radial velocity values from the medium-resolution spectra and our high-resolution GRACES spectra is shown in  Fig.~\ref{fig:rv_comp}.  There has been significant improvement in the radial velocity determinations between our initial MRS results \citep[in \textit{Pristine} VI,  ][]{Aguado19Pristine} and the most recent MRS analysis \citep[in \textit{Pristine} X,  ][]{Sestito19Pristine}.
The most recent MRS radial velocities show a mean offset of only $\Delta$(RV) $\sim 6$ km/s, with a dispersion of 12 km/s, relative to our HRS values.

\section{Chemical Abundances}

Chemical abundances are determined for each star using the stellar parameters discussed above and a classical model atmospheres analysis.  Model atmospheres were generated using the most up-to-date models on the 
MARCS website\footnote{MARCS model atmospheres at 
https://marcs.astro.uu.se/} \citep[][with additions by B. Plez]{Gustafsson08}; the OSMARCS spherical models are used when log\,$g<3.5$.
Spectral lines of iron were selected from the recent metal-poor stars analyses by \citet{Norris2017} and \citet{Monty2020}.
Atomic data was adopted from the \textit{linemake}\footnote{\textit{linemake} contains laboratory atomic data (transition probabilities, hyperfine
and isotopic substructures) published by the Wisconsin Atomic Physics
and the Old Dominion Molecular Physics groups. These lists and accompanying line list assembly software have been developed by C. Sneden and are curated by V. Placco at https://github.com/vmplacco/linemake.
} atomic and molecular line database. 
Chemical abundances are compared to the Sun using the \citet{Asplund09} solar data.

\begin{figure}
	\includegraphics[width=\columnwidth]{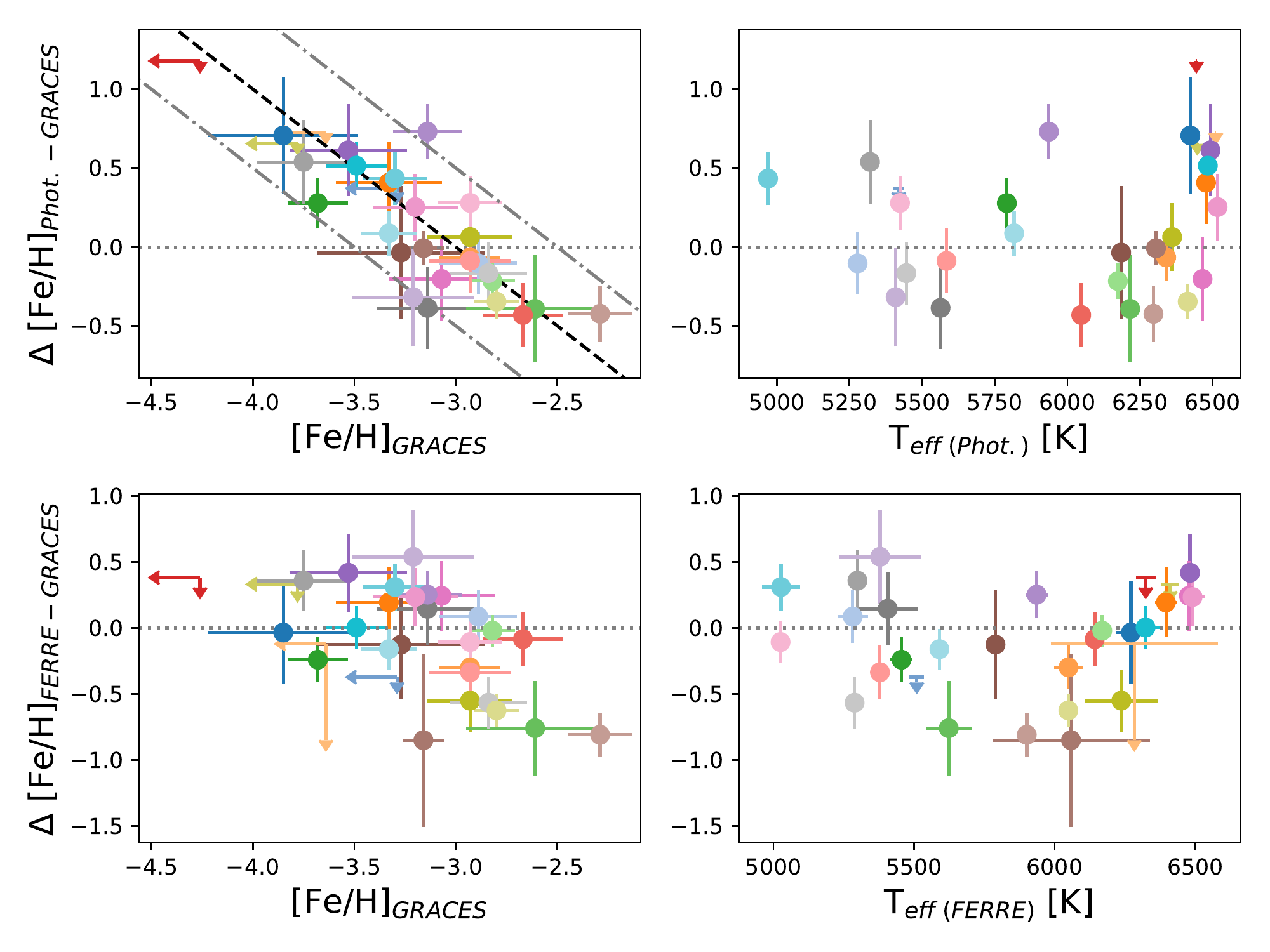}
    \caption{Comparison of [Fe/H] derived from photometry, MRS INT spectra, and our GRACES spectra.  Dashed lines of fixed [Fe/H] = $-2.5, -3.0$, and $-3.5$ are shown in the top left panel, which suggest that our Pristine survey photometric metallicities may be limited to [Fe/H] $\gtrsim -3$. Y-axes are shared between left and right panels.}
    \label{fig:fe_comp}
\end{figure}

\begin{figure*}
	\includegraphics[width=\linewidth]{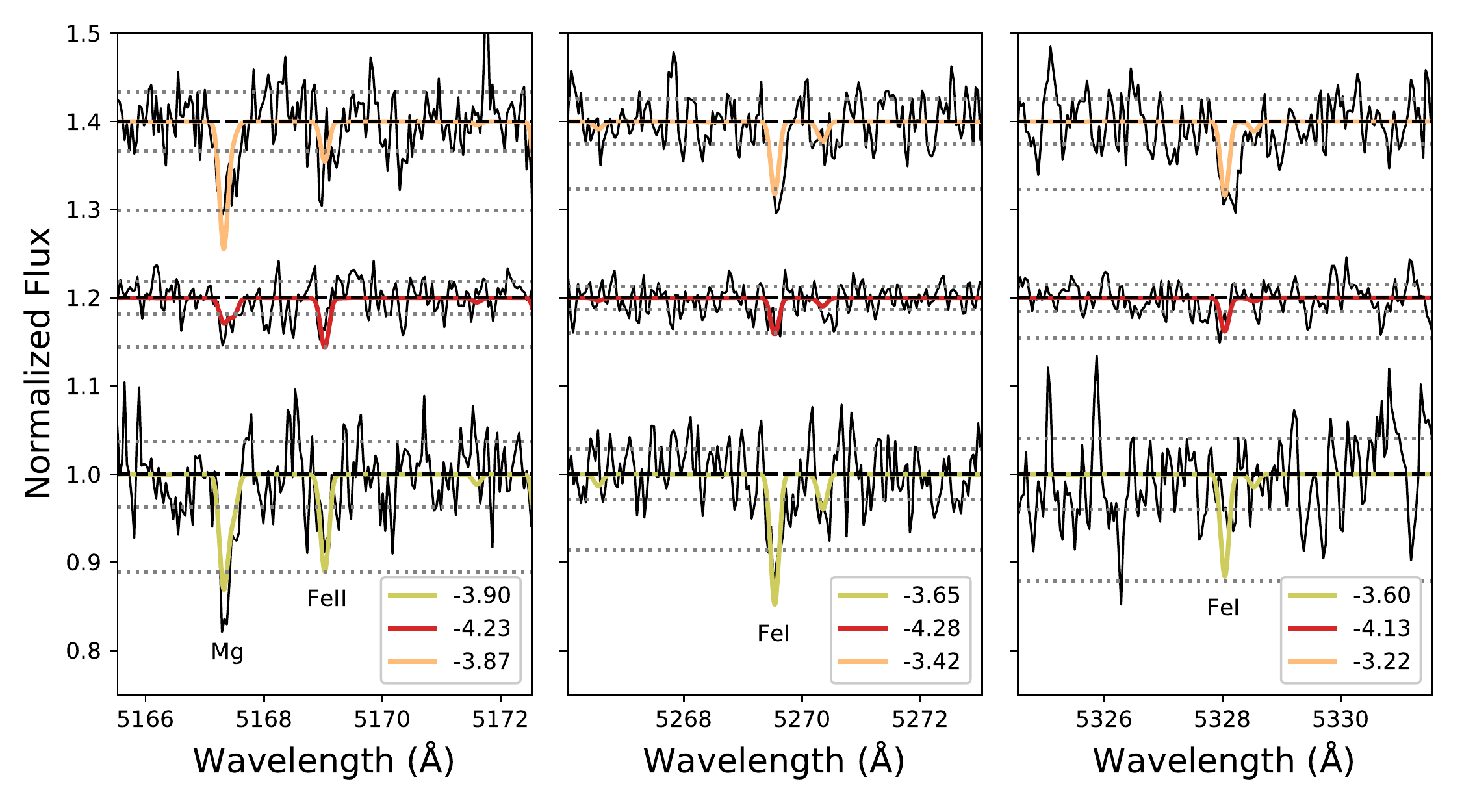}
    \caption{Synthesized Fe I and Fe II lines for the three stars with only Fe upper limits. Arranged top to bottom in order of hottest to coolest are P258+40 (\teff = 6554K), P237+12 (\teff = 6148K), and P246+08 (\teff = 6140K). Three Fe lines are shown (\ion{Fe}{II} 5169.028\AA, \ion{Fe}{I} 5269.537\AA, and \ion{Fe}{I} 5328.039\AA), chosen as they provide the tightest constraints on [Fe/H] for most stars in our sample.
    The GRACES spectra (solid black line) are compared to synthesized spectra (solid coloured lines that match the colour labels in Fig.~\ref{fig:gaia_isochrone} per star). The [Fe/H] abundances used in the synthetic spectra are noted in each panel.   The continuum placement (dashed black line) and the $\pm 1\sigma$ and $-3\sigma$ noise levels (grey dotted lines) are shown.}
    \label{fig:Fe_ULS}
\end{figure*}

Elemental abundances are calculated in a three step process: (1) the list of well known iron lines was examined in each spectrum for an initial iron abundance, and the metallicity of the model atmosphere was updated.  This process was run twice with the updated metallicities to ensure convergence.  (2) A new synthesis of all elements was generated including line abundances and upper limits for all of the clean spectral lines (e.g., see Figs. \ref{fig:Fe_ULS}, \ref{fig:P184_P207_MgNaCa}, \ref{fig:P184_P192_MgNaBa}).  (3) For the more challenging spectral features (i.e., those that were more severely blended or required hyperfine structure corrections), then a full spectrum synthesis was carried out using \textit{linemake} to find all features and components within $\pm$10 \AA\ of the feature of interest.
When a spectral feature was well fit, then the abundance was measured.  If not, a 3$\sigma$ upper limit was calculated.  Each synthetic spectrum was broadened in MOOG using a Gaussian smoothing kernel with FWHM=0.15, other than two stars (P198+08 and P016+28) which required more broadening (FWHM=0.25).
When there were multiple spectral lines for a given element, then the average of the measured abundances, or the lowest (most constrained) upper limit, was adopted.  The number of lines synthesized per element ranged from 1 (\ion{Eu}{II}) to 25 (\ion{Cr}{I}).  
Blends, isotopic corrections, and hyperfine structure corrections were taken from \textit{linemake} and included in the spectrum syntheses for lines of \ion{Li}{I}, \ion{O}{I} \ion{Sc}{II}, \ion{Mn}{I}, \ion{Cu}{I}, \ion{Zn}{I}, \ion{Y}{II}, \ion{Zr}{II}, \ion{Ba}{II}, \ion{La}{II}, and \ion{Eu}{II}.  A sample line list is available in Appendix~\ref{app:appendixC}; the full line list is available online.

Abundance errors are determined in two ways: (1) the line to line variations that represent measurement errors, e.g., in the continuum placement and/or due to the local SNR (see Tables \ref{tab:graces_params_fe} to \ref{tab:graces_neutron}), and (2) systematic uncertainties due to the stellar parameters (see Tables \ref{tab:graces_li} to \ref{tab:graces_Ba_sys}).  
The final abundance uncertainties are calculated by adding the line-to-line scatter ($\sigma$EW) in quadrature with the uncertainties imposed by the stellar parameter errors ($\sigma$T$_{\rm eff}$, $\sigma$log\,$g$, and $\sigma$[Fe/H]).  These final uncertainties are used in the abundance plots (i.e., Figures \ref{fig:fe_comp}, \ref{fig:fe_lg_comp}-\ref{fig:good_elems},  \ref{fig:ul_abunds}, \ref{fig:O_NLTE}).

The EMP standard star HD\,122563 has been analysed from an archival very high SNR ($>$200) GRACES spectrum.  We determined its stellar parameters using the same methods as for the GRACES targets, and find \teff = $4750 \pm89$~K, \logg = $1.12\pm0.11$, after adopting an initial metallicity of [Fe/H] = $-2.7$.  
These results are in fair agreement with the literature; e.g.,  \citet{Collet2018} revisit the analysis of this star with high resolution 3D model atmospheres.  They adopt  \teff= $4600\pm47$~K based on the IR flux method by \citet{Casagrande2014}, and they use the Stefan-Boltzman formula to determine surface gravity \logg= $1.61\pm0.07$, based on the available parameters at that time (e.g., they adopted a larger {\sl Hipparcos} parallax $\pi$'= $4.22\pm0.35$ mas vs the Gaia DR2 value of $\pi$'= $3.444\pm0.063$ with zero point offset of $-0.029$ mas).  
Their results are consistent with other photometric, interferometric, and spectroscopic methods, whereas ours are both hotter and brighter based on the new Gaia DR2 parallax measurement. 
Nevertheless, our abundance results for HD\,122563 are similar to those in the literature and to other Galactic standard stars (see Figs.~\ref{fig:good_elems}, ~\ref{fig:Li}, and \ref{fig:ul_abunds}.)

\subsection{Iron-group \label{sect:iron}}

Iron abundances are calculated from the average of the synthetic fits to each spectral feature identified in Table~\ref{tab:sample_lines}.  Iron is calculated from \ion{Fe}{I} and \ion{Fe}{II} independently, where [Fe/H] in Table~\ref{tab:graces_params_fe} is the weighted average.

A comparison of the iron abundances between our high-resolution GRACES spectra, the MRS analysis, and the \textit{Pristine} photometric estimates are shown in  Fig.~\ref{fig:fe_comp}.   In the top left panel, lines of constant metallicity are shown at [Fe/H] = $-2.5$, $-3.0$, and $-3.5$.   This plot suggests that the precision in the current \textit{Pristine} photometric metallicities may be limited to [Fe/H] $\ge-3$, particularly for the hotter stars that dominate our sample. 
%%The \textit{Pristine} team is working on complementary machine-learning methods to derive photometric metallicities, but it is yet unclear if these will result in better accuracy in the extremely metal-poor regime.
%
In the top right panel, our metallicities from the GRACES analyses suggest that the \textit{Pristine} photometric metallicities near [Fe/H]=$-3$ have an intrinsic dispersion of $\Delta$[Fe/H] $\sim\pm$0.5.  This dispersion appears to be even larger for the FERRE results, as seen in the bottom left panel.  The latter may also be temperature dependent, in that stars near T$_{\rm eff}$ (FERRE) $= 6000$\,K result in [Fe/H]$_{\rm MRS}$ (FERRE) metallicities that are too low, as seen in the bottom right panel.

Samples of the spectrum synthesis of three Fe lines (and one Mg line) are shown in Fig.~\ref{fig:Fe_ULS} for three stars: P258+40 (\teff = 6554 K), P237+12 (\teff = 6148 K), and P246+08 (\teff = 6140 K).   Note that
P237+12 is the lowest metallicity star in our sample, and we determine an upper limit of [Fe/H] $\le-4.26$, despite a very high SNR spectrum (SNR$\sim$95 at 6000~\AA). 
This is partially due to the limited (red) wavelength region available in the GRACES spectra and lack of strong \ion{Fe}{II} lines in this wavelength region.  
Examination of all three spectral lines shown in Fig.~\ref{fig:Fe_ULS} provide the mean 1DLTE 3$\sigma$ upper limit for P237+12.   A follow-up VLT-UVES spectrum of this star confirms this low metallicity from more and bluer iron features (Lardo et al., {\sl in prep.}). 

\begin{figure}
	\includegraphics[width=\columnwidth]{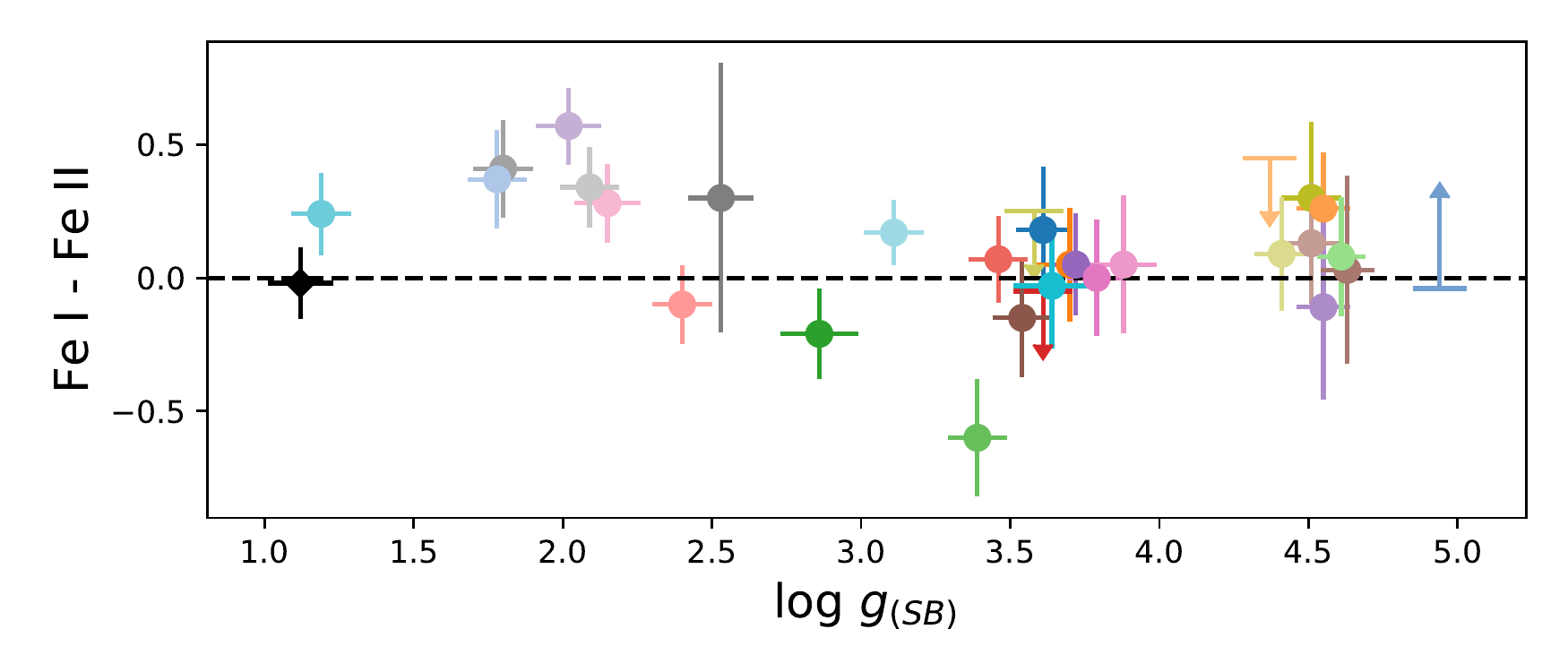}
	\includegraphics[width=\columnwidth]{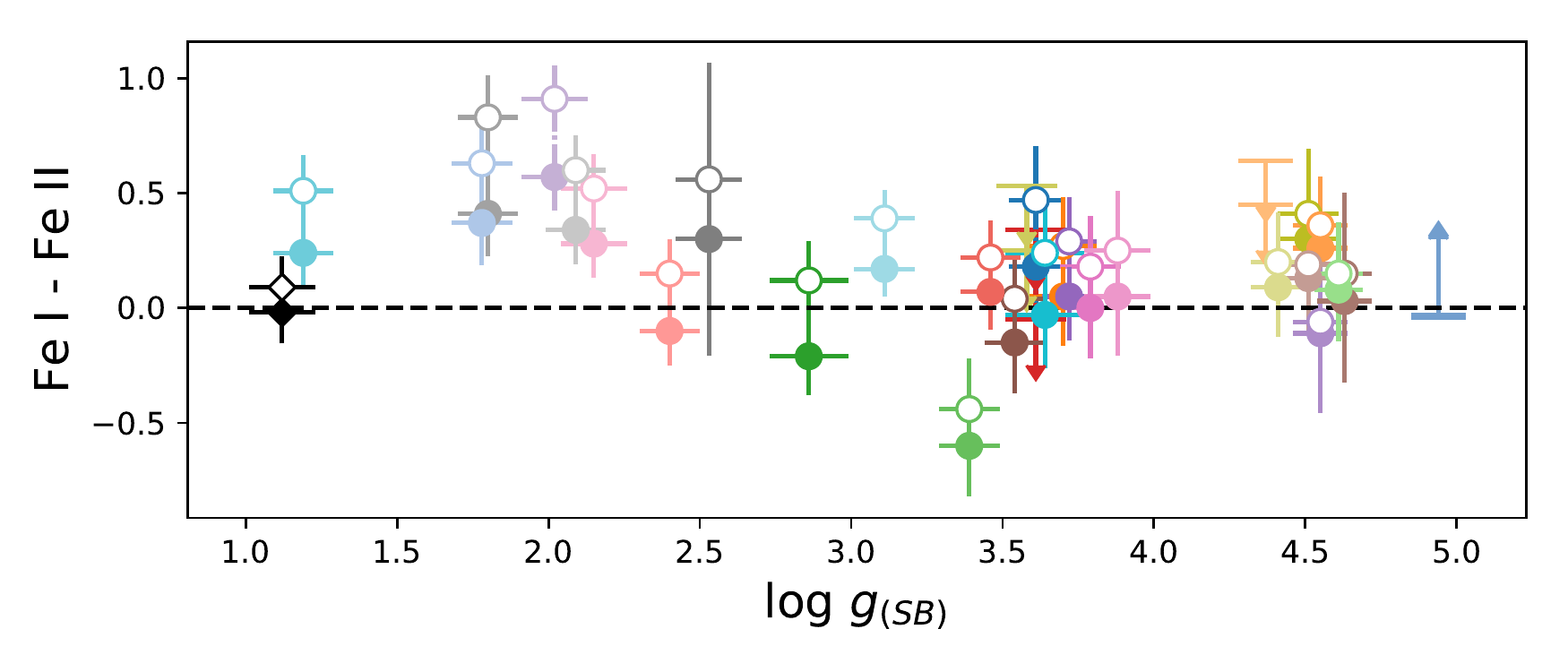}
    \caption{Comparison of [Fe I/H] - [Fe II/H] vs. surface gravity derived from the Stefan-Boltzmann law.  Upper panel are 1DLTE abundances, and lower panel includes NLTE corrections (open circles; see text). [Fe II/H] is lower than [Fe I/H] for a majority of the stars. The dashed black line at [Fe I/H] - [Fe II/H] = 0 represents ionization equilibrium - a spectroscopic check for the accuracy of the surface gravity. Reference star HD 122563 is shown as the black diamond.
    }
    \label{fig:fe_lg_comp}
\end{figure}

Departures from LTE are known to overionize the \ion{Fe}{I} atoms due to the impact of the stellar radiation field on the level populations, particularly in hotter stars and metal-poor giants. These non-LTE
(NLTE) effects can be significant in our stellar parameter range.
To investigate the impact of NLTE corrections on the iron abundances, the [Fe~I/H]$-$[Fe~II/H] differences are compared to the surface gravities in Fig.~\ref{fig:fe_lg_comp}.
We find that FeI and FeII are in excellent agreement for the dwarfs and subgiants (\logg $\gtrsim2.5$), whereas FeII tends to be lower than FeI for the giants (by $>$1$\sigma$).
This is {\it not} the typical signature of neglected NLTE effects.  NLTE corrections are examined from two databases; (1) INSPECT\footnote{INSPECT NLTE corrections are available at http://inspect-stars.net.} provides NLTE corrections for some of our \ion{Fe}{I} and \ion{Fe}{II} lines \citep{Bergemann2012, Lind2012}, and (2) the MPIA NLTE grid\footnote{MPIA NLTE corrections are available at http://nlte.mpia.de.} which provides NLTE correction for several additional lines \citep{Bergemann2012, Kovalev2019}.  The NLTE corrections for our \ion{Fe}{II} lines are negligible throughout.
However, when the \ion{Fe}{I} NLTE corrections are applied (bottom panel), the differences increase such that the giants show $<$[\ion{Fe}{I}/H]$_{\rm NLTE}-$[\ion{Fe}{II}/H]$> \sim +0.5$.
The average NLTE corrections per star are summarized in Table~\ref{tab:graces_params_fe}.  

Offsets in stellar parameters due to enhancements of certain chemical elements have been predicted to affect the isochrones for old, metal-poor stars \citep{Vandenberg2012}.  Specifically, enhanced [Mg/Fe] or [Si/Fe] by +0.4 relative to scaled-solar has been predicted to move a 12 Gyr isochrones with [Fe/H]=$-2$ to lower temperatures and higher gravities.  These changes in \teff and \logg would impact the \ion{Fe}{I} and \ion{Fe}{II} line abundances, and move \ion{Fe}{I} closer to \ion{Fe}{II}, providing a possible resolution to our results.
It is not clear how/if this effect is incorporated into the MB2020 calibrations, but regardless, this chemical effect will make any calibrations of the stellar parameters of old, metal-poor upper RGB stars subject to larger uncertainties.   
We note that our upper RGB stars do show a significant range in [Mg/Fe]$_{\rm RGB}$ = +0.1 to +0.8, with $\sigma$[Mg/Fe] = 0.1 to 0.4.
We also notice that the offsets from \ion{Fe}{I} = \ion{Fe}{II} are small for our stars on the lower RGB and seem to rise on the upper RGB, consistent with the predictions from the chemically-dependent isochrone analysis.
Regardless of these observations, we do not attempt a purely spectroscopic analysis of the stellar parameters for our RGB stars since we have very few \ion{Fe}{II} lines ($<5$) in these GRACES spectra.  

\begin{figure*}
	\includegraphics[width=\textwidth]{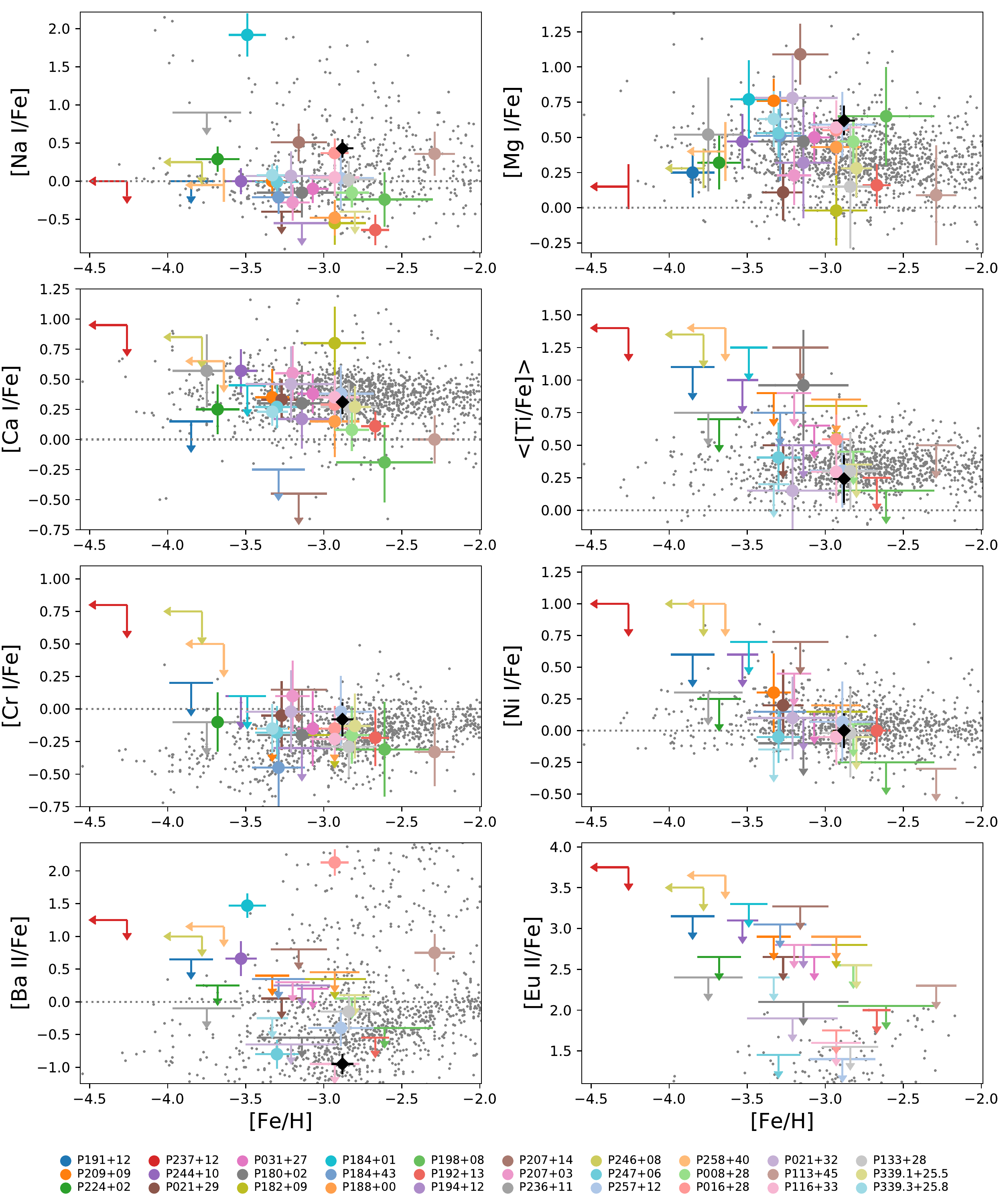}
    \caption{Elements with good spectral line detections, and \ion{Eu}{II} . $<$[Ti/Fe]$>$ is the average of [Ti I/Fe] and [Ti II/Fe] when a detection was made for both species, otherwise it is the highest upper limit between the two species. Our analysis of the GRACES spectrum for HD 122563 is shown as the black diamond. The Galactic reference star abundances are taken from the literature  \citep{Venn04, Aoki13, Yong13, Roederer14, Frebel14}.
    \label{fig:good_elems}
    }
\end{figure*}

The other iron-group elements Cr and Ni were also synthesized, but only 1-4 spectral lines of each element are available in our GRACES spectra. 
% \ion{Cr}{I} 5208.4 and \ion{Ni}{I} 5476.  
Their abundances are provided in Table~\ref{tab:graces_light} and shown in Fig.~\ref{fig:good_elems}.  When Cr or Ni are calculated, their 1DLTE abundances are in excellent agreement with Fe, which is similar to other metal-poor stars in the Galaxy.   
Examination of the NLTE corrections suggests that [Cr/Fe] may be increase by up to $\sim0.4$. This would be important for the detailed interpretation of the chemistry of each star; however, we do not include the NLTE effects in Fig.~\ref{fig:good_elems} since the Galactic comparison stars are not also corrected.

\subsection{Carbon \label{carbon}}
As the GRACES spectra are restricted to redder wavelengths, we do not have any carbon features to analyse.
However, [C/Fe] is determined in the FERRE analysis of our MRS from the G-band and features below 4300 \AA.  In Table~\ref{tab:ferre_params}, we identify our targets that were reported to have C-enhancements by \citet{Aguado19Pristine}.
A total of eight of the 30 stars in our sample 
were identified by FERRE to have [C/Fe] $>+1.0$ (P016+28, P021+29, P113+45, P133+28, P184+01, P188+00, P224+02, P339.1+25.5).  
Unfortunately, the FERRE pipeline struggles to determine C when the SNR of the MRS is low ($\lesssim 25$) in the temperature range of our targets.
%, which is the case for most of our targets.
%\citep[see][and Table \ref{tab:ferre_params}]{Aguado19Pristine}.
%
%The FERRE results can also be degenerate between carbon and surface gravity, and the temperature is difficult to determine when the SNR is quite low - all of which further complicate any interpretations of C from the MRS of our targets.
%
Closer inspection of the MRS shows that five of these are likely upper limits to non-existent G-bands.
%(at least at the SNR of our MRS spectra and the temperatures of those targets).  

Only three stars appear to have noticeable G bands in the MRS spectra, and in each of these cases the FERRE analysis found them to be very carbon-rich, with [C/Fe]$ > +2$ (P016+28, P184+01, P224+02).  Only P016+28 is flagged to have a reliable carbon abundance.  The other two do appear to be C-enhanced, but their specific [C/Fe] values are quite uncertain.   
%, and five stars are slightly carbon-rich with [C/Fe]$ > +1$ (P021+29, P113+45, P133+28, P188+00, P339.1+25.5).  
The slight changes in our stellar parameters from the MB2020 calibrations/SB law are also likely to affect the final C abundances.

In summary, we find the C abundances for most of our targets are not sufficiently reliable from the MRS analysis; nevertheless, one star is clearly C-rich (P016+28), and two others are very likely C-enhanced (P184+01, P224+02).
%Another 7 targets are within 1$\sigma$ of the CNO-cycle corrected upper-limit for C-enhancement of $+0.7 <$ [C/Fe] $< +1.0$.

\subsection{Alpha elements}\label{alpha}

Alpha elements in extremely metal-poor stars form primarily though He-capture during various stages in the evolution of massive stars, and dispersed through SNe II events.  Thus, the [$\alpha$/Fe] ratio is a key tracer of the relative contributions of SN II to SN Ia products in a star forming region.
In this paper, the $\alpha$-elements include Na, Mg, Ca, and Ti.  We have included Na because it typically scales linearly with Mg in metal-poor stars in the Galaxy \citep[e.g.,][]{Pilachowski1996, Venn04, Norris2013}.  We have also included Ti since it too scales with other $\alpha$-elements even though the dominant isotope $^{48}$Ti forms primarily through Si-burning in massive stars \citep[e.g.,][]{Woosley2002}.

\begin{figure}
	\includegraphics[width=\linewidth]{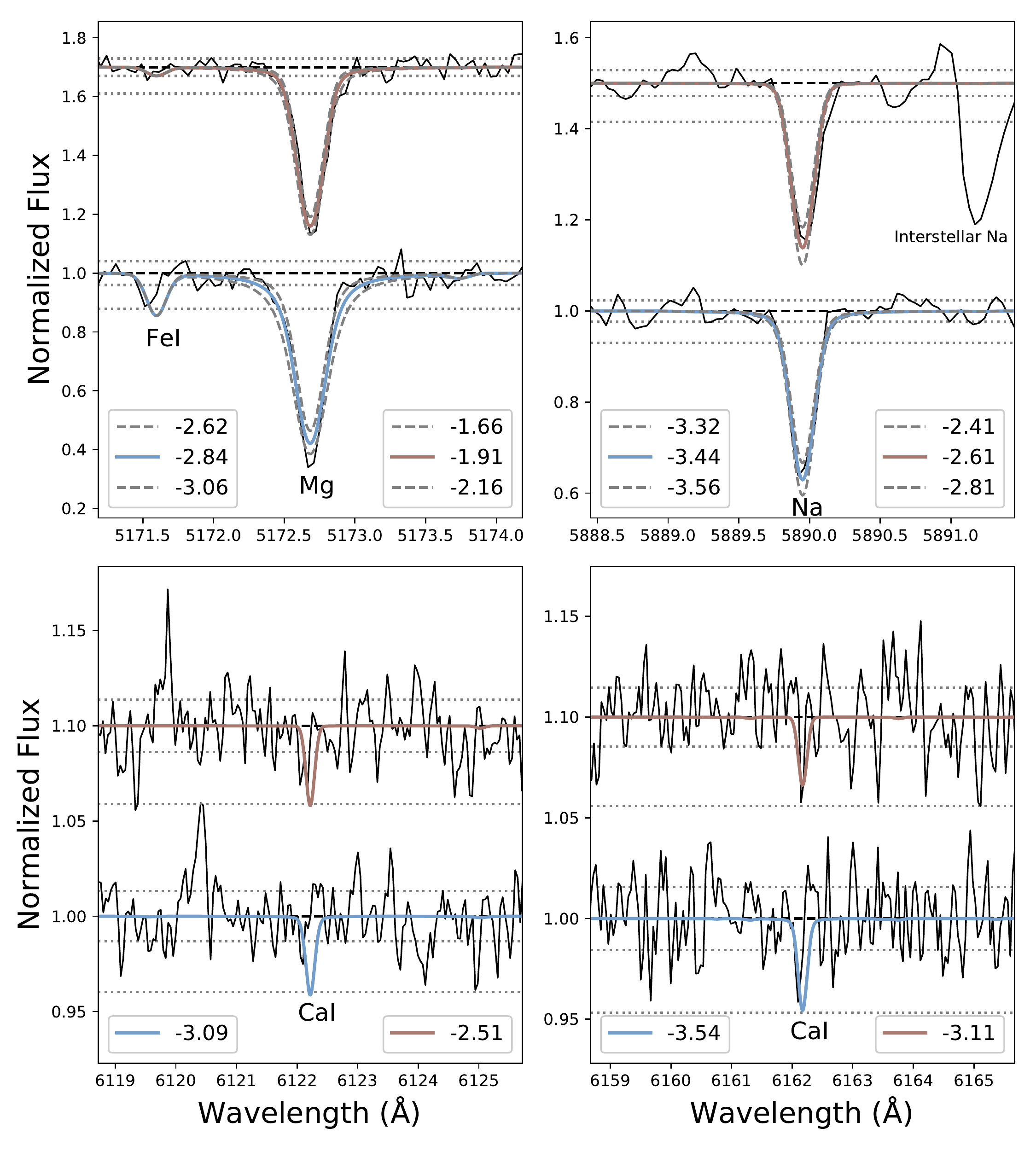}
    \caption{Spectrum syntheses for two stars near [Fe/H]$=-3$ that are notably Mg-rich, but Ca-poor.  Syntheses of the Mg I 5172.68\AA\ and Na I 5889.95\AA\ (top panels), and two Ca I lines (6122.21\AA, 6162.17\AA; bottom panels) are shown, for P207+14 (brown, top) and P184+43 (blue, bottom). These absorption lines provide the best abundances or the tightest constraints for these two stars. 
    The GRACES spectra are the solid black lines, and the synthesized spectra are the solid coloured lines with the [X/H] measurements in the legend. Spectra are offset for clarity.  Dashed grey lines represent the $\pm 1\sigma$ synthesis. The dashed black line represents the continuum placement and the dotted grey lines are $\pm 1\sigma$ and $-3\sigma$, where $\sigma$ is defined as the measured scatter in the continuum.
    % See Fig.~\ref{fig:P184_P192_MgNaBa} for additional label information.
     }
    \label{fig:P184_P207_MgNaCa}
\end{figure}

Na abundances or upper limits are from the two strong \ion{Na}{D} lines.  These lines are clear and present in most of the stars in our sample, and easily separated from any interstellar lines. NLTE effects can be significant for these resonance lines, however corrections in this analysis are small (ranging from $\Delta$(Na) = $-0.1$ to $-0.2$ for most stars), as shown in Table~\ref{tab:graces_nlte}.
Mg is from 2-4 lines of \ion{Mg}{I} in all stars, even P237+12 which has only an iron upper-limit.  NLTE corrections are small (typically $\Delta$Mg $\le+0.2$).
Ca is from 1-10 lines of \ion{Ca}{I} in most stars, or the \ion{Ca}{I} 6122.217 and \ion{Ca}{I} 6162.173 lines are used to estimate an upper limit (e.g., see Fig.~\ref{fig:P184_P207_MgNaCa}). NLTE corrections are moderate (typically $\Delta$Ca $\le+0.3$).  The \ion{Ca}{II} triplet is also examined, however we do not use those results in our analysis (especially without NLTE corrections).
Ti is from 1-4 lines of \ion{Ti}{I} and 1-6 lines of \ion{Ti}{II}. For many stars, upper limits only were available and estimated from \ion{Ti}{I} lines.   The average NLTE corrections for \ion{Ti}{I} can be large ($\Delta$Ti $\le+0.6$), however the NLTE corrections to \ion{Ti}{II} are negligible.
Again, we do not include the NLTE corrections in Fig.~\ref{fig:good_elems} since abundances for the Galactic comparison stars are not also corrected.

\begin{figure}
	\includegraphics[width=\linewidth]{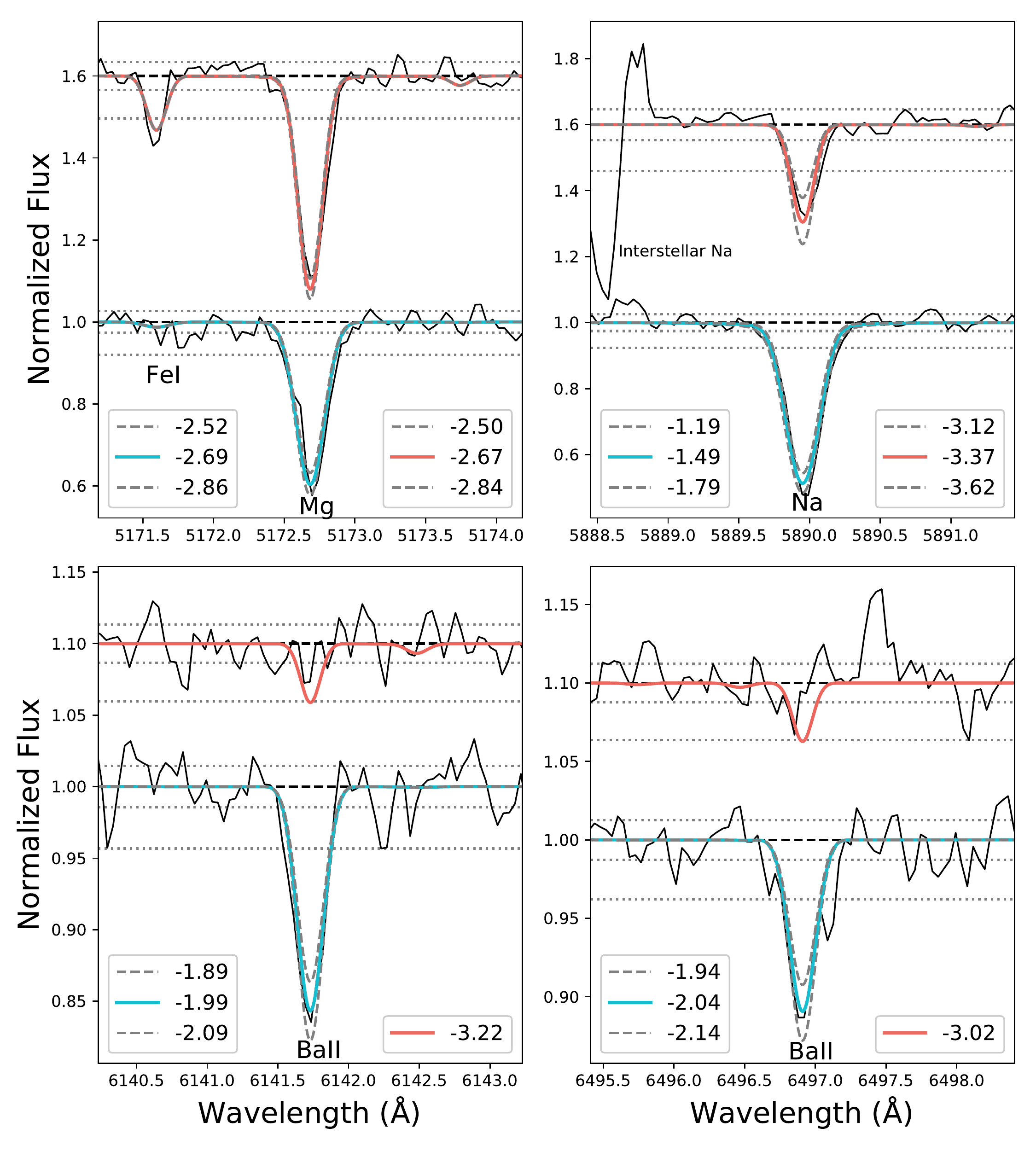}
    \caption{Spectrum syntheses for one Ba-rich star (P184+01, blue) and one Ba-poor star (P192+13, red).  Our spectrum syntheses for the two Ba II lines (6141.73\AA, 6496.91\AA) are shown (bottom panels), and our best fit syntheses for Mg I 5172.68\AA\ and Na I 5889.95\AA\ (top panels) for comparison.  See Fig.~\ref{fig:P184_P207_MgNaCa} for additional label information.   
%The GRACES spectrum is the solid black line, and the synthesized spectra are the solid colored lines with the [X/H] measurements in the legend. Dashed grey lines represent the $\pm 1\sigma$ synthesis. The dashed black line represents the continuum placement and the dotted grey lines are $\pm 1\sigma$ and $-3\sigma$, where $\sigma$ is defined as the measured scatter in the continuum.
        }
    \label{fig:P184_P192_MgNaBa}
\end{figure}

The majority of these newly discovered extremely metal-poor stars show 1DLTE abundances that are within 1$\sigma$ uncertainties of the Galactic comparison stars, especially given our results for the EMP standard star, HD~122563 (black point in each abundance plot).  Only a handful of stars have $\alpha$-element abundances that are statistically lower than the Galactic comparison stars.
These include; 
\begin{enumerate}
% \item
% The three most metal-rich stars in the sample (P113+45, P192+13, P198+08), all show scaled-solar ratios of [Ca/Fe], rather than the higher plateau value near [Ca/Fe]=+0.4. This Ca-depletion may be supported by low [Na/Fe], but not (clearly) by [Mg/Fe], though  NLTE corrections may lower these ratios. To demonstrate that these low abundances are real, we show the observed and synthetic spectra around the \ion{Mg}{I} 5172.68 and \ion{Na}{I} 5889.95 lines in P192+13  in
% Fig.~\ref{fig:P184_P192_MgNaBa}.  

\item
Two stars with [Fe/H] $<-3$ (P184+43 and P207+14), and a third star near [Fe/H] $=-2.6$ (P198+08), show sub-solar [Ca/Fe]. This is an unusual abundance signature when compared with the Galactic halo sample, especially as all three are enriched in [Mg/Fe] ($>+0.5$).  Sample spectra and our 1DLTE synthesis for lines of \ion{Mg}{I}, \ion{Na}{I}, and \ion{Ca}{I} are shown in Fig.~\ref{fig:P184_P207_MgNaCa}.  NLTE corrections are small and do not affect these trends (see Table~\ref{tab:graces_nlte}). 
This abundance pattern has been seen in only a few EMP stars \citep[see][]{Sitnova2019},
and is discussed further below.

\item
One star near [Fe/H] $=-3$ (P182+09) shows a low, solar-like [Mg/Fe] value, yet high values of [Ca/Fe] $\sim+0.8$. This pattern has been seen for stars in dwarf galaxies, and is typically attributed to an effectively truncated upper IMF, loss of gas from high mass supernova events, and/or inhomogeneous mixing of supernova yields in the dwarf galaxy's interstellar medium \citep[e.g.,][]{Tolstoy09, McWilliam13, Kobayashi2015, FN15}.
% We notice that [Na/Fe] is also low in two of these stars (P182+09 and P021+28).

\item
One star near [Fe/H] $=-3.5$ (P184+01) shows an enrichment in [Na/Fe].  This star will be discussed further below (see Section~\ref{CEMP}).

% particularly one star that was highlighted by \citet{Cohen2007}, HE1424-0241, which has a similarly high [Mg/Ca] = +0.83, mostly driven by deficient Ca, but at a slightly lower metallicity [Fe/H]$=-4$.
% %
% This abundance pattern is interpreted as contributions to stellar Mg and Ca abundances from only a small number of SN~II explosions, i.e., where the nucleosynthetic yield for explosive alpha-burning nuclei like Ca was very low compared to that for the hydrostatic alpha-burning element Mg, 
% %
% \citet{Sitnova2019} note that this is a rare type of star, comprising $<$10\% of stars with [Fe/H] < −3 and that do not reveal carbon enhancement.

\end{enumerate}

Oxygen is examined from the \ion{O}{I} 7770~\AA\ triplet feature, however this resulted in upper limit abundances for most of our stars (see Appendix~\ref{app:appendixA}.  Oxygen could be measured in only six stars, and NLTE corrections were applied (see Table~\ref{tab:graces_nlte}.  
Three stars (P116+33, P198+08, and P339.1+25.5), showing [O/Fe] $\sim +0.5$, consistent with [Mg/Fe] after oxygen NLTE corrections are applied.  
Three other stars (P184+01, P207+14, P224+02) show [O/Fe] $> +1.5$, which is much higher than the other $\alpha$-elements in those stars.   We note that two of these stars are CEMP (Section~\ref{carbon} and the third is one of our high [Mg/Ca] stars.    
The other CEMP and high [Mg/Ca] stars in our sample have high [O/Fe] {\it upper limits} only that do not constrain oxygen.  
These oxygen abundances support our identifications above that these are stars of special interest.

\subsection{Neutron-capture elements}

Elemental abundances or upper limits are determined for six neutron capture elements: Y, Zr, La, Nd, Ba, and Eu.   These formed in massive stars and core collapse supernovae through rapid neutron capture reactions, and those other than Eu also form via slow neutron captures during the thermal pulsing AGB phase in intermediate-mass stars.   
The specific details and yields from these nucleosynthetic processes is a dynamic field of current research.  For the core collapse SNe, new models and calculations of their yields include details of the SN explosion energies, explosion
symmetries, early rotation rates, and metallicity distributions
\citep[e.g.,][]{Kratz2007, Nishimura2015, Tsujimoto2015, Kobayashi2020}, as well as exploration of contributions from compact binary mergers as a (or as the most) significant site for the r-process \citep[e.g.,][]{Fryer12, Korobkin2012, Cote16, Emerick2018}.
Similarly, predictive yields from AGB stars by mass, age, metallicity distributions, and details of convective-reactive mixing are also an active field of research \citep[e.g.,][]{Lugaro2012, Cristallo2015, Pignatari2016}.

Only Ba and Eu are discussed in this section.  For all stars, we have 1-2 \ion{Ba}{II} lines (6141.73 \AA, 6496.91 \AA).  Hyperfine structure and isotopic splitting are taken into account using the atomic data in \textit{linemake}.  For Eu, the GRACES spectra only permit studies of the \ion{Eu}{II} 6645 \AA\ line, which is too weak to be observed in any of our spectra.  The upper limits for Eu from this line are also too high to be scientifically useful in testing for pure r-process enrichment in these stars.  It would be important to examine the much stronger \ion{Eu}{II} 4129 \AA\ line to constrain the pure r-process contributions in these stars.

Two stars appear to be Ba-rich (P184+01 and P016+28); as seen in Fig.~\ref{fig:good_elems}, both appear to have [Ba/Fe] $\sim +1$ with [Fe/H] $\lesssim -3$. The spectrum synthesis of the two \ion{Ba}{II} lines in P184+01 is shown in Fig.~\ref{fig:P184_P192_MgNaBa}, where it is clear that these lines are clear, strong, and well measured.  
NLTE corrections were calculated for these two Ba-rich stars following the methods in \citet{Mashonkina1999} and \citet{Mashonkina2019}, using models representing their specific stellar atmospheres and high LTE Ba abundances. The NLTE corrections for P184+01 increase [Ba/Fe] by 0.09 dex, when averaged between the individual corrections for each of the two lines (6141 and 6496\AA).
Alternatively, [Ba/Fe] decreases by 0.44 dex when NLTE corrections are calculated for P016+28.
Including these corrections, the NLTE Ba abundances are still significantly higher than the 1DLTE results in similar EMP stars in the Galaxy.

Six of our stars appear to be Ba-poor (Fig.~\ref{fig:good_elems}).  The spectrum synthesis for P192+13 shows that the \ion{Ba}{II} 6141 and 6497\AA\ lines are not present  (Fig.~\ref{fig:P184_P192_MgNaBa}.  Using the 3-$\sigma$ line depth, we find upper limits of [Ba/Fe] $<-0.4$ near [Fe/H] $=-2.6$.   
For our six Ba-poor stars, the [Ba/Fe] abundances and upper limits are within the lower envelope of Ba abundances for stars in the Galaxy.

% , but more common in ultra faint dwarf galaxies, such as Segue~I, Com~Ber, and Hercules \citep[e.g.,][]{Koch13, Frebel14, Ji16}.

The other neutron capture elements with spectral features in the GRACES wavelength regions are examined in Appendix \ref{app:uls}.  Only upper limits could be determined, and they did not provide useful scientific constraints.   

\begin{figure}
	\includegraphics[width=\columnwidth]{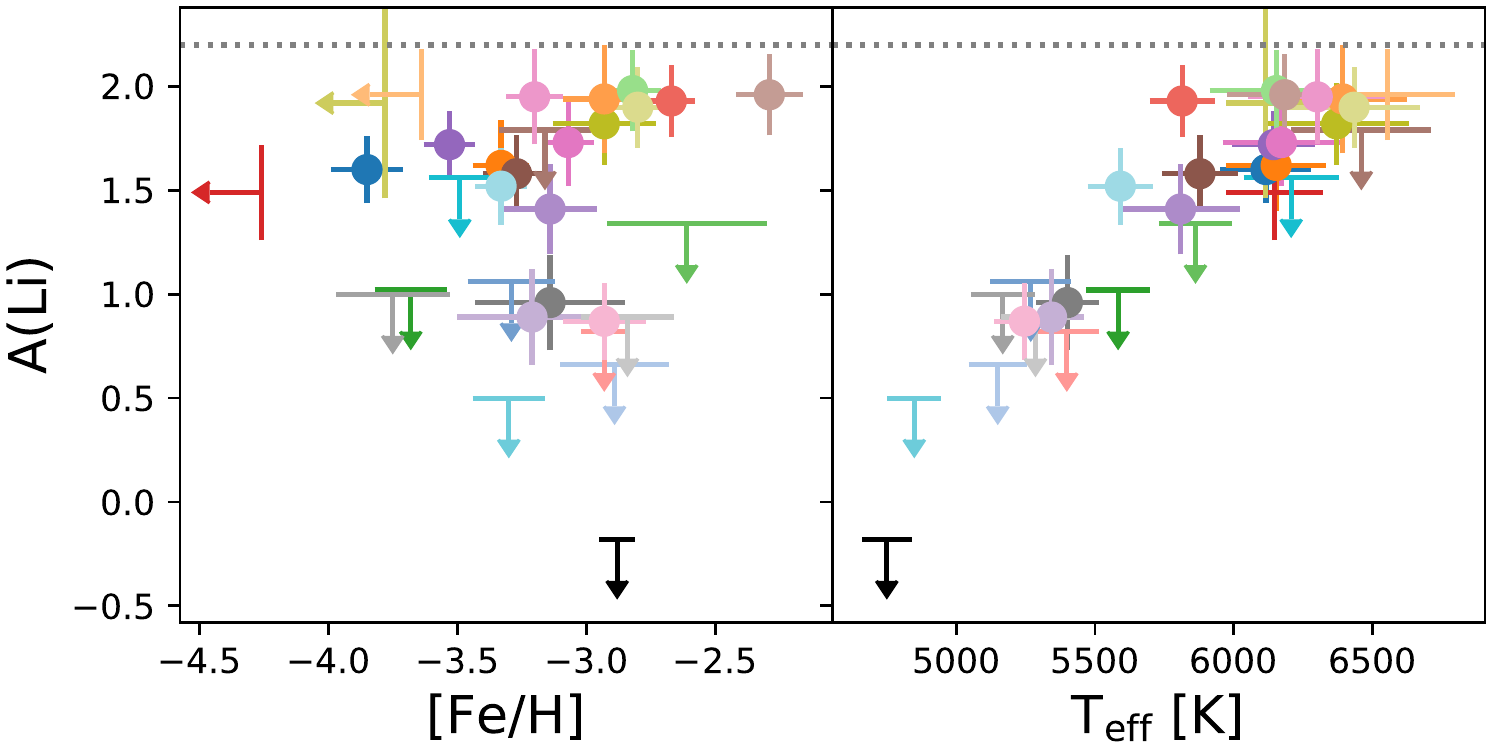}
    \caption{Lithium abundances from \ion{Li}{I} 6707 \AA.  Stellar labels are the same as in Fig.~\ref{fig:good_elems}. 
    The Spite plateau \citep{Spite82, Sbordone2010} is shown as the dashed gray line. 
    \label{fig:Li} }
\end{figure}

\subsection{Lithium}

The \ion{Li}{I} 6707 spectral line is present in the spectra of over half of our targets.   Our spectrum syntheses for Li included hyperfine structure and isotopic splitting, with atomic data taken from \textit{linemake}.   
The study of lithium in EMP stars is an active topic of discussion due to the links between EMP stars and the chemistry of the early Universe.
The cosmological lithium problem refers to the discrepancy between the amount of Li predicted from Big Bang nucleosynthesis \citep[A(Li) = 2.67 to 2.74;][]{Cyburt16, Coc2017} and the highest Li abundances measured in the atmospheres of unevolved metal-poor stars \citep[A(Li) $\sim2.2$, the Spite Plateau;][]{Spite82, Bonifacio2007a, GonzalezHernandez2008, Aoki09, Sbordone2010}.
Finding unevolved EMP stars with detectable Li provides strong constraints on the lithium problem.

A majority of our sample shows expected trends between A(Li), \teff, and metallicity (Fig. \ref{fig:Li}). Most of the hotter and higher metallicity stars are found near the Spite plateau. As metallicity decreases, a higher degree of scatter in A(Li) is observed, consistent with the meltdown of the Spite plateau observed by \citet{Sbordone2010, Bonifacio2012}. 
The RGB stars in our sample show lower lithium abundances (or upper limits), as expected since Li is destroyed through surface convection in cooler stars \citep{Spite82}.
Five stars, however, are notable: P191+12, P224+10, P237+12, P246+08, P258+40. P191+12 and P224+10 have metallicities of [Fe/H] $=-3.85$ and $-3.68$, respectively, and the other three stars only have metallicity upper-limits with [Fe/H] $<-3.5$, but all four have detectable Li at the 3$\sigma$ level. Their measured A(Li) places them at the Spite plateau
%with A(Li) = $1.92\pm0.22$ for P237+12 ([Fe/H]$<-3.88$), $2.08\pm0.36$ for P46+08 ([Fe/H]$<-3.60$), and $2.04\pm0.20$ for P258+40 ([Fe/H]$<-3.52$) 
(see Table \ref{tab:graces_li}). 
Similar in \teff, \logg, metallicity, and A(Li) to the primary star of the spectroscopic binary CS22876-032 \citep{GonzalezHernandez2008,GonzalezHernandez2019}, these main-sequence and turn-off stars are excellent candidates for follow-up studies related to the cosmological lithium problem.
% The slightly depleted value for Li in our hotter stars is also consistent with slight depletions during early (or pre-) main sequence evolution.  These results are consistent with comparisons of solar twins to the Sun \citep[e.g.,][]{Thevenin2017, Melendez2020}.

\section{Discussion}

Using the chemical abundances determined in this paper, we discuss our EMP stars in terms of the accretion history and chemical evolution of the Galaxy.  

\subsection{New CEMP candidates \label{CEMP}}

In this paper, we have analysed three
stars with slight to large carbon enhancements.
%, but please note that these abundances might change with updated stellar parameters (see Section~\ref{carbon}). 
As their precise [C/Fe] results are quite uncertain, we regard these stars simply as carbon-enhanced metal-poor (CEMP) candidates, rather than confirmed CEMP stars.
At the lowest metallicities, stars are often found to be enhanced in carbon \citep{Beers92, Norris1997}, typically comprising 40\% of the EMP stars 
%and an even higher percentage at lower metallicities 
\citep[see][]{yong2013CEMP, Lee13, Placco14}, though recently those percentages have been lowered through considerations of the carbon 3DNLTE corrections \citep{Norris2019}.

Different types of CEMP stars have been identified and defined by \citet{BC05}, where the two main classes are the
CEMP-s stars, which show additional enhancement in s-process elements (such that [C/Fe] $>+0.7$ and [Ba/Fe] $>+1.0$), and the CEMP-no stars, which do not show any s-process enhancements.  
These initial definitions have been further refined by \citet{Yoon16}, based on the trends observed between [Fe/H] and [C/Fe]. 
The C excess in CEMP-no stars is generally attributed to nucleosynthetic pathways associated with the very first stars to be born in the universe \citep{Iwamoto2005, Meynet2006}.
%``Group I" represents the stars with a large degree of scatter in A(C) vs. [Fe/H], and is primarily attributed to the CEMP$-s/rs$ stars (though CEMP-no stars are still found in this population), while ``Group II" and ``Group III" contain the majority of the CEMP-no stars.
%Stars in Group II typically have lower carbonicities than Group III stars and show a clear trend between A(C) and [Fe/H], while Group III stars are found at the lowest metallicities and have carbonicities clustered around A(C)$\sim 0.6$ \citep{Yoon16}.

Of our three new CEMP candidates, we find that two (P016+28
%P113+45, 
and P184+01) are enriched in barium, with [Ba/Fe] $>+1$ (to within their 1$\sigma$ uncertainties), which suggests that they may belong to the CEMP-s sub-class.
%and most likely associated with Group I.
Examining P016+28 further, 
%if we accept the carbonicity 
the carbon abundance from FERRE is [C/Fe] $=+2.42$, for an absolute carbon abundance of A(C) $=7.76$.  If this carbon abundance is accurate, this star would be amongst the high-C/Group I population in \citet[][A(C) =$7.96\pm 0.42$]{Yoon16}.
Meanwhile, P184+01 has [Fe/H] $=-3.49$ and a very uncertain carbon abundance of A(C) $=7.26$, which places it between Groups I, II, and III. 
Another way to test the CEMP-s hypothesis for these two stars is through radial velocity monitoring, as most CEMP-s stars are found in a binary systems \citep[e.g.,][and references therein]{Lucatello05, Hansen16s, Starkenburg14}. 
This property has contributed to the theory that CEMP-s stars have received their carbon and s-process enhancements through mass-transfer with an asymptotic giant branch (AGB) star in a binary system \citep{Abate13}.
We do not search for radial velocity variations in our data though, since the GRACES spectra were rarely taken over several epochs, and the medium-resolution INT spectra do not have sufficient precision.

The remaining CEMP-no candidate, P224+02, has a low barium upper-limit.
% (ranging from $-1.0<$[Ba/Fe]$<+0.8$). 
As this star has [Fe/H] $=-3.68$ (and a tentative A(C) = 7.29), it falls between the A(C)-metallicity groups in \citet{Yoon16}.
%then they are most clearly identified with the Group I stars by \citet{Yoon16}.  
Groups II and III are dominated by CEMP-no stars, but recent analysis of Group I CEMP stars has shown that 14\% are CEMP-no \citep{Norris2019}. While generally low in s-process elements, the Group I CEMP-no stars also show higher [Sr/Ba] abundances than the majority of Group I CEMP-s/rs stars.  To explain the [Sr/Ba] ratios, \citet{Norris2019} speculated that Group I CEMP-no stars may experience some mass exchange with massive AGB stars or rapidly-rotating "spinstars" in  binary systems (both produce less s-process material).
% some of which may be in binary systems and experience mass exchange with massive AGB or highly-rotating "spinstars".  The latter is significant as spinstars are not expected to produce significant s-process elements \citep{Meynet2006, Frischknecht2010, Frischknecht2012}.  Nevertheless, the Group I CEMP-no 
Unfortunately, we do not determine [Sr/Fe] as the \ion{Sr}{II} lines are at blue wavelengths, not reached by our GRACES spectra.   
Alternatively, we examine the $\alpha$-elements (Na, Mg, Ca) since some CEMP-no stars can show enrichments in these elements; P224+02 shows normal halo abundances for these elements.
As discussed by \citet{Meynet2015}, the predictions for these elements depend on mixing in massive stars though, and the predictions can vary widely. 
\citet{FN15} showed that enhanced alpha-elements only occur in about half of their CEMP-no sample, and even less when [Fe/H] $\lesssim-3$.

The exact origin(s) of the CEMP-no stars is not yet clear, however it has also been proposed that CEMP-no stars may form in dwarf galaxies, and thereby may be associated with accreted systems \citep{Yuan2020, Limberg2020}.  To examine this further, we compute the orbits of our 30 GRACES stars below.

% One of the newly discovered EMP stars (P133+28) is a CEMP-no candidate. It shows C-enrichment [C/Fe]$\sim+2$ and [Ba/Fe]$>1.5$, at [Fe/H]$<-2.8$, and is dynamically similar to Gaia-Sequoia.   
% Two EMP stars (P016+28, P184+01) are CEMP-s candidates.  They show both C and Ba enrichments.  Neither show interesting dynamical properties within the MWG potential.
% One star (P113+45) is Ba-rich, where [Fe/H]=-2.20 $\pm$0.12, and [Ba/Fe]=+0.78 $\pm$0.20.  This star is not carbon rich, though, therefore is not  CEMP-s.  The upper limits for all other heavy elements are quite high, and provide no further constraints.  It appears to be a normal halo star.

\begin{figure}
	\includegraphics[width=\columnwidth]{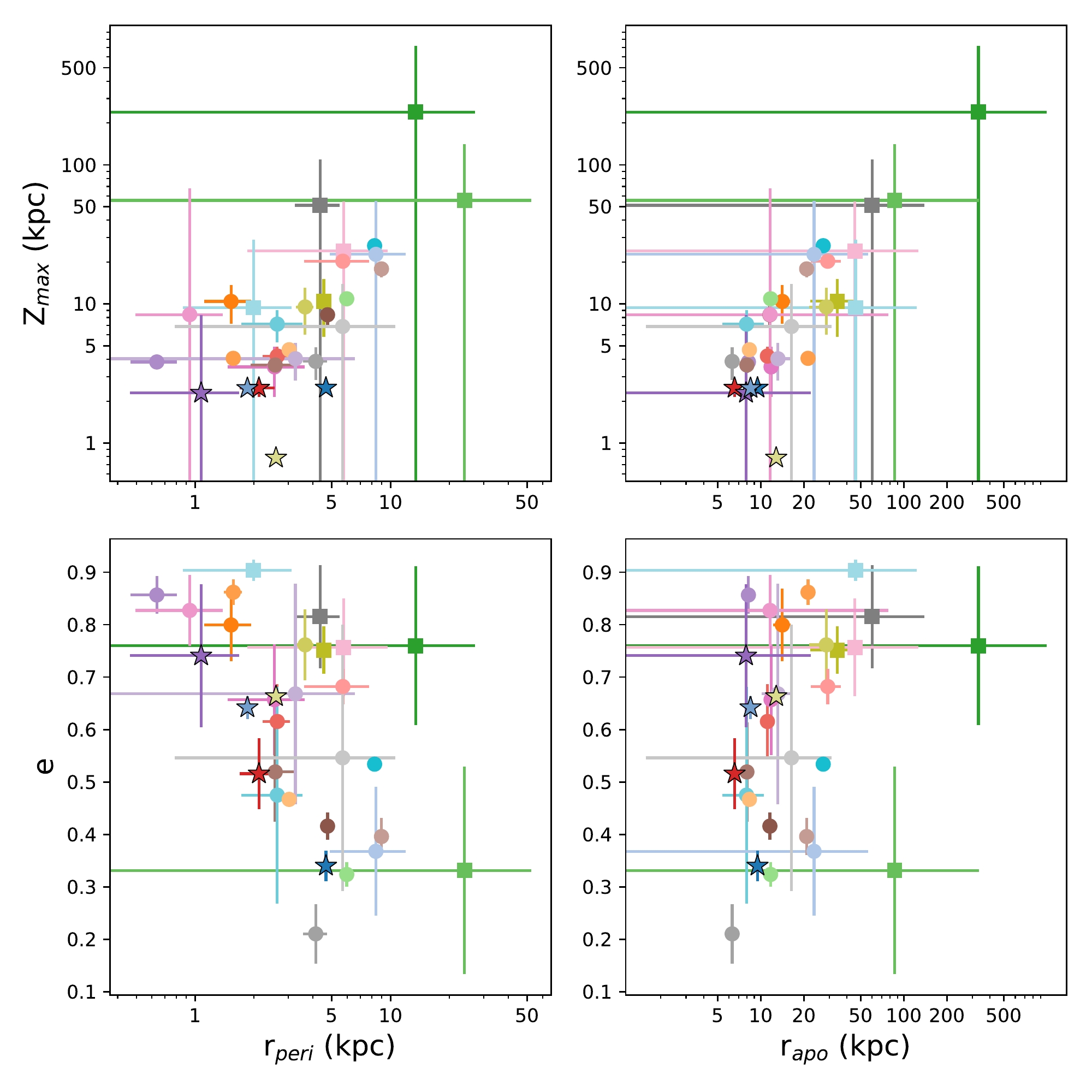}
    \caption{Orbital elements for stars from 2018A-2019B datasets. Star symbols represent stars in planar disk orbits, squares are stars that reach the outer Galactic halo.
    \label{fig:Orbits}  }
\end{figure}

\begin{figure*}
	\includegraphics[width=\textwidth]{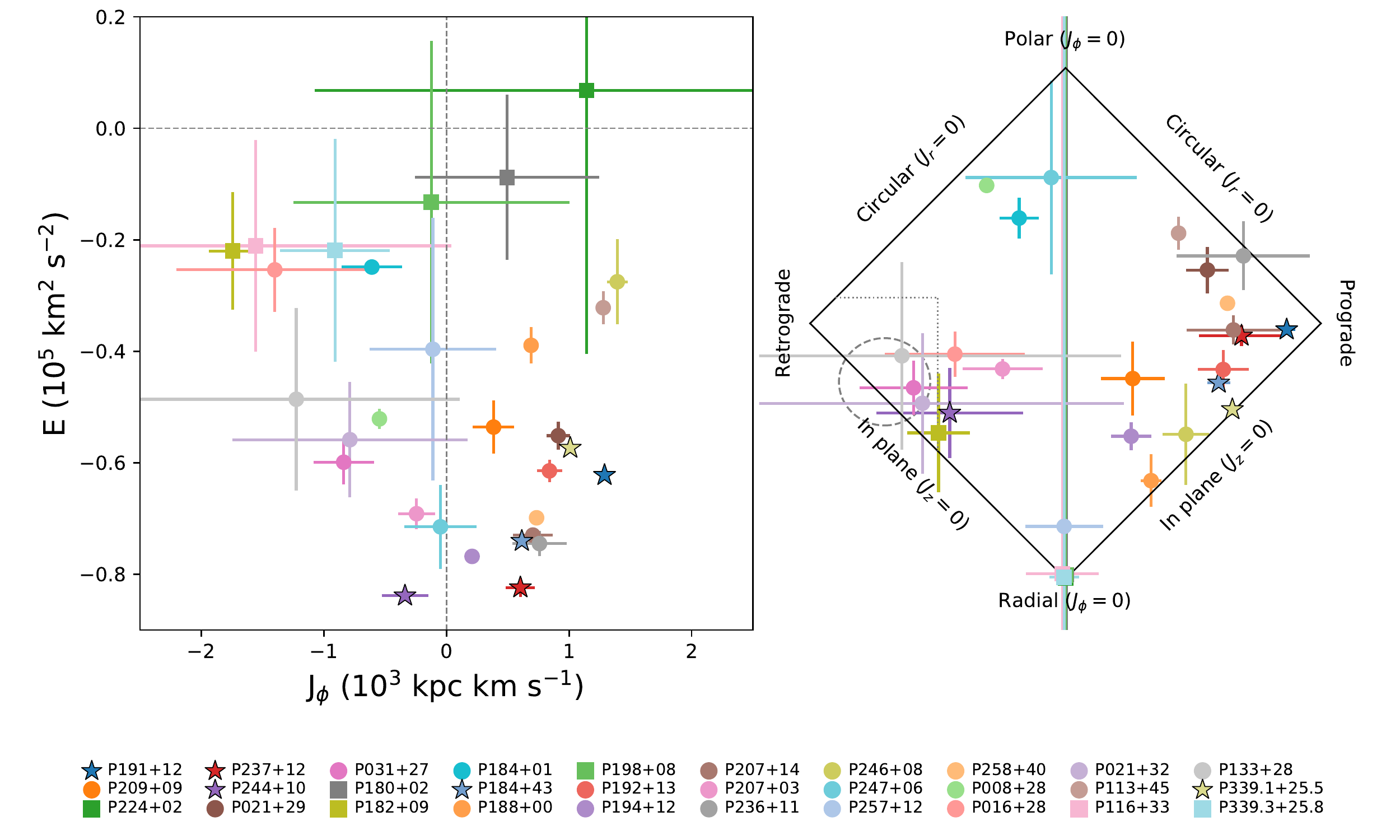}
    \caption{Action vectors and energies for stars from 2018A-2019B datasets. Same symbols as in Fig.~\ref{fig:Orbits}. The dotted grey box in the right panel is the dynamical cut for Gaia-Sequoia from \citet{Myeong19} ($e\sim$0.6 with $J_{\phi}/J_{tot} < -0.5$ and $(J_z - J_R)/J_{tot}<0.1)$ and the dashed grey ellipse represents the mean dynamical properties of the associated Gaia-Seqouia stars from \citet{Limberg2020}.
    \label{fig:Action} }
\end{figure*}

\subsection{Other stars with interesting chemistries \label{Others}}

Highlighted in Section \ref{alpha}, three stars (P184+43, P198+08, and P207+14) show very low [Ca/Fe] ($\leq-0.3$) and large [Mg/Fe] ($>+0.5$). This results in [Mg/Ca] $\gtrsim+0.8$.
This rare abundance pattern had been seen in only a few EMP stars, e.g., 
one star (HE1424-0241) at [Fe/H]=$-4$ was highlighted by \citet{Cohen2007} for its high [Mg/Ca] $=+0.83$, mostly driven by its low Ca.
This abundance pattern cannot be explained by uncertainties in the stellar parameters (including any possible systematic errors in the surface gravity, e.g., if these were on the horizontal branch); instead, it is interpreted as
contributions to stellar Mg and Ca abundances from only a small number of SN~II explosions, i.e., where the nucleosynthetic yield for explosive alpha-burning nuclei like Ca was very low compared to that for the hydrostatic alpha-burning element Mg \citep[see][]{Sitnova2019}.  These results are further supported by our high [O/Fe]$\_{\rm NLTE}$ values (see Appendix~\ref{app:appendixA}.

Two similar stars in the Hercules dwarf galaxy with [Mg/Ca] =$+0.58$ and $+0.94$ dex were studied by \citet{Koch08}. 
\citet{Koch08} argued that such high ratios can be attributed to enrichment from high mass ($\sim 35M_{\odot}$) Type II SNe, based on yields from \citet{Woosley1995}.
Furthermore, their chemical evolution models for Hercules-like dwarf galaxies indicate that the observed [Mg/Ca] ratios can only be reproduced in 10\% of the systems that are enriched by only a few (1-3) Type II events.
Clearly these are chemically unique objects which reflect the chemical evolution of rare environments.
\citet{Sitnova2019} also note that these exceptional stars comprise $<$10\% of stars with [Fe/H]$<-3$ and do not typically reveal carbon enhancement, consistent with our results.    

\subsection{Orbit calculations}

\textit{Gaia} DR2 proper motions and astrometry have dramatically accelerated the fields of Galactic Archaeology and near-field cosmology by providing the data needed to calculate the detailed orbits of nearby stars.  The stars in this study were also selected to have small parallax errors, and therefore precise distances.  When combined with the precision radial velocities from our high-resolution GRACES spectra, then we are able to estimate the orbits for these stars to within the accuracy of our assumptions on the MW potential. 

Orbital parameters for the stars in this paper are calculated with \textit{Galpy} \citep{Galpy}, using 
the parallax distances, our radial velocities from the GRACES spectra, and the \textit{Gaia} DR2 proper motions.
The \textit{MWPotential14}\footnote{This potential is three component model composed of a power law, exponentially cut-off bulge, Miyamoto
Nagai Potential disc, and Navarro, Frenk \& White (1997) dark matter halo.} was adopted, though a more massive halo of $1.2x10^{12}$\,M$\odot$ was chosen following \citet{Sestito19UMP}.
Errors have been propagated from the uncertainties in the proper motions, RVs, and distances via Monte-Carlo sampling of the Gaussian distributions of the input quantities.

The apocentric and pericentric distances 
(R$_{\rm apo}$ and R$_{\rm peri}$), 
perpendicular distance from the Galactic plane
(Z$_{\rm max}$), and eccentricity ($e$) of the calculated orbits for our stars are shown in Fig.~\ref{fig:Orbits}.
Following \citet{Sestito19UMP}, stars with R$_{\rm apo} <15$~kpc and |Z|$_{\rm max}<3$~kpc are considered to be confined to the Galactic plane, while stars with R$_{\rm apo}>30$~kpc are considered to be members of the outer halo.
The orbital energy (E) and action parameters 
(vertical J$_z$, azimuthal J$_{\phi}$) 
for the sample are also calculated with \textit{Galpy}, and shown in Fig. \ref{fig:Action}.  All targets appear to be bound to the Milky Way, to within their uncertainties. 

\subsection{Stars with Interesting Orbits}

Five of our new EMP stars (P184+43, P191+12, P237+12, P244+10, and P339.1+25.5) have distinctly planar-like orbits (|Z|$_{\rm max}<3$ kpc). All five have [Fe/H] $<-3$ and somewhat elliptical orbits ($e= 0.3$ to 0.7). 
P237+12 is the most metal-poor star in the GRACES sample with a 1DLTE metallicity upper-limit of [Fe/H] $<-4.26$. This star has Mg below the canonical MW halo plateau values, though the abundances, when available, are generally within the regime of normal Galactic halo stars. The implications of finding planar-like EMP stars, in the context of Galactic evolution, is discussed further in the following section.
P339.1+25.5 stands out dynamically as it has the lowest |Z|$_{\rm max}$ ($<1$\,kpc) of the stars in this sample, suggesting it may be coincident with the prograde Galactic thin disk.
However, its somewhat elliptical orbit ($e\sim0.7$) is uncharacteristic of typical thin disk stars.
%This is significant as this star stands out chemically; it is a new CEMP-no candidate.  Revised - its C is only an upper limit.

The star P184+43 with a prograde, planar orbit also has an unusual chemistry, as it is enriched in magnesium, [Mg/Fe] $=+0.6$, and yet has a very low upper-limit on calcium, [Ca/Fe] $<-0.2$. 
In  Section~\ref{Others}, we speculated that this peculiar abundance pattern may suggest a lack of contributions from lower mass stars and SN~Ia.  
If so, then this star could have formed in a dwarf galaxy that was accreted by the Milky Way at early times, before SN~Ia could enrich it.  
This speculation is discussed further below (Section~\ref{sec:MWG}).

Five high-energy, highly-retrograde stars (P016+28, P021+32, P031+27, P133+28, P182+09) appear to be dynamically related to the ``Gaia-Sequoia" accretion event \citep{Myeong19}.
Sequoia is the population of high-energy retrograde halo stars that are presumed to be associated with an ancient accretion event of a counter-rotating progenitor dwarf galaxy \citep{Myeong19, Matsuno2019, Monty2020, Cordoni2020, Yuan2020, Limberg2020}.
\citet{Myeong19} identifies stars with $e\sim0.6$, $J_{\phi}/J_{tot}<-0.5$, and $(J_z-J_R)/J_{tot}<0.1$ to be linked to Gaia-Sequoia.
These five stars also meet the more recent and stricter membership criteria by 
\citet[][see Fig. \ref{fig:Action}]{Limberg2020},
to within their errors.
We acknowledge that these simple action vector cuts do not account for background stars which are non-coincidental with Sequoia. \citet{Limberg2020} explored the effect of background contamination via a membership clustering algorithm and found that up to $\sim50\%$ of their Sequoia sample may indeed be contamination.
Due to the small size of our sample, we do not explore this further and only offer these stars as Gaia-Sequoia candidates to be confirmed/rejected in later studies.  We note that all five have [Fe/H] $\sim-3$, and normal halo [$\alpha$/Fe] ratios.
These add to the few extremely metal-poor stars now found associated with Gaia-Sequoia \citep[see also][]{Monty2020, Cordoni2020, Limberg2020}.

The action vectors computed for the orbit of P244+10 suggests that it may also be associated with Gaia-Sequoia; however, its low energy planar-like orbit is uncharacteristic of other Sequoia members.  P244+10 is more dynamically similar to stars associated with the Thamnos event \citep{Helmi2017, Koppelman2018, Limberg2020}. 
Thamnos is also believed to be a lower metallicity - higher $\alpha$-abundance population than Gaia-Seqouia \citep{Koppelman2018, Monty2020, Limberg2020}, and our chemical abundance study would support that.   We find P244+10 has [Fe/H] =$-3.5\pm0.3$, compared to the five Gaia-Sequoia targets (above) with $<$[Fe/H]$>=-3.0 \pm0.2$, and highlight that
P244+10 is an interesting star for follow-up investigations.

\begin{figure}
	\includegraphics[width=\linewidth]{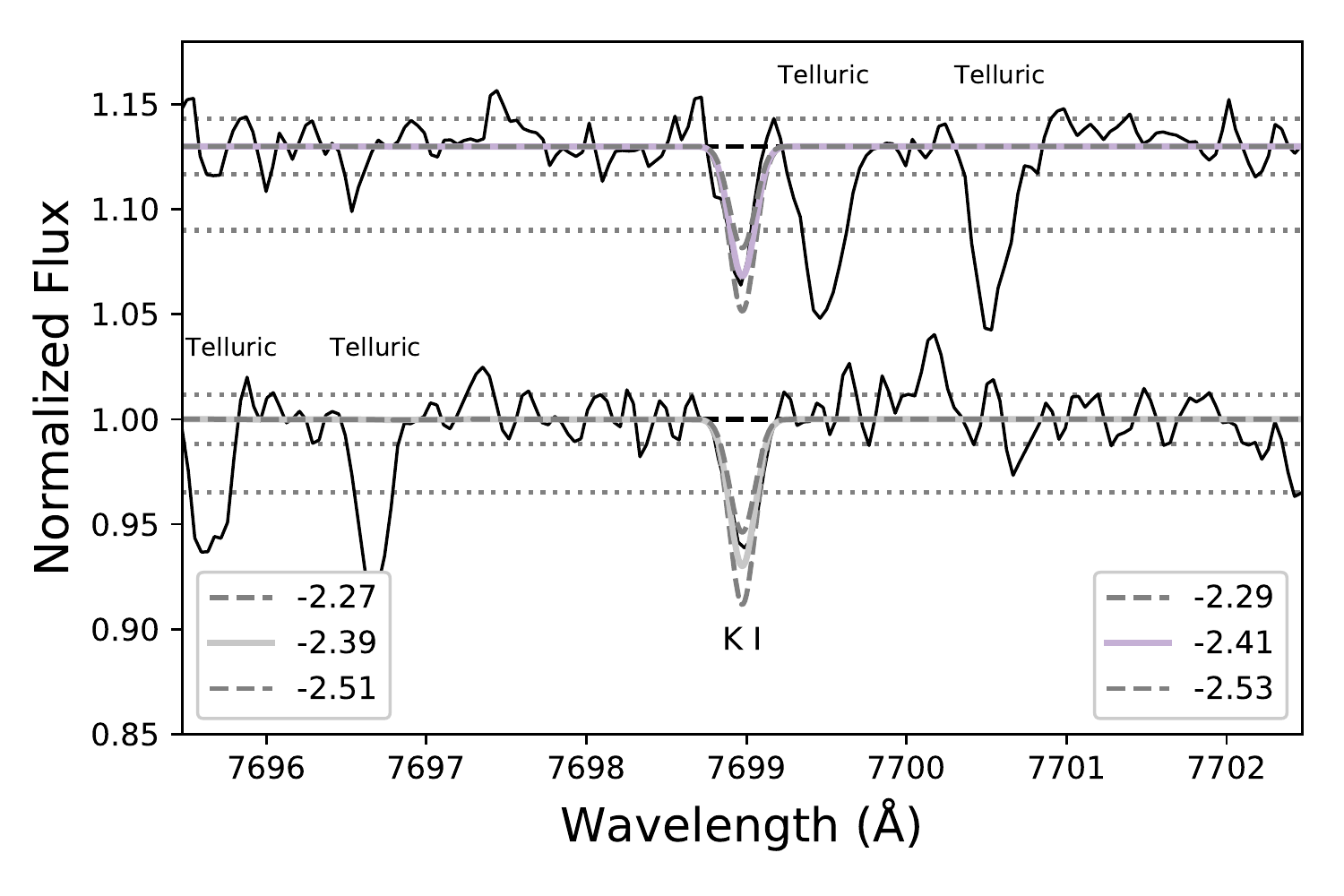}
    \caption{Syntheses of \ion{K}{I} at 7698.97 \AA. P133+28 (light grey, bottom) and P021+32 (light purple, top). Both objects are K-rich (e.g. Fig. \ref{fig:ul_abunds}) and appear to be dynamically related to the Gaia-Sequoia accretion event (e.g. Fig. \ref{fig:Action}). Telluric absorption lines are identified, however telluric subtraction was not performed as no spectral standards were observed in tandem with these science acquisitions. See Fig. \ref{fig:P184_P192_MgNaBa} for additional label information.
        }
    \label{fig:P133_P021_K}
\end{figure} 

\subsection{Formation and Early Chemical Evolution of the Galaxy} \label{sec:MWG}

This work is consistent with the results by \citet{Sestito19UMP} who found that a large number of ultra metal-poor stars in the literature are on planar orbits in the Galaxy.  We have found five new EMP stars on planar-like orbits (or 16\% of our sample).  Four are on prograde orbits and one (P244+10) is on a retrograde orbit. These ratios are comparable to our earlier results from an analysis of CFHT spectra for 115 metal-poor candidates which resulted in 16 very metal-poor stars ([Fe/H]$<-2.5$) confined to the Galactic plane (R$_{\rm apo}<15$~kpc, Z$_{\rm max}<3$~kpc), including one on an extremely retrograde orbit \citep{Venn2020}.   
In comparison with the larger \textit{Pristine} database of medium-resolution INT spectra, \citet{Sestito19Pristine} found that the ratio of metal-poor stars ([Fe/H] $<-2.5)$ with prograde versus retrograde orbits is 1.28, based on 358 stars.  We find a similar ratio of 1.23, for our sample of 29 stars with [Fe/H] $<-2.5$.
When compared to the Numerical Investigation of a Hundred Astronomical Objects \citep[NIHAO][]{Wang2015} hydrodynamic simulations of galaxy formation, specifically the NIHAO-UHD simulations \citep{Buck2020}, these simulations show that such a population of stars is ubiquitous among these Milky Way-like galaxies as investigated by \citet{Sestito2020}.

In Section~\ref{Others}, we speculated that P184+43, with its planar orbit and unusual chemistry ([Mg/Ca] $>+0.8$), may have formed in a dwarf galaxy that was accreted at early times.  The age estimate is based on its low metallicity [Fe/H] =$-3.3\pm0.3$ and chemical pattern, which suggests a lack of contributions from lower mass stars and SN~Ia.
It is possible that this star and its host dwarf galaxy were amongst the original building blocks that formed the proto Milky Way, as suggested from an analysis of NIHAO hydrodynamic simulations by \citet{Sestito2020}.

The chemistry of the Gaia-Sequoia candidates P016+28, P021+32, and P031+27 are largely consistent with previous studies, which show typical $\alpha$-enhancement at low metallicities \citep{Matsuno2019, Monty2020, Limberg2020, Yuan2020}. \citet{Cordoni2020} find that this $\alpha$-abundances pattern span a metallicity range from $-3.6<$ [Fe/H] $<-2.4$.
However, amongst our two other Gaia-Sequoia candidate stars, P133+28 and P182+09, the former shows low [Mg/Fe] and [Ca/Fe], whereas the latter has solar-like [Mg/Fe] and enriched [Ca/Fe], such that P182+09 has [Mg/Ca] = $-0.8$. 
\citet{Monty2020} also found one Gaia-Sequoia candidate (G184-007) with low [Mg/Fe] and [Ca/Fe] though at a much higher metallicity ([Fe/H]=$-1.67$ for G184-007 vs. [Fe/H] = $-2.84$ and $-2.93$ for P133+28 and P182+09, respectively).
If the location of the ``knee" in the [Mg/Fe] vs. [Fe/H] plane, as identified by \citet{Monty2020} at [Fe/H]=$-1.6$ or $-2.3$, is an accurate reflection of the chemical evolution of Gaia-Sequoia, then our low [Mg/Fe] ratios in P133+28 and P182+09 would be difficult to reconcile.  Either they are not Gaia-Sequoia members, or 
alternatively, if Gaia-Sequoia had episodes of star formation that stochastically enriched its interstellar medium, there could be a range in elements like Mg. 
Similar abundance patterns have been seen in EMP stars in dwarf galaxies, such as Carina and Sextans \citep{Norris2017, deBoer2014, Theler2020, Lucchesi20}. 
Furthermore, two of our Gaia-Sequoia candidates (P021+32 and P133+28) may be enriched in potassium (see Figs. \ref{fig:P133_P021_K} and \ref{fig:ul_abunds}).  An anti-correlation of stars that are Mg-poor but K-rich was discovered in the outer globular cluster NGC~2419 \citep{Cohen2011, Cohen2012}. This describes P133+28 well, though our Mg abundance for P021+32 is quite high and would not fit this pattern well.  Other elements (Sc, and to a lesser extent Si and Ca) also showed variations, such that a more detailed analysis of these two Gaia-Sequoia candidates, especially with broader spectra that can reach more elements, would be interesting.

Finally, one of our new CEMP candidates, P016+28, appears to be \textit{dynamically} associated with the Gaia-Sequoia event. 
We note that \citealt{Yuan2020} identify one CEMP-no candidate with Gaia-Sequoia (CS29514-007, from \citealt{Roederer14}).  
CS29514-007 ([Fe/H] $=-2.8$) has a similar metallicity to P016+28 ([Fe/H] $=-2.93\pm0.20$), but significantly different barium (CS29514-007 has [Ba/Fe] $\sim0$, whereas we find P016+28 is rich in neutron-capture elements (Ba, La, Nd) by $\ge +1.5$; see Appendix \ref{app:uls}).
It would be interesting if these two stars probe the most metal-poor regime of their (former) host and could constrain its early chemical evolution,
%\citep[also see the discussion on EMP stars in Gaia-Sequoia by ][]{Monty2020}.
%

\section{Conclusions}

We present detailed spectral analyses for 30 new metal-poor stars found within the Pristine survey and followed-up with Gemini GRACES high-resolution spectroscopy.  All of these stars were previously observed with INT medium-resolution spectra.

\begin{itemize}
\item 
We confirm that 19 of our targets are EMP with [Fe/H]$<-3.0$ (63\%), three of which only have iron upper-limits.   If we consider their 1$\sigma$(Fe) errors, then we confirm 24 are EMP stars (80\%).   The most metal-poor star in the sample is P237+12, with [Fe/H]$<-4.26$. 

\item
The INT medium-resolution spectra showed that three
%eight
of our targets may be carbon-enhanced.  We find that one is a CEMP-no candidate based on low [Ba/Fe] upper-limits, while the other two appear to be CEMP-s candidates.

\item
Three stars (P184+43, P198+08, and P207+14) are found to be deficient in Ca, yet Mg enriched, yielding [Mg/Ca] $\gtrsim+0.8$.  This is a rare abundance signature, interpreted as the yields from a small number of SN~II that underproduce Ca in explosive alpha-element production compared to Mg from hydrostatic nucleosynthesis.

\item 
Five stars (P184+43, P191+12, P237+12, P244+10, P339.1+25.5) orbit in the Galactic plane, including the most metal-poor star in our sample (P237+12).
%; all but P339.1+25.5 are EMP.
We suggest they were brought in with one or more dwarf galaxies that were building blocks that formed the Galactic plane.  As additional support one of these stars (P244+10) has a retrograde planar orbit.
This star overlaps in eccentricity and action with the Gaia-Sequoia accreted dwarf galaxy; however, its low energy and low metallicity ([Fe/H]=$-3.5 \pm0.3$) are in better agreement with the Thamnos event.

\item
Five stars are new candidates for the accreted stellar population from Gaia-Sequoia, based on their positions in the eccentricity-action-energy phase space (P016+28, P021+32, P031+27, P133+28, and P182+09).  We find that P016+28, P021+32, P031+27 are enhanced in Mg and Ca, consistent with previous chemical studies of the Sequoia population; however, P133+28 and P182+09 both show low [Mg/Fe].  This could imply these stars are non-members; however, it is also possible that Gaia-Sequoia had episodes of star formation that stochastically enriched its interstellar medium.  These two stars show different [Ca/Fe] from the rest of the population (one is Ca-poor and the other Ca-rich), and two Gaia-Sequoia members 
P133+28 and P021+32) are enriched in K.  Also, we have found that P016+28 is a CEMP-s candidate, showing enhancements of C, Ba, La, and Nd.

\end{itemize}

This work shows that the Pristine survey has been highly successful in finding new and interesting metal-poor stars, especially when combined with Gaia DR2 parallaxes and proper motions for testing stellar population and galaxy formation models.  We look forward to the upcoming large spectroscopic surveys that will be able to tackle statistically large samples of these metal-poor stars for a detailed chemo-dynamical evaluation of the metal-poor components of our Galaxy.

\section*{Acknowledgements}

We wish to thank the referee Mike Bessell for his expert advice in exploring the stellar parameter determinations for our metal-poor stars.  We also wish to thank Keith Hawkins for helpful discussions (and warnings) on systematic errors in stellar spectral analyses and on the chemical evolution of the metal-poor Galaxy.

This work is based on observations obtained with Gemini Remote Access to CFHT ESPaDOnS Spectrograph (GRACES), as part of the Gemini Large and Long Program, GN-X-LP-102 (where X includes semesters 2018B to 2020A), and also the PI Kielty program, GN-2018BA-Q-117. 
Gemini Observatory is operated by the Association of Universities for Research in Astronomy, Inc., under a cooperative agreement with the NSF on behalf of the Gemini partnership: the National Science Foundation (United States), the National Research Council (Canada), CONICYT (Chile), Ministerio de Ciencia, Tecnolog\'{i}a e Innovaci\'{o}n Productiva (Argentina), Minist\'{e}rio da Ci\^{e}ncia, Tecnologia e Inova\c{c}\~{a}o (Brazil), and Korea Astronomy and Space Science Institute (Republic of Korea).
CFHT is operated by the National Research Council of Canada, the Institut National des Sciences de l'Univers of the Centre National de la Recherche Scientique of France, and the University of Hawai'i. ESPaDOnS is a collaborative project funded by France (CNRS, MENESR, OMP, LATT), Canada (NSERC), CFHT, and the European Space Agency. 
Data was reduced using the CFHT developed OPERA data reduction pipeline.

This work has also made use of data from the European Space Agency mission Gaia (https://www.cosmos.esa.int/gaia), processed by the Gaia Data Processing and Analysis Consortium (DPAC, https://www.cosmos.esa.int/web/gaia/dpac/consortium). Funding for the DPAC has been provided by national institutions, in particular the institutions participating in the Gaia Multilateral Agreement. This research has made use of use of the SIMBAD database, operated at CDS, Strasbourg, France (Wenger et al. 2000).

This research made use of NASA's Astrophysics Data System, the SIMBAD astronomical database, operated at CDS, Strasbourg, France. 
This work also made extensive use of Astropy,\footnote{http://www.astropy.org} a community-developed core Python package for Astronomy \citep{Astropy1, Astropy2}.

CLK and KAV are grateful for funding through the National Science and Engineering Research Council Discovery Grants program and the CREATE training program on New Technologies for Canadian Observatories. 
ES gratefully acknowledge funding by the Emmy Noether program from the Deutsche Forschungsgemeinschaft (DFG).
FS and NFM gratefully acknowledge support from the French National Research Agency (ANR) funded project ``Pristine" (ANR-18-CE31-0017) along with funding from CNRS/INSU through the Programme National Galaxies et Cosmologie and through the CNRS grant PICS07708. FS thanks the Initiative dExcellence IdEx from the University of Strasbourg and the Programme Doctoral International PDI for funding his PhD. This work has been published under the framework of the IdEx Unistra and benefits from funding from the state managed by the French National Research Agency (ANR) as part of the investments for the future program. 

The authors thank the International Space Science Institute (ISSI) in Bern, Switzerland, for providing financial support and meeting facilities to the international team Pristine.

The authors wish to recognize and acknowledge the very significant cultural role and reverence that the summit of Maunakea has always had within the Native Hawaiian community.  We are very fortunate to have had the opportunity to conduct observations from this mountain.

\section*{Data Availability}
The data underlying this article and online supplementary material will be shared on reasonable requests to the corresponding author.


\bibliographystyle{mnras}
\bibliography{bibliography.bib}

%%%%%%%%%%%%%%%%% APPENDICES %%%%%%%%%%%%%%%%%%%%%

\appendix

\section{Spectra and Line Data}
\label{app:appendixA}

\begin{figure*}
	\includegraphics[width=\textwidth]{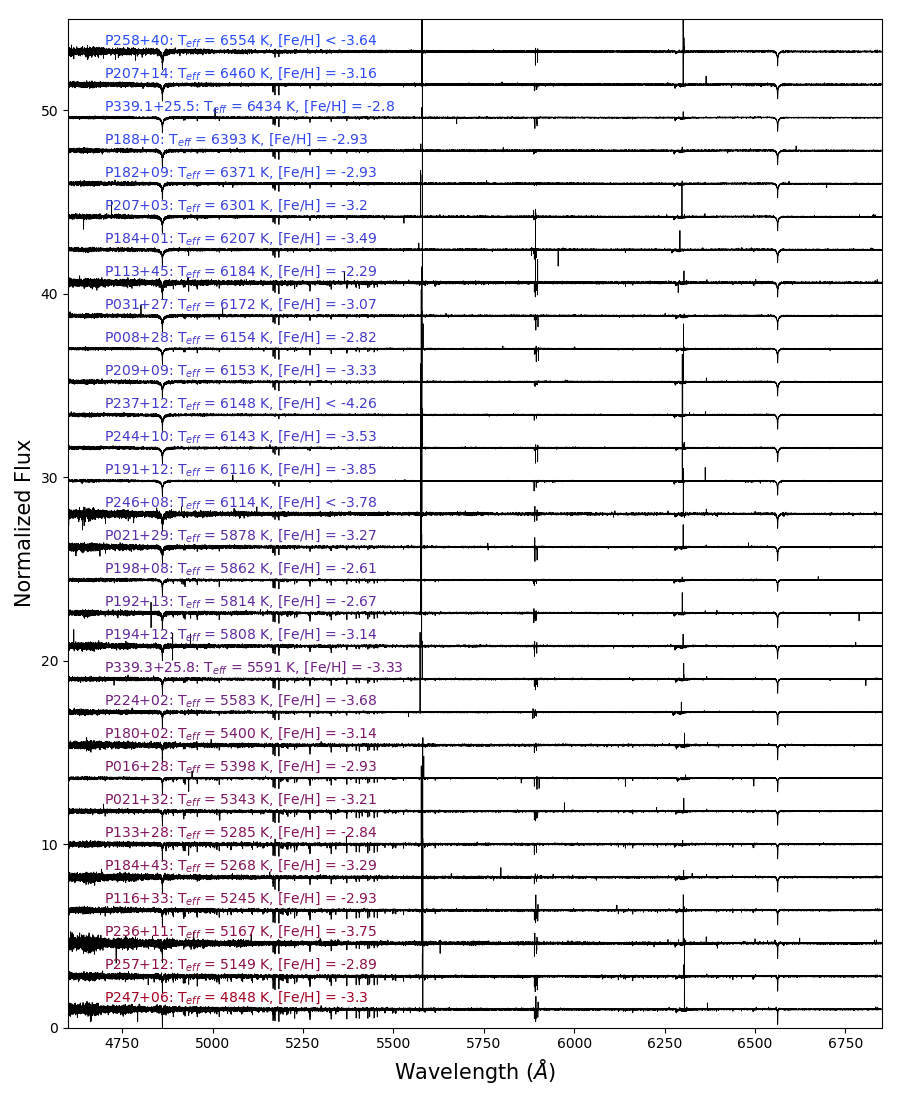}
    \caption{Portion of the 1D spectra for our GRACES data, sorted by temperature. This region was selected for our detailed analyses as it has the highest SNR, is mostly free of telluric and sky lines, and contains spectral lines for many elements of interest. }
    \label{fig:zoom_spectra}
\end{figure*}

\subsection{The 1D Spectra}

The reduced 1D spectra for a subset of the stars in this paper (in \teff and [Fe/H]) are shown in Figure \ref{fig:zoom_spectra}.   Only the wavelength region used in this paper for the chemical analysis is shown; a zoomed in region from 470 to 680 nm.   Each star and its metallicity are labelled, and the objects are sorted by effective temperature.  The Balmer lines (H$\alpha$ and H$\beta$ are clear, as well as the Mgb lines.  The atmospheric bands near 5850 and 6300 are clear in most stars, as well as some sky emission lines from an imperfect sky subtraction.
These stars were not telluric cleaned due to inconsistent observations of telluric standards. 

\subsection{The world of (mostly) upper limits.}\label{app:uls}

\begin{figure*}
	\includegraphics[width=\textwidth]{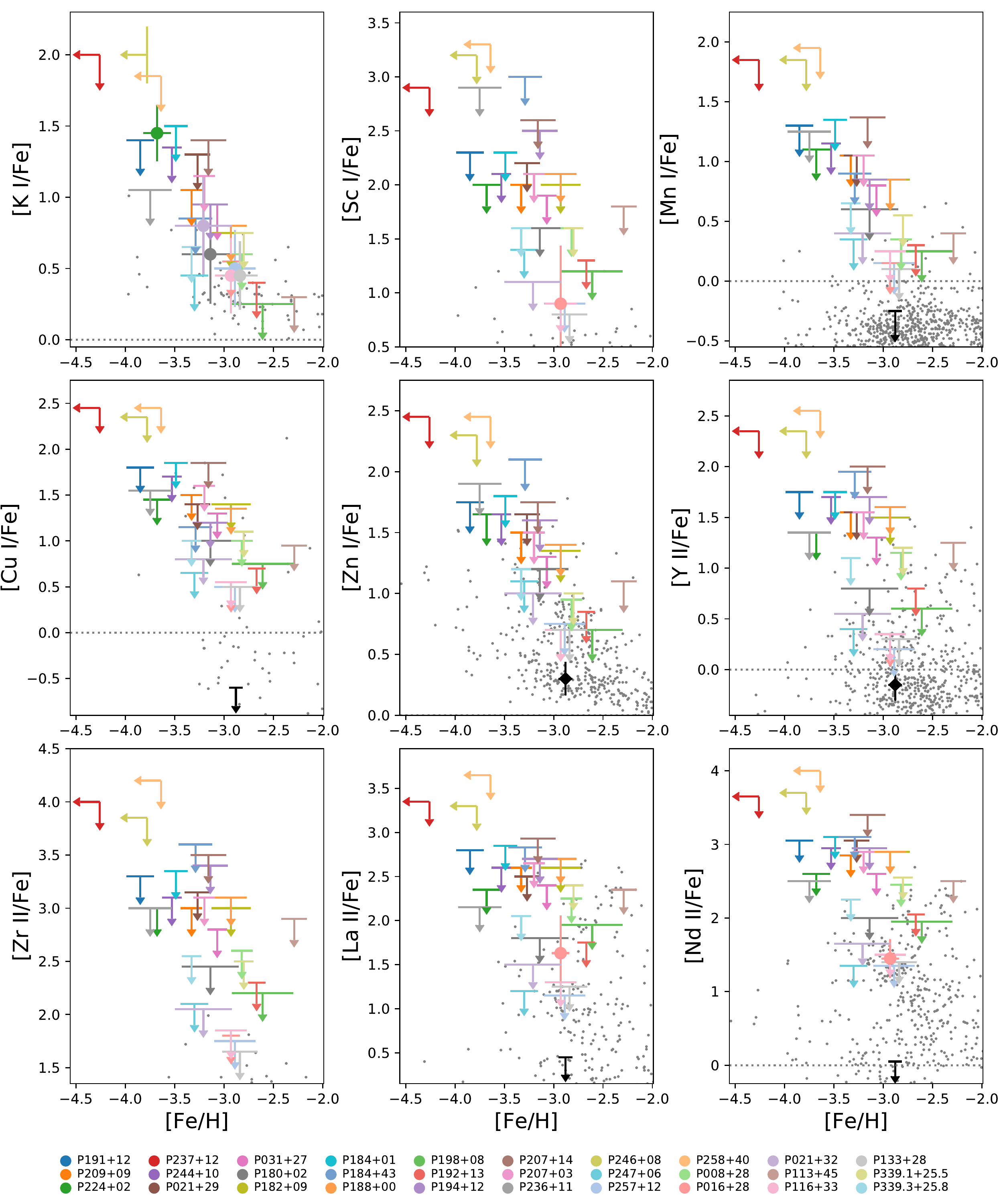}
    \caption{Elemental abundances for the chemical elements with mainly upper limit in our analysis.  See Fig.~\ref{fig:good_elems} for more information on the labels and Galactic comparison stars. }
    \label{fig:ul_abunds}
\end{figure*}

\begin{figure}
	\includegraphics[width=\linewidth]{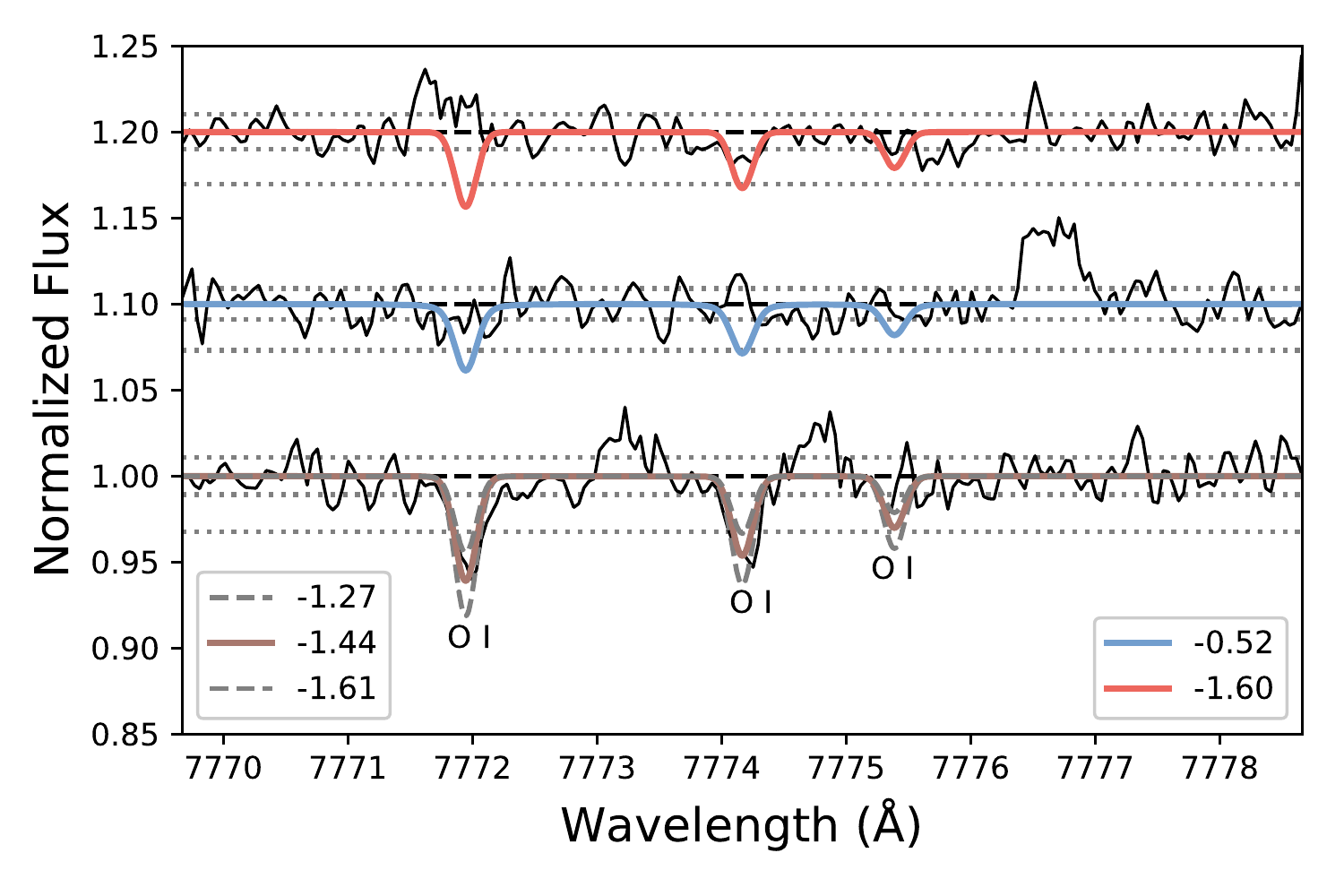}
    \caption{Synthesized \ion{O}{I} lines for P207+14 (brown, bottom), P184+43 (blue, middle), and P192+13 (red, top). Labels the same as in Fig.~\ref{fig:good_elems}.
    \label{fig:O_trip}
    }
\end{figure}

\begin{figure}
	\includegraphics[width=\columnwidth]{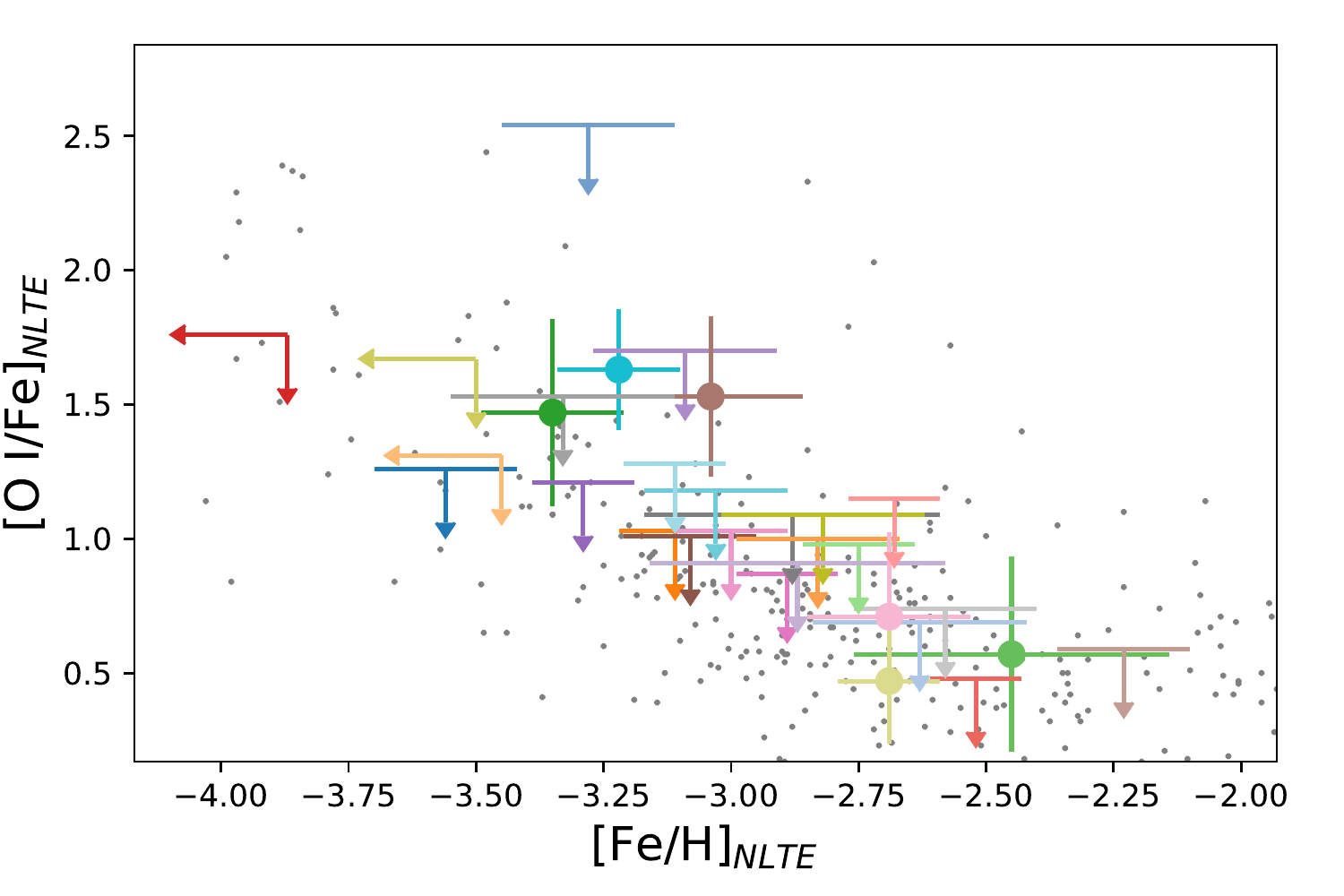}
    \caption{NLTE abundances for oxygen and iron, [\ion{O}{I}/Fe] vs. [Fe/H].  Labels the same as in Fig.~\ref{fig:good_elems}.}
    \label{fig:O_NLTE}
\end{figure}

%The lovely world of upper limits that do not provide constraints.

In higher metallicity stars, several spectral lines of additional elements exist in the GRACES wavelength region.   We examine those regions, to provide additional abundances, but mostly upper limits only.   The upper limits in this section do not provide useful constrains for nucleosynthetic interpretations, thus they are collected here in an Appendix, for completeness.   These elements include the odd-elements (K, Sc, Mn, Cu), also Zn, and the neutron capture elements (Y, Zr, La, Nd).   See Table \ref{tab:sample_lines} for our line list and their adopted atomic data, including isotopic and/or hyperfine structure information.

Sc and K abundances were measured in 1 and 5 stars, respectively.  It is unclear if the slight increase in K with decreasing metallicity is astrophysically significant as the uncertainties for each star are large, but certainly some stars appear to be rich in K.
The upper limits on Zn, Y, Nd, and La are in good agreement with the distribution of [X/Fe] in metal-poor stars in the Galaxy.   One r-process rich star (P016+28; CEMP-s, discussed in the main text) also has enriched, and therefore measureable, La and Nd abundances.

Finally, oxygen is included here since we could analyse the strong $\lambda$7770 triplet (see Fig.~\ref{fig:O_trip}); however, these lines are known to form over many layers in a stellar atmosphere, being very sensitive to small uncertainties in stellar parameters, and especially NLTE corrections. They are also in a region of significant telluric contamination.  The O abundances for six stars where we could measure the \ion{O}{I} triplet are shown in Fig.~\ref{fig:O_NLTE}.

\subsection{Line List}

A sample line list is provided in Table \ref{tab:sample_lines}, including the element, wavelength, excitation potential ($\chi$ in eV), and oscillator strengths (log\,$gf$).   The majority of the analysis of our GRACES data was carried out using spectrum syntheses, thus the line abundance from each spectral feature is listed (rather than an equivalent width).  These line abundances have been averaged together for the final abundance, per element, per star.    

  %%%%%%%%%%%%%%%%%%%% TABLES %%%%%%%%%%%%%%%%%%
\section{Data Tables}
\label{app:appendixC}

\input{Tables/photometry.txt}

\input{Tables/ferre_params.txt}

\begin{landscape}
\input{Tables/graces_params_fe.txt}
\end{landscape}

\input{Tables/graces_alpha.txt}

\begin{landscape}
\input{Tables/graces_light_elems.txt}
\end{landscape}

%\begin{landscape}
\input{Tables/graces_neutron.txt}
%\end{landscape}

\input{Tables/graces_lithium.txt}

%%% Systematic Tables %%%

%\begin{landscape}
\input{Tables/graces_alpha_systematics_1.txt}
%\end{landscape}

%\begin{landscape}
\input{Tables/graces_alpha_systematics_2.txt}
%\end{landscape}

%\begin{landscape}
\input{Tables/graces_light_systematics.txt}
%\end{landscape}

\begin{landscape}
\input{Tables/graces_fe_peak_systematics.txt}
\end{landscape}

\input{Tables/graces_Ba_systematics.txt}

\input{Tables/graces_P016_sys.txt}

\input{Tables/graces_nlte_corrections.txt}

%\begin{landscape}
\input{Tables/sample_linelist.txt}
%\end{landscape}

\begin{landscape}
    \input{Tables/inv_par_orbits.txt}
\end{landscape}

% Don't change these lines
\bsp	% typesetting comment
\label{lastpage}
\end{document}

%% file: Tables/photometry.txt
\begin{table*}
\centering
\caption{Stellar identifications, positions, and CaHK and SDSS dereddened magnitudes for our GRACES targets, selected from the \textit{Pristine} survey \citep{Starkenburg17b, Aguado19Pristine}.  E(B-V) values from \citet{Schlegel1998}, and the observer V magnitudes from the SDSS conversions.  Total exposure times (number of exposures), observation dates, and final SNR (at 6000~\AA) are also provided. }
\begin{tabular}{cccccccccccl}
\hline
ID & RA & Dec & 
\textit{V} & E(\textit{B-V}) & 
\textit{CaHK$_0$} & 
\textit{g$_0$} & \textit{r$_0$} & \textit{i$_0$} & 
$t_{\rm exp}$ & \textit{SNR} & Obs. Dates \\
& (deg) & (deg) & 
(mag) & &
(mag) & 
(mag) & (mag) & (mag) &
(s, \#) & & \\
\hline

\textit{2018A}:\\
P191.8535+12.0508 & 191.8535 & 12.0508 & 15.21 & 0.03 & 15.52 & 15.26 & 15.05 & 14.98 & 5580 (3) & 121 & 6/16, 6/17/2018 \\
P209.0986+09.8244 & 209.0986 & 9.8244 & 15.50 & 0.03 & 15.79 & 15.51 & 15.32 & 15.25 & 6300 (3) & 75 & 6/15/2018 \\
P224.8444+02.3043 & 224.8444 & 2.3043 & 15.21 & 0.05 & 15.64 & 15.27 & 14.91 & 14.76 & 4500 (3) & 91 & 4/24/2018 \\
P237.8589+12.5660 & 237.8589 & 12.5660 & 15.58 & 0.04 & 15.84 & 15.58 & 15.37 & 15.30 & 8100 (4) & 95 & 6/15, 6/16/2018 \\
P244.8986+10.9310 & 244.8986 & 10.9310 & 15.58 & 0.07 & 15.76 & 15.48 & 15.28 & 15.22 & 7200 (4) & 86 & 4/25, 6/15/2018 \\

\textit{2018B}:\\
P021.6938+29.0039 & 21.6938 & 29.0039 & 15.78 & 0.08 & 16.00 & 15.72 & 15.47 & 15.36 & 9600 (4) & 61 & 1/16, 1/17/2019 \\
P031.9938+27.7363 & 31.9938 & 27.7363 & 15.55 & 0.05 & 15.77 & 15.54 & 15.34 & 15.27 & 5400 (3) & 74 & 1/17/2019 \\
P180.3206+02.5788 & 180.3206 & 2.5788 & 15.71 & 0.02 & 16.30 & 15.89 & 15.46 & 15.28 & 9600 (4) & 50 & 1/18, 1/20/2019 \\
P182.5866+09.8940 & 182.5866 & 9.8940 & 15.23 & 0.02 & 15.60 & 15.30 & 15.07 & 15.00 & 5400 (3) & 87 & 1/19/2019 \\
P184.1783+01.0664 & 184.1783 & 1.0664 & 15.81 & 0.02 & 16.11 & 15.85 & 15.66 & 15.59 & 7200 (3) & 75 & 1/16/2019 \\
P184.2997+43.4721 & 184.2997 & 43.4721 & 15.92 & 0.01 & 16.70 & 16.16 & 15.68 & 15.49 & 7200 (3) & 46 & 1/19/2019 \\
P188.0262+00.2055 & 188.0262 & 0.2055 & 15.51 & 0.02 & 15.86 & 15.58 & 15.34 & 15.27 & 7200 (4) & 76 & 1/19, 1/20/2019 \\
P198.0851+08.9428 & 198.0851 & 8.9428 & 15.67 & 0.03 & 16.05 & 15.74 & 15.47 & 15.38 & 7200 (3) & 114 & 1/20/2019 \\

\textit{2019A}:\\
P192.3242+13.3956 & 192.3242 & 13.3956 & 15.61 & 0.03 & 16.03 & 15.69 & 15.38 & 15.28 & 9600 (4) & 68 & 3/27/2019 \\
P194.9935+12.0585 & 194.9935 & 12.0585 & 15.25 & 0.03 & 15.81 & 15.36 & 15.02 & 14.91 & 6000 (3) & 53 & 3/27, 3/28/2019 \\
P207.3454+14.1268 & 207.3454 & 14.1268 & 16.45 & 0.03 & 16.77 & 16.50 & 16.28 & 16.18 & 14400 (6) & 54 & 6/22, 6/28/2019 \\
P207.9290+03.2767 & 207.9290 & 3.2767 & 15.50 & 0.03 & 15.79 & 15.53 & 15.34 & 15.27 & 9600 (4) & 81 & 3/26/2019 \\
P236.9604+11.6155 & 236.9604 & 11.6155 & 15.63 & 0.06 & 16.26 & 15.73 & 15.24 & 15.01 & 8800 (4) & 34 & 3/29/2019 \\
P246.9682+08.5360 & 246.9682 & 8.5360 & 15.57 & 0.06 & 15.77 & 15.51 & 15.30 & 15.24 & 8800 (4) & 44 & 3/30/2019 \\
P247.2115+06.6348 & 247.2115 & 6.6348 & 15.39 & 0.06 & 16.33 & 15.55 & 14.92 & 14.65 & 6600 (3) & 47 & 3/31/2019 \\
P257.3131+12.8939 & 257.3131 & 12.8939 & 15.31 & 0.12 & 15.82 & 15.24 & 14.71 & 14.49 & 6600 (3) & 53 & 3/29/2019 \\
P258.1048+40.5405 & 258.1048 & 40.5405 & 16.06 & 0.04 & 16.34 & 16.07 & 15.88 & 15.82 & 31200 (13) & 53 & 3/31, 6/22/2019 \\

\textit{2019B}:\\
P008.5638+28.1855 & 8.5638 & 28.1855 & 15.31 & 0.05 & 15.65 & 15.34 & 15.06 & 14.97 & 8100 (3) & 105 & 11/14/2019 \\
P016.2907+28.3957 & 16.2907 & 28.3957 & 14.93 & 0.06 & 15.49 & 15.01 & 14.59 & 14.41 & 5400 (3) & 112 & 11/08/2019 \\
P021.9576+32.4131 & 21.9576 & 32.4131 & 15.75 & 0.05 & 16.36 & 15.89 & 15.42 & 15.21 & 8100 (3) & 78 & 11/11, 11/12/2019 \\
P113.8240+45.1863 & 113.8240 & 45.1863 & 15.43 & 0.07 & 15.71 & 15.39 & 15.13 & 15.06 & 7200 (3) & 40 & 11/09/2019 \\
P116.9657+33.5337 & 116.9657 & 33.5337 & 15.52 & 0.06 & 16.16 & 15.59 & 15.13 & 14.91 & 7200 (3) & 55 & 11/11/2019 \\
P133.0683+28.7219 & 133.0683 & 28.7219 & 15.39 & 0.03 & 16.08 & 15.56 & 15.11 & 14.90 & 7200 (3) & 66 & 01/18/2020 \\
P339.1417+25.5503 & 339.1417 & 25.5503 & 14.77 & 0.04 & 15.01 & 14.76 & 14.54 & 14.47 & 5400 (3) & 112 & 11/08/2019 \\
P339.3203+25.8764 & 339.3203 & 25.8764 & 15.37 & 0.06 & 15.78 & 15.40 & 15.04 & 14.90 & 7200 (3) & 66 & 11/09, 11/11/2019 \\

\hline
\end{tabular}
\label{tab:photometry}
\end{table*}

%% file: Tables/ferre_params.txt
\begin{table*}
\centering
\caption{Stellar parameters from the Pristine photometric survey (PRIS, \citealt{Starkenburg17a}) and follow-up medium resolution spectroscopy (MRS) analyzed with FERRE \citet{Aguado19Pristine} are shown.  Carbonicity is also determined from the medium resolution spectra, with the exception of P207.3454+14.1268 which was observed with a medium-resolution ESO 3.6m-EFOSC spectrum \citep{Buzzoni1984}. The C-rich flag denotes whether the star has [C/Fe]$>1.0$, and an asterisk denotes that the SNR was too low for a reliable determination. Radial velocities from medium resolution spectroscopy (from \citealt{Sestito19Pristine}) are compared to our more precise GRACES values. 
}
\begin{tabular}{lccccccccc}
\hline

ID &
PRIS &
PRIS &
MRS  & 
MRS & MRS & MRS &
MRS & MRS 
& GRACES \\

&
T$_{\rm phot}$ &
[Fe/H]$_{\rm phot}$ &
$SNR$  & 
T$_{\rm eff}$ & log\,g & [Fe/H] &
C-rich? & RV 
& RV \\

&
(K) &
(dex) & 
&
(K) & (dex) & (dex) & 
& (km s$^{-1}$)
& (km s$^{-1}$) \\

\hline
P008.5638+28.1855 & 6173 & -3.04 & 34 & 6169 $\pm$ 17  & 5.00 $\pm$ 0.07  & -2.84 $\pm$ 0.05  & N* & -266.4 $\pm$ 7.4 & -272.3 $\pm$ 0.5 \\
P016.2907+28.3957 & 5583 & -3.02 & 21 & 5378 $\pm$ 24  & 1.13 $\pm$ 0.10  & -3.27 $\pm$ 0.04  & Y & -384.5 $\pm$ 10.8 & -374.3 $\pm$ 0.5 \\
P021.6938+29.0039 & 6184 & -3.31 & 25 & 5789 $\pm$ 20  & 1.09 $\pm$ 0.09  & -3.40 $\pm$ 0.04  & Y* & -91.0 $\pm$ 8.4 & -85.4 $\pm$ 0.5 \\
P021.9576+32.4131 & 5408 & -3.53 & 55 & 5379 $\pm$ 147  & 2.40 $\pm$ 0.84  & -2.67 $\pm$ 0.19  & N* & --- & -152.6 $\pm$ 0.1 \\
P031.9938+27.7363 & 6465 & -3.27 & 25 & 6477 $\pm$ 35  & 4.93 $\pm$ 0.04  & -2.83 $\pm$ 0.04  & N* & -215.1 $\pm$ 11.1 & -209.4 $\pm$ 0.6 \\
P113.8240+45.1863 & 6296 & -2.71 & 27 & 5900 $\pm$ 12  & 5.00 $\pm$ 0.00  & -3.10 $\pm$ 0.03  & Y* & -126.9 $\pm$ 7.6 & -135.9 $\pm$ 0.6 \\
P116.9657+33.5337 & 5423 & -2.65 & 20 & 5025 $\pm$ 15  & 1.01 $\pm$ 0.01  & -3.04 $\pm$ 0.02  & N & -171.5 $\pm$ 11.6 & -180.9 $\pm$ 0.4 \\
P133.0683+28.7219 & 5445 & -3.01 & 8 & 5288 $\pm$ 29  & 1.02 $\pm$ 0.01  & -3.41 $\pm$ 0.05  & Y* & 6.8 $\pm$ 20.7 & 11.9 $\pm$ 0.6 \\
P180.3206+02.5788 & 5563 & -3.53 & 25 & 5406 $\pm$ 107  & 1.84 $\pm$ 1.14  & -2.99 $\pm$ 0.11  & N* & -212.3 $\pm$ 9.7 & -204.9 $\pm$ 0.4 \\
P182.5866+09.8940 & 6360 & -2.87 & 34 & 6237 $\pm$ 131  & 5.00 $\pm$ 0.44  & -3.48 $\pm$ 0.11  & N* & 139.5 $\pm$ 8.3 & 95.3 $\pm$ 0.2 \\
P184.1783+01.0664 & 6483 & -2.97 & 33 & 6323 $\pm$ 48  & 4.36 $\pm$ 0.14  & -3.49 $\pm$ 0.06  & Y* & 407.2 $\pm$ 10.8 & 397.6 $\pm$ 1.4 \\
P184.2997+43.4721 & 5419 & -2.92 & 24 & 5509 $\pm$ 26  & 4.97 $\pm$ 0.02  & -3.66 $\pm$ 0.04  & N* & -104.4 $\pm$ 12.0 & -110.3 $\pm$ 0.6 \\
P188.0262+00.2055 & 6339 & -3.00 & 16 & 6049 $\pm$ 52  & 4.76 $\pm$ 0.18  & -3.23 $\pm$ 0.07  & Y* & 114.4 $\pm$ 8.7 & 93.3 $\pm$ 0.3 \\
P191.8535+12.0508 & 6422 & -3.14 & 33 & 6270 $\pm$ 53  & 4.75 $\pm$ 0.12  & -3.88 $\pm$ 0.11  & N* & 18.8 $\pm$ 9.9 & 30.6 $\pm$ 0.4 \\
P192.3242+13.3956 & 6047 & -3.10 & 22 & 6142 $\pm$ 34  & 4.96 $\pm$ 0.03  & -2.75 $\pm$ 0.04  & N* & --- & 61.1 $\pm$ 0.7 \\
P194.9935+12.0585 & 5936 & -2.41 & 28 & 5935 $\pm$ 39  & 4.95 $\pm$ 0.05  & -2.89 $\pm$ 0.06  & N* & -34.8 $\pm$ 7.8 & -42.7 $\pm$ 3.0 \\
P198.0851+08.9428 & 6215 & -3.00 & 7 & 5623 $\pm$ 81  & 1.26 $\pm$ 0.25  & -3.37 $\pm$ 0.11  & N* & --- & 105.1 $\pm$ 0.1 \\
P207.3454+14.1268 & 6304 & -3.17 & 55 & 6058 $\pm$ 280  & 4.14 $\pm$ 0.89  & -4.01 $\pm$ 0.65  & N* & --- & -78.8 $\pm$ 0.1 \\
P207.9290+03.2767 & 6516 & -2.95 & 18 & 6490 $\pm$ 46  & 4.89 $\pm$ 0.10  & -2.96 $\pm$ 0.07  & N* & 126.4 $\pm$ 7.1 & 109.3 $\pm$ 0.3 \\
P209.0986+09.8244 & 6477 & -2.92 & 32 & 6395 $\pm$ 38  & 4.85 $\pm$ 0.09  & -3.14 $\pm$ 0.05  & N* & -108.8 $\pm$ 8.9 & -116.7 $\pm$ 0.1 \\
P224.8444+02.3043 & 5791 & -3.40 & 19 & 5455 $\pm$ 39  & 1.23 $\pm$ 0.18  & -3.92 $\pm$ 0.08  & Y* & 196.3 $\pm$ 24.3 & 189.2 $\pm$ 0.8 \\
P236.9604+11.6155 & 5320 & -3.21 & 23 & 5298 $\pm$ 29  & 1.10 $\pm$ 0.08  & -3.39 $\pm$ 0.03  & N & -47.8 $\pm$ 9.8 & -53.5 $\pm$ 0.9 \\
P237.8589+12.5660 & 6443 & -3.08 & 37 & 6323 $\pm$ 34  & 4.53 $\pm$ 0.09  & -3.88 $\pm$ 0.08  & N* & 6.5 $\pm$ 8.0 & -3.8 $\pm$ 0.9 \\
P244.8986+10.9310 & 6492 & -2.92 & 26 & 6481 $\pm$ 28  & 4.97 $\pm$ 0.03  & -3.11 $\pm$ 0.05  & N* & -182.1 $\pm$ 11.6 & -176.7 $\pm$ 0.3 \\
P246.9682+08.5360 & 6446 & -3.13 & 34 & 6411 $\pm$ 32  & 4.70 $\pm$ 0.07  & -3.45 $\pm$ 0.05  & N* & -55.2 $\pm$ 9.8 & -58.0 $\pm$ 0.5 \\
P247.2115+06.6348 & 4969 & -2.87 & 9 & 5026 $\pm$ 68  & 1.96 $\pm$ 0.36  & -2.99 $\pm$ 0.08  & N* & --- & -200.8 $\pm$ 0.4 \\
P257.3131+12.8939 & 5277 & -2.99 & 19 & 5281 $\pm$ 53  & 2.50 $\pm$ 0.48  & -2.80 $\pm$ 0.06  & N & -129.4 $\pm$ 7.0 & -129.4 $\pm$ 0.4 \\
P258.1048+40.5405 & 6510 & -2.92 & 79 & 6283 $\pm$ 297  & 3.95 $\pm$ 0.90  & -3.76 $\pm$ 0.73  & N* & --- & -169.4 $\pm$ 1.5 \\
P339.1417+25.5503 & 6413 & -3.15 & 27 & 6048 $\pm$ 30  & 5.00 $\pm$ 0.00  & -3.42 $\pm$ 0.06  & Y* & -53.5 $\pm$ 7.8 & -78.9 $\pm$ 0.1 \\
P339.3203+25.8764 & 5816 & -3.24 & 22 & 5589 $\pm$ 31  & 4.99 $\pm$ 0.01  & -3.49 $\pm$ 0.07  & N* & -141.5 $\pm$ 9.0 & -157.7 $\pm$ 0.3 \\
\end{tabular}
\label{tab:ferre_params}
\end{table*}

%% file: Tables/graces_params_fe.txt
\begin{table}
\centering
\caption{Stellar parameters and iron abundances for our GRACES sample.
Effective temperatures are from the \citet{MB2020} calibration and surface gravities are calculated from the Stefan-Boltzmann law (see text). 
Short ID's are used to identify stars throughout the text and in the figures.
Iron abundance errors represent the total combined systematic error due to the stellar parameters and the line-to-line scatter (see Table \ref{tab:graces_fe_peak_sys}). The number of lines used to calculate the average [Fe/H] are given in parentheses. $\Delta$\ion{Fe}{I}$_{\rm NLTE}$ is the averaged NLTE correction for \ion{Fe}{I} lines with known NLTE calculations, in the sense that \ion{Fe}{I}$_{\rm NLTE}$ = \ion{Fe}{I}$_{\rm LTE}$ + $\Delta$\ion{Fe}{I}$_{\rm NLTE}$.  \textit{Slope} is the slope of the line fit to A(Fe I) vs. excitation potential, which can serve as a spectroscopic check of the effective temperature when more than 10 lines are available.
}
\begin{tabular}{lccccccccr}
\hline

ID &
Short ID&
T$_{\rm eff}$ (K) & 
$\log g$ (dex) &
[Fe I/H] &
[Fe II/H] &
[Fe/H] &
FeI$-$FeII &
$\Delta$\ion{Fe}{I}$_{\rm NLTE}$ &
\textit{Slope} \\

\hline

P008.5638+28.1855 & P008+28 & 6154 $\pm$ 241 & 4.61 $\pm$ 0.08 & -2.80 $\pm$ 0.23 (22) & -2.88 $\pm$ 0.12 (3) & -2.82 $\pm$ 0.11 (25) & 0.08 & 0.07 (13) & -0.06 \\
P016.2907+28.3957 & P016+28 & 5398 $\pm$ 116 & 2.40 $\pm$ 0.10 & -2.95 $\pm$ 0.26 (50) & -2.85 $\pm$ 0.28 (6) & -2.93 $\pm$ 0.20 (56) & -0.10 & 0.25 (26) & -0.02 \\
P021.6938+29.0039 & P021+29 & 5878 $\pm$ 137 & 3.54 $\pm$ 0.10 & -3.29 $\pm$ 0.95 (12) & -3.14 $\pm$ 0.45 (4) & -3.27 $\pm$ 0.41 (16) & -0.15 & 0.19 (7) & 0.10 \\
P021.9576+32.4131 & P021+32 & 5343 $\pm$ 116 & 2.02 $\pm$ 0.11 & -3.05 $\pm$ 0.13 (45) & -3.62 $\pm$ 0.13 (3) & -3.21 $\pm$ 0.30 (48) & 0.57 & 0.34 (23) & -0.03 \\
P031.9938+27.7363 & P031+27 & 6172 $\pm$ 212 & 3.79 $\pm$ 0.10 & -3.07 $\pm$ 0.78 (13) & -3.07 $\pm$ 0.27 (4) & -3.07 $\pm$ 0.26 (17) & 0.00 & 0.18 (8) & 0.01 \\
P113.8240+45.1863 & P113+45 & 6184 $\pm$ 210 & 4.51 $\pm$ 0.09 & -2.28 $\pm$ 0.27 (27) & -2.41 $\pm$ 0.18 (4) & -2.29 $\pm$ 0.16 (31) & 0.13 & 0.06 (13) & 0.03 \\
P116.9657+33.5337 & P116+33 & 5245 $\pm$ 108 & 2.15 $\pm$ 0.11 & -2.86 $\pm$ 0.11 (62) & -3.14 $\pm$ 0.13 (5) & -2.93 $\pm$ 0.16 (67) & 0.28 & 0.24 (32) & -0.03 \\
P133.0683+28.7219 & P133+28 & 5285 $\pm$ 123 & 2.09 $\pm$ 0.10 & -2.75 $\pm$ 0.09 (68) & -3.09 $\pm$ 0.13 (6) & -2.84 $\pm$ 0.19 (74) & 0.34 & 0.26 (36) & -0.06 \\
P180.3206+02.5788 & P180+02 & 5400 $\pm$ 113 & 2.53 $\pm$ 0.11 & -2.90 $\pm$ 0.26 (7) & -3.20 $\pm$ 0.31 (3) & -3.14 $\pm$ 0.25 (10) & 0.30 & 0.26 (5) & {\it -0.03} \\
P182.5866+09.8940 & P182+09 & 6371 $\pm$ 260 & 4.51 $\pm$ 0.10 & -2.90 $\pm$ 0.19 (7) & -3.20 $\pm$ 0.21 (3) & -2.93 $\pm$ 0.21 (10) & 0.30 & 0.11 (5) & {\it -0.03} \\
P184.1783+01.0664 & P184+01 & 6207 $\pm$ 171 & 3.64 $\pm$ 0.13 & -3.50 $\pm$ 0.43 (3) & -3.47 $\pm$ 0.16 (3) & -3.49 $\pm$ 0.15 (6) & -0.03 & 0.27 (1) & {\it 1.85} \\
P184.2997+43.4721 & P184+43 & 5268 $\pm$ 146 & 4.94 $\pm$ 0.09 & -3.31 $\pm$ 0.28 (14) & $<$-3.27 & -3.29 $\pm$ 0.28 (17) & $>$-0.04 & 0.01 (6) & -0.08 \\
P188.0262+00.2055 & P188+00 & 6393 $\pm$ 230 & 4.55 $\pm$ 0.09 & -2.86 $\pm$ 0.10 (10) & -3.12 $\pm$ 0.10 (3) & -2.93 $\pm$ 0.15 (13) & 0.26 & 0.10 (4) & -0.05 \\
P191.8535+12.0508 & P191+12 & 6116 $\pm$ 164 & 3.61 $\pm$ 0.09 & -3.78 $\pm$ 1.01 (7) & -3.96 $\pm$ 0.39 (3) & -3.85 $\pm$ 0.37 (10) & 0.18 & 0.29 (2) & {\it 0.11} \\
P192.3242+13.3956 & P192+13 & 5814 $\pm$ 117 & 3.46 $\pm$ 0.10 & -2.66 $\pm$ 0.56 (42) & -2.73 $\pm$ 0.21 (4) & -2.67 $\pm$ 0.20 (46) & 0.07 & 0.15 (25) & -0.01 \\
P194.9935+12.0585 & P194+12 & 5808 $\pm$ 214 & 4.55 $\pm$ 0.09 & -3.15 $\pm$ 0.40 (8) & -3.04 $\pm$ 0.18 (2) & -3.14 $\pm$ 0.17 (10) & -0.11 & 0.05 (5) & {\it -0.02} \\
P198.0851+08.9428 & P198+08 & 5862 $\pm$ 132 & 3.39 $\pm$ 0.10 & -2.66 $\pm$ 0.29 (24) & -2.06 $\pm$ 0.19 (6) & -2.61 $\pm$ 0.34 (30) & -0.60 & 0.16 (14) & 0.02 \\
P207.3454+14.1268 & P207+14 & 6460 $\pm$ 253 & 4.63 $\pm$ 0.09 & -3.16 $\pm$ 0.11 (2) & -3.19 $\pm$ 0.25 (2) & -3.16 $\pm$ 0.10 (4) & 0.03 & 0.12 (1) & {\it -0.89} \\
P207.9290+03.2767 & P207+03 & 6301 $\pm$ 251 & 3.88 $\pm$ 0.11 & -3.19 $\pm$ 0.68 (8) & -3.24 $\pm$ 0.22 (3) & -3.20 $\pm$ 0.21 (11) & 0.05 & 0.20 (4) & {\it -0.02} \\
P209.0986+09.8244 & P209+09 & 6153 $\pm$ 180 & 3.70 $\pm$ 0.11 & -3.32 $\pm$ 0.78 (8) & -3.37 $\pm$ 0.28 (3) & -3.33 $\pm$ 0.26 (11) & 0.05 & 0.22 (4) & {\it 0.04} \\
P224.8444+02.3043 & P224+02 & 5583 $\pm$ 116 & 2.86 $\pm$ 0.13 & -3.72 $\pm$ 0.13 (7) & -3.51 $\pm$ 0.16 (3) & -3.68 $\pm$ 0.15 (10) & -0.21 & 0.33 (4) & {\it 0.03} \\
P236.9604+11.6155 & P236+11 & 5167 $\pm$ 115 & 1.80 $\pm$ 0.10 & -3.52 $\pm$ 0.11 (11) & -3.93 $\pm$ 0.21 (3) & -3.75 $\pm$ 0.23 (14) & 0.41 & 0.42 (5) & 0.12 \\
P237.8589+12.5660 & P237+12 & 6148 $\pm$ 175 & 3.61 $\pm$ 0.10 & $<$-4.28 & $<$-4.23 & $<$-4.26 & $<$-0.05 & 0.39 (1) & {\it 0.07} \\
P244.8986+10.9310 & P244+10 & 6143 $\pm$ 150 & 3.72 $\pm$ 0.09 & -3.52 $\pm$ 0.82 (6) & -3.57 $\pm$ 0.31 (3) & -3.53 $\pm$ 0.29 (9) & 0.05 & 0.24 (3) & {\it 0.00} \\
P246.9682+08.5360 & P246+08 & 6114 $\pm$ 142 & 3.58 $\pm$ 0.10 & $<$-3.65 & $<$-3.90 & $<$-3.78 & $<$0.25 & 0.28 (9) & {\it 0.33} \\
P247.2115+06.6348 & P247+06 & 4848 $\pm$ 97 & 1.19 $\pm$ 0.10 & -3.26 $\pm$ 0.12 (45) & -3.50 $\pm$ 0.18 (4) & -3.30 $\pm$ 0.16 (49) & 0.24 & 0.27 (22) & -0.06 \\
P257.3131+12.8939 & P257+12 & 5149 $\pm$ 105 & 1.78 $\pm$ 0.10 & -2.85 $\pm$ 0.07 (63) & -3.22 $\pm$ 0.13 (3) & -2.89 $\pm$ 0.19 (66) & 0.37 & 0.26 (33) & -0.05 \\
P258.1048+40.5405 & P258+40 & 6554 $\pm$ 245 & 4.37 $\pm$ 0.09 & $<$-3.42 & $<$-3.87 & $<$-3.64 & $<$0.45 & 0.19 (8) & {\it -0.26} \\
P339.1417+25.5503 & P339.1+25.5 & 6434 $\pm$ 237 & 4.41 $\pm$ 0.09 & -2.78 $\pm$ 0.25 (19) & -2.87 $\pm$ 0.11 (4) & -2.80 $\pm$ 0.11 (23) & 0.09 & 0.11 (13) & 0.00 \\
P339.3203+25.8764 & P339.3+25.8 & 5591 $\pm$ 118 & 3.11 $\pm$ 0.10 & -3.28 $\pm$ 0.14 (23) & -3.45 $\pm$ 0.17 (3) & -3.33 $\pm$ 0.14 (26) & 0.17 & 0.22 (12) & -0.01 \\
HD 122563 & & 4749 $\pm$ 89 & 1.12 $\pm$ 0.11 & -2.89 $\pm$ 0.10 (54) & -2.87 $\pm$ 0.09 (9) & -2.88 $\pm$ 0.07 (63) & -0.02 & 0.11 (48) & -0.08 \\

\hline
\end{tabular}
\label{tab:graces_params_fe}
\end{table}

%% file: Tables/graces_alpha.txt
\begin{table*}

\caption{LTE abundances for the $\alpha$-elements. Errors represent the total combined systematic error due to the stellar parameters and line-to-line scatter (Table \ref{tab:graces_alpha_sys_1} and \ref{tab:graces_alpha_sys_2}). The number of lines used is given in parentheses. }

\begin{tabular}{lcccccc}

\hline

ID &
[O I/Fe]&
[Mg I/Fe]&
[Ca I/Fe] &
[Ca II/Fe] & 
[Ti I/Fe] &
[Ti II/Fe] \\

\hline

P008.5638+28.1855 & $<$1.05 & 0.47 $\pm$ 0.13 (4) & 0.08 $\pm$ 0.15 (3) & 0.78 $\pm$ 0.15 (3) & $<$0.45 & $<$0.55 \\
P016.2907+28.3957 & $<$1.40 & 0.55 $\pm$ 0.11 (4) & 0.29 $\pm$ 0.13 (10) & 1.17 $\pm$ 0.13 (3) & 0.48 $\pm$ 0.15 (4) & 0.61 $\pm$ 0.11 (6) \\
P021.6938+29.0039 & $<$1.20 & 0.11 $\pm$ 0.18 (2) & 0.33 $\pm$ 0.15 (4) & 0.92 $\pm$ 0.20 (3) & $<$1.10 & $<$0.50 \\
P021.9576+32.4131 & $<$1.25 & 0.78 $\pm$ 0.29 (3) & 0.46 $\pm$ 0.30 (7) & 0.42 $\pm$ 0.32 (3) & $<$0.20 & 0.15 $\pm$ 0.34 (1) \\
P031.9938+27.7363 & $<$1.05 & 0.50 $\pm$ 0.16 (3) & 0.38 $\pm$ 0.14 (4) & 0.95 $\pm$ 0.12 (3) & $<$0.80 & $<$0.65 \\
P113.8240+45.1863 & $<$0.70 & 0.09 $\pm$ 0.34 (2) & 0.00 $\pm$ 0.17 (5) & 0.30 $\pm$ 0.21 (1) & $<$0.55 & $<$0.50 \\
P116.9657+33.5337 & 0.95 $\pm$ 0.30 (1) & 0.57 $\pm$ 0.18 (3) & 0.35 $\pm$ 0.17 (8) & 0.25 $\pm$ 0.25 (3) & 0.35 $\pm$ 0.19 (3) & 0.24 $\pm$ 0.18 (2) \\
P133.0683+28.7219 & $<$1.00 & 0.15 $\pm$ 0.44 (2) & 0.27 $\pm$ 0.20 (10) & 0.43 $\pm$ 0.25 (3) & 0.30 $\pm$ 0.27 (2) & 0.30 $\pm$ 0.22 (1) \\
P180.3206+02.5788 & $<$1.35 & 0.47 $\pm$ 0.19 (3) & 0.30 $\pm$ 0.20 (6) & 0.60 $\pm$ 0.21 (3) & 0.52 $\pm$ 0.27 (2) & 1.40 $\pm$ 0.27 (1) \\
P182.5866+09.8940 & $<$1.20 & -0.02 $\pm$ 0.22 (2) & 0.80 $\pm$ 0.28 (1) & 0.73 $\pm$ 0.22 (3) & $<$0.80 & $<$0.95 \\
P184.1783+01.0664 & 1.90 $\pm$ 0.20 (2) & 0.77 $\pm$ 0.26 (2) & $<$0.45 & 1.20 $\pm$ 0.12 (2) & $<$1.30 & $<$1.25 \\
P184.2997+43.4721 & $<$2.55 & 0.51 $\pm$ 0.28 (2) & $<$-0.25 & 0.11 $\pm$ 0.17 (3) & $<$0.75 & $<$1.35 \\
P188.0262+0.2055 & $<$1.10 & 0.43 $\pm$ 0.17 (3) & 0.15 $\pm$ 0.29 (2) & 0.88 $\pm$ 0.21 (3) & $<$0.90 & $<$0.85 \\
P191.8535+12.0508 & $<$1.55 & 0.25 $\pm$ 0.16 (2) & $<$0.15 & 1.27 $\pm$ 0.16 (3) & $<$1.35 & $<$1.10 \\
P192.3242+13.3956 & $<$0.70 & 0.16 $\pm$ 0.13 (3) & 0.11 $\pm$ 0.10 (5) & 0.62 $\pm$ 0.26 (2) & $<$0.25 & 0.35 $\pm$ 0.20 (1) \\
P194.9935+12.0585 & $<$1.75 & 0.32 $\pm$ 0.36 (2) & 0.17 $\pm$ 0.19 (2) & 0.53 $\pm$ 0.35 (2) & $<$0.50 & $<$1.05 \\
P198.0851+08.9428 & 0.81 $\pm$ 0.36 (2) & 0.65 $\pm$ 0.34 (2) & -0.19 $\pm$ 0.32 (4) & 1.13 $\pm$ 0.33 (3) & $<$0.15 & 0.47 $\pm$ 0.33 (2) \\
P207.3454+14.1268 & 1.65 $\pm$ 0.26 (1) & 1.09 $\pm$ 0.15 (3) & $<$-0.45 & 1.06 $\pm$ 0.17 (2) & $<$1.35 & $<$1.25 \\
P207.9290+03.2767 & $<$1.23 & 0.23 $\pm$ 0.19 (2) & 0.55 $\pm$ 0.21 (3) & 1.04 $\pm$ 0.14 (3) & $<$0.95 & $<$0.90 \\
P209.0986+09.8244 & $<$1.25 & 0.76 $\pm$ 0.12 (4) & 0.35 $\pm$ 0.22 (2) & 0.96 $\pm$ 0.13 (3) & $<$1.05 & $<$0.90 \\
P224.8444+02.3043 & 1.80 $\pm$ 0.34 (1) & 0.32 $\pm$ 0.17 (3) & 0.25 $\pm$ 0.19 (1) & 1.39 $\pm$ 0.22 (3) & $<$0.95 & $<$0.70 \\
P236.9604+11.6155 & $<$1.95 & 0.52 $\pm$ 0.40 (2) & 0.57 $\pm$ 0.30 (2) & 0.64 $\pm$ 0.25 (3) & $<$1.00 & $<$0.75 \\
P237.8589+12.5660 & $<$2.15 & $<$0.15 & $<$0.95 & $<$1.25 & $<$1.40 & $<$1.40 \\
P244.8986+10.9310 & $<$1.45 & 0.47 $\pm$ 0.18 (4) & 0.57 $\pm$ 0.16 (3) & 1.16 $\pm$ 0.12 (3) & $<$1.10 & $<$1.00 \\
P246.9682+08.5360 & $<$1.95 & $<$0.28 & $<$0.85 & $<$1.22 & $<$1.40 & $<$1.35 \\
P247.2115+06.6348 & $<$1.45 & 0.53 $\pm$ 0.16 (3) & 0.27 $\pm$ 0.17 (5) & 0.24 $\pm$ 0.36 (2) & 0.38 $\pm$ 0.18 (4) & 0.43 $\pm$ 0.23 (4) \\
P257.3131+12.8939 & $<$0.95 & 0.59 $\pm$ 0.21 (4) & 0.38 $\pm$ 0.22 (8) & 0.37 $\pm$ 0.23 (3) & 0.39 $\pm$ 0.25 (3) & 0.22 $\pm$ 0.22 (5) \\
P258.1048+40.5405 & $<$1.50 & $<$0.40 & $<$0.65 & $<$1.46 & $<$1.40 & $<$1.40 \\
P339.1417+25.5503 & 0.70 $\pm$ 0.22 (2) & 0.28 $\pm$ 0.19 (2) & 0.27 $\pm$ 0.15 (5) & 0.92 $\pm$ 0.15 (3) & 0.60 $\pm$ 0.25 (1) & $<$0.35 \\
P339.3203+25.8764 & $<$1.50 & 0.63 $\pm$ 0.20 (2) & 0.23 $\pm$ 0.12 (4) & 0.64 $\pm$ 0.14 (3) & $<$0.20 & $<$0.45 \\
HD 122563 & --- & 0.62 $\pm$ 0.09 (4) & 0.31 $\pm$ 0.08 (14) & --- & 0.15 $\pm$ 0.16 (1) & 0.33 $\pm$ 0.07 (10) \\
\hline
\end{tabular}
\label{tab:graces_alpha}

\end{table*}

%% file: Tables/graces_light_elems.txt
\begin{table}
\centering
\caption{LTE abundances for light elements and Fe-peak elements. Errors represent the total combined systematic error due to the stellar parameters and line-to-line scatter (Table \ref{tab:graces_light_sys_1} and \ref{tab:graces_fe_peak_sys}). The number of lines used is given in parentheses. P016.2907+28.3957 is the only star in the sample to have a detectable Sc abundance. The systematic errors for P016.2907+28.3957 are given in Table \ref{tab:P016_sys}.}
\begin{tabular}{lccccccccc}

\hline

ID &
[Na I/Fe] &
[K I/Fe] &
[Sc II/Fe] &
[Cr I/Fe] & 
[Mn I/Fe] &
[Ni I/Fe] &
[Cu I/Fe] &
[Zn I/Fe] \\

\hline

P008.5638+28.1855 & -0.15 $\pm$ 0.17 (1) & $<$0.60 & $<$1.60 & -0.20 $\pm$ 0.20 (1) & $<$0.35 & $<$0.05 & $<$1.00 & $<$0.95 \\
P016.2907+28.3957 & 0.37 $\pm$ 0.18 (2) & $<$0.55 & 0.90 $\pm$ 0.53 (1) & -0.15 $\pm$ 0.16 (4) & $<$0.15 & -0.05 $\pm$ 0.16 (1) & $<$0.50 & $<$0.70 \\
P021.6938+29.0039 & $<$-0.40 & $<$1.30 & $<$2.20 & -0.05 $\pm$ 0.25 (1) & $<$1.05 & 0.20 $\pm$ 0.27 (1) & $<$1.40 & $<$1.65 \\
P021.9576+32.4131 & 0.07 $\pm$ 0.31 (2) & 0.80 $\pm$ 0.33 (1) & $<$1.10 & -0.02 $\pm$ 0.31 (3) & $<$0.40 & 0.10 $\pm$ 0.32 (1) & $<$0.80 & $<$1.00 \\
P031.9938+27.7363 & -0.10 $\pm$ 0.17 (1) & $<$0.95 & $<$1.90 & -0.15 $\pm$ 0.27 (1) & $<$0.80 & $<$0.15 & $<$1.30 & $<$1.30 \\
P113.8240+45.1863 & 0.36 $\pm$ 0.27 (2) & $<$0.30 & $<$1.80 & -0.33 $\pm$ 0.24 (2) & $<$0.40 & $<$-0.30 & $<$0.95 & $<$1.10 \\
P116.9657+33.5337 & 0.05 $\pm$ 0.24 (2) & 0.45 $\pm$ 0.25 (1) & $<$0.90 & -0.23 $\pm$ 0.21 (3) & $<$0.25 & -0.05 $\pm$ 0.21 (1) & $<$0.55 & $<$0.70 \\
P133.0683+28.7219 & 0.02 $\pm$ 0.24 (2) & 0.45 $\pm$ 0.24 (1) & $<$0.80 & -0.29 $\pm$ 0.23 (3) & $<$0.10 & -0.02 $\pm$ 0.35 (2) & $<$0.50 & $<$0.70 \\
P180.3206+02.5788 & -0.15 $\pm$ 0.22 (2) & 0.60 $\pm$ 0.25 (1) & $<$1.60 & -0.20 $\pm$ 0.24 (1) & $<$0.60 & $<$-0.10 & $<$1.00 & $<$1.20 \\
P182.5866+09.8940 & -0.55 $\pm$ 0.26 (2) & $<$0.75 & $<$2.00 & $<$-0.20 & $<$0.85 & $<$0.15 & $<$1.40 & $<$1.35 \\
P184.1783+01.0664 & 1.92 $\pm$ 0.27 (2) & $<$1.50 & $<$2.30 & $<$0.10 & $<$1.35 & $<$0.70 & $<$1.85 & $<$1.80 \\
P184.2997+43.4721 & -0.21 $\pm$ 0.16 (2) & $<$0.85 & $<$3.00 & -0.45 $\pm$ 0.27 (1) & $<$0.90 & $<$0.15 & $<$1.15 & $<$2.10 \\
P188.0262+0.2055 & -0.48 $\pm$ 0.22 (2) & $<$0.80 & $<$2.10 & $<$-0.15 & $<$0.85 & $<$0.20 & $<$1.35 & $<$1.40 \\
P191.8535+12.0508 & $<$0.00 & $<$1.40 & $<$2.30 & $<$0.20 & $<$1.30 & $<$0.60 & $<$1.80 & $<$1.75 \\
P192.3242+13.3956 & -0.64 $\pm$ 0.19 (2) & $<$0.40 & $<$1.30 & -0.22 $\pm$ 0.21 (2) & $<$0.30 & 0.00 $\pm$ 0.16 (1) & $<$0.70 & $<$0.85 \\
P194.9935+12.0585 & $<$-0.55 & $<$0.95 & $<$2.50 & $<$-0.30 & $<$0.85 & $<$0.10 & $<$1.20 & $<$1.60 \\
P198.0851+08.9428 & -0.24 $\pm$ 0.35 (2) & $<$0.25 & $<$1.20 & -0.31 $\pm$ 0.36 (2) & $<$0.25 & $<$-0.25 & $<$0.75 & $<$0.70 \\
P207.3454+14.1268 & 0.51 $\pm$ 0.19 (2) & $<$1.40 & $<$2.60 & $<$0.15 & $<$1.37 & $<$0.70 & $<$1.85 & $<$1.75 \\
P207.9290+03.2767 & -0.28 $\pm$ 0.22 (2) & $<$1.15 & $<$2.10 & 0.10 $\pm$ 0.25 (2) & $<$1.05 & $<$0.45 & $<$1.60 & $<$1.50 \\
P209.0986+09.8244 & 0.00 $\pm$ 0.18 (2) & $<$1.05 & $<$2.00 & $<$-0.15 & $<$1.05 & 0.30 $\pm$ 0.29 (1) & $<$1.50 & $<$1.50 \\
P224.8444+02.3043 & 0.29 $\pm$ 0.14 (2) & 1.45 $\pm$ 0.18 (1) & $<$2.00 & -0.10 $\pm$ 0.21 (1) & $<$1.10 & $<$0.25 & $<$1.45 & $<$1.65 \\
P236.9604+11.6155 & $<$0.90 & $<$1.05 & $<$2.90 & $<$-0.10 & $<$1.25 & $<$0.30 & $<$1.55 & $<$1.90 \\
P237.8589+12.5660 & $<$0.00 & $<$2.00 & $<$2.90 & $<$0.80 & $<$1.85 & $<$1.00 & $<$2.45 & $<$2.45 \\
P244.8986+10.9310 & 0.00 $\pm$ 0.15 (2) & $<$1.35 & $<$2.10 & $<$0.10 & $<$1.15 & $<$0.60 & $<$1.70 & $<$1.65 \\
P246.9682+08.5360 & $<$0.25 & $<$2.00 & $<$3.20 & $<$0.75 & $<$1.85 & $<$1.00 & $<$2.35 & $<$2.30 \\
P247.2115+06.6348 & -0.01 $\pm$ 0.21 (2) & $<$0.45 & $<$1.40 & -0.18 $\pm$ 0.20 (4) & $<$0.35 & -0.05 $\pm$ 0.20 (1) & $<$0.65 & $<$1.10 \\
P257.3131+12.8939 & 0.04 $\pm$ 0.41 (2) & 0.50 $\pm$ 0.25 (1) & $<$0.90 & -0.02 $\pm$ 0.26 (3) & $<$0.15 & 0.07 $\pm$ 0.31 (2) & $<$0.50 & $<$0.75 \\
P258.1048+40.5405 & $<$-0.05 & $<$1.85 & $<$3.30 & $<$0.50 & $<$1.95 & $<$1.00 & $<$2.45 & $<$2.45 \\
P339.1417+25.5503 & $<$-0.40 & $<$0.75 & $<$1.60 & -0.13 $\pm$ 0.24 (2) & $<$0.55 & $<$-0.05 & $<$1.10 & $<$1.00 \\
P339.3203+25.8764 & 0.08 $\pm$ 0.14 (2) & $<$0.65 & $<$1.60 & -0.15 $\pm$ 0.19 (1) & $<$0.65 & $<$-0.15 & $<$1.00 & $<$1.20 \\
HD 122563 & 0.43 $\pm$ 0.11 (2) & --- & --- & -0.08 $\pm$ 0.12 (8) & $<$-0.25 & 0.00 $\pm$ 0.13 (1) & $<$-0.60 & 0.30 $\pm$ 0.13 (1) \\
\hline
\end{tabular}
\label{tab:graces_light}

\end{table}

%% file: Tables/graces_neutron.txt
\begin{table*}

\caption{LTE abundances for neutron-capture elements. Errors represent the total combined systematic error due to the stellar parameters and line-to-line scatter (Table \ref{tab:graces_Ba_sys}). P016.2907+28.3957 is the only star in the sample to have detectable La and Nd abundances. The systematic errors for P016.2907+28.3957 are given in Table \ref{tab:P016_sys}.
}

\begin{tabular}{lcccccc}

\hline

ID &
[Y II/Fe]&
[Zr II/Fe]&
[Ba II/Fe] &
[La II/Fe] & 
[Nd II/Fe] &
[Eu II/Fe] \\

\hline

P008.5638+28.1855 & $<$1.15 & $<$2.60 & $<$0.05 & $<$2.25 & $<$2.45 & $<$2.55 \\
P016.2907+28.3957 & $<$0.30 & $<$1.80 & 2.13 $\pm$ 0.19 (2) & 1.63 $\pm$ 0.42 (1) & 1.45 $\pm$ 0.26 (1) & $<$1.75 \\
P021.6938+29.0039 & $<$1.55 & $<$3.15 & $<$0.05 & $<$2.50 & $<$3.05 & $<$2.65 \\
P021.9576+32.4131 & $<$0.55 & $<$2.05 & $<$-0.65 & $<$1.50 & $<$1.65 & $<$1.90 \\
P031.9938+27.7363 & $<$1.30 & $<$2.80 & $<$0.20 & $<$2.40 & $<$2.60 & $<$2.65 \\
P113.8240+45.1863 & $<$1.25 & $<$2.90 & 0.75 $\pm$ 0.27 (2) & $<$2.35 & $<$2.50 & $<$2.30 \\
P116.9657+33.5337 & $<$0.35 & $<$1.85 & $<$-0.95 & $<$1.30 & $<$1.50 & $<$1.60 \\
P133.0683+28.7219 & $<$0.30 & $<$1.65 & -0.15 $\pm$ 0.29 (2) & $<$1.25 & $<$1.40 & $<$1.55 \\
P180.3206+02.5788 & $<$0.80 & $<$2.45 & $<$-0.55 & $<$1.80 & $<$2.00 & $<$2.10 \\
P182.5866+09.8940 & $<$1.50 & $<$3.00 & $<$0.35 & $<$2.60 & $<$2.90 & $<$2.80 \\
P184.1783+01.0664 & $<$1.75 & $<$3.35 & 1.47 $\pm$ 0.16 (2) & $<$2.85 & $<$3.10 & $<$3.30 \\
P184.2997+43.4721 & $<$1.95 & $<$3.60 & $<$0.35 & $<$2.83 & $<$3.10 & $<$3.05 \\
P188.0262+0.2055 & $<$1.60 & $<$3.10 & $<$0.45 & $<$2.70 & $<$2.90 & $<$2.90 \\
P191.8535+12.0508 & $<$1.75 & $<$3.30 & $<$0.65 & $<$2.80 & $<$3.05 & $<$3.15 \\
P192.3242+13.3956 & $<$0.80 & $<$2.30 & $<$-0.55 & $<$1.75 & $<$2.05 & $<$2.00 \\
P194.9935+12.0585 & $<$1.70 & $<$3.40 & $<$0.25 & $<$2.70 & $<$2.95 & $<$2.80 \\
P198.0851+08.9428 & $<$0.60 & $<$2.20 & $<$-0.40 & $<$1.95 & $<$1.95 & $<$2.05 \\
P207.3454+14.1268 & $<$2.00 & $<$3.50 & $<$0.80 & $<$2.93 & $<$3.40 & $<$3.27 \\
P207.9290+03.2767 & $<$1.55 & $<$3.10 & $<$0.30 & $<$2.65 & $<$2.90 & $<$2.80 \\
P209.0986+09.8244 & $<$1.55 & $<$3.00 & $<$0.40 & $<$2.60 & $<$2.85 & $<$2.90 \\
P224.8444+02.3043 & $<$1.35 & $<$3.00 & $<$0.25 & $<$2.35 & $<$2.60 & $<$2.65 \\
P236.9604+11.6155 & $<$1.35 & $<$3.00 & $<$-0.10 & $<$2.15 & $<$2.50 & $<$2.40 \\
P237.8589+12.5660 & $<$2.35 & $<$4.00 & $<$1.25 & $<$3.35 & $<$3.65 & $<$3.75 \\
P244.8986+10.9310 & $<$1.70 & $<$3.10 & 0.66 $\pm$ 0.25 (2) & $<$2.60 & $<$2.95 & $<$3.10 \\
P246.9682+08.5360 & $<$2.35 & $<$3.85 & $<$1.00 & $<$3.30 & $<$3.70 & $<$3.50 \\
P247.2115+06.6348 & $<$0.40 & $<$2.10 & -0.80 $\pm$ 0.21 (2) & $<$1.20 & $<$1.35 & $<$1.45 \\
P257.3131+12.8939 & $<$0.20 & $<$1.75 & -0.40 $\pm$ 0.26 (2) & $<$1.15 & $<$1.35 & $<$1.40 \\
P258.1048+40.5405 & $<$2.55 & $<$4.20 & $<$1.15 & $<$3.65 & $<$4.00 & $<$3.65 \\
P339.1417+25.5503 & $<$1.20 & $<$2.50 & $<$0.10 & $<$2.40 & $<$2.55 & $<$2.55 \\
P339.3203+25.8764 & $<$1.10 & $<$2.55 & $<$-0.25 & $<$2.05 & $<$2.25 & $<$2.40 \\
HD 122563 & -0.15 $\pm$ 0.15 (4) & --- & -0.95 $\pm$ 0.15 (2) & $<$0.45 & $<$0.05 & $<$0.60 \\
\hline
\end{tabular}
\label{tab:graces_neutron}

\end{table*}

%% file: Tables/graces_lithium.txt
\begin{table*}
\centering
\caption{LTE lithium abundances from the Li doublet at 6707 \AA.
The $\sigma$ represents the line measurement error due to the continuum placement. $\Delta$T and $\Delta$g are the systematic errors in the stellar parameters T$_{\rm eff}$ and $\log g$, while $\Delta$Fe is due to the uncertainty in [Fe/H] (see Table~\ref{tab:graces_params_fe}). $\Delta$ is the total combined error of these added in quadrature. }
\begin{tabular}{lcccccc}
\hline
ID &
A(Li) & 
$\sigma$ &
$\Delta$T &
$\Delta$g &
$\Delta$Fe &
$\Delta$
\\

\hline

P008.5638+28.1855 & 1.98 & 0.14 & -0.13 & -0.08 & -0.09 & 0.23 \\
P016.2907+28.3957 & $<$0.82 & & & & &  \\
P021.6938+29.0039 & 1.58 & 0.14 & -0.52 & -0.42 & -0.46 & 0.82 \\
P021.9576+32.4131 & 0.89 & 0.22 & 0.08 & 0.13 & 0.11 & 0.29 \\
P031.9938+27.7363 & 1.73 & 0.18 & -0.42 & -0.37 & -0.38 & 0.70 \\
P113.8240+45.1863 & 1.96 & 0.18 & -0.17 & -0.12 & -0.12 & 0.30 \\
P116.9657+33.5337 & 0.87 & 0.20 & -0.05 & -0.05 & -0.04 & 0.21 \\
P133.0683+28.7219 & $<$0.89 & & & & &  \\
P180.3206+02.5788 & 0.96 & 0.22 & -0.09 & -0.04 & -0.09 & 0.26 \\
P182.5866+09.8940 & 1.82 & 0.16 & -0.08 & -0.08 & -0.10 & 0.22 \\
P184.1783+01.0664 & $<$1.56 & & & & &  \\
P184.2997+43.4721 & $<$1.06 & & & & &  \\
P188.0262+00.2055 & 1.94 & 0.18 & 0.07 & 0.12 & 0.11 & 0.25 \\
P191.8535+12.0508 & 1.60 & 0.16 & -0.43 & -0.43 & -0.41 & 0.75 \\
P192.3242+13.3956 & 1.93 & 0.14 & -0.25 & -0.20 & -0.20 & 0.40 \\
P194.9935+12.0585 & 1.41 & 0.18 & -0.32 & -0.32 & -0.32 & 0.58 \\
P198.0851+08.9428 & $<$1.34 & & & & &  \\
P207.3454+14.1268 & $<$1.79 & & & & &  \\
P207.9290+03.2767 & 1.95 & 0.14 & -0.14 & -0.14 & -0.11 & 0.27 \\
P209.0986+09.8244 & 1.62 & 0.22 & -0.39 & -0.29 & -0.30 & 0.61 \\
P224.8444+02.3043 & $<$1.02 & & & & &  \\
P236.9604+11.6155 & $<$1.00 & & & & &  \\
P237.8589+12.5660 & 1.49 & 0.20 & -0.48 & -0.38 & -0.38 & 0.75 \\
P244.8986+10.9310 & 1.72 & 0.16 & -0.38 & -0.33 & -0.30 & 0.61 \\
P246.9682+08.5360 & 1.92 & 0.27 & -0.11 & 0.04 & 0.04 & 0.30 \\
P247.2115+06.6348 & $<$0.50 & & & & &  \\
P257.3131+12.8939 & $<$0.66 & & & & &  \\
P258.1048+40.5405 & 1.96 & 0.18 & -0.13 & -0.08 & -0.08 & 0.25 \\
P339.1417+25.5503 & 1.90 & 0.14 & -0.11 & -0.11 & -0.10 & 0.23 \\
P339.3203+25.8764 & 1.52 & 0.14 & 0.03 & 0.03 & 0.05 & 0.16 \\
HD122563 & $<$-0.18 & & & & &  \\

\hline
\end{tabular}
\label{tab:graces_li}

\end{table*}

%% file: Tables/graces_alpha_systematics_1.txt
\begin{table*}

\caption{Systematic errors for the $\alpha$-elements (pt. 1). The  $\sigma$ represents the line-to-line scatter for a given element, added in quadrature with errors imposed by the continuum placement for each line used. When the number of lines $N_X > 5$ for species X, then $\sigma$ is reduced by $1/\sqrt{N_X}$. $\Delta$T and $\Delta$g are the systematic errors in the stellar parameters T$_{\rm eff}$ and $\log g$, while $\Delta$Fe is due to the uncertainty in [Fe/H] (see Table~\ref{tab:graces_params_fe}). $\Delta$ is the total combined error of these added in quadrature.}

\begin{tabular}{lccccccccccccccc}

\hline

ID &
\multicolumn{5}{c}{[O I/H]} & 
\multicolumn{5}{c}{[Mg I/H]} &
\multicolumn{5}{c}{[Ca I/H]}  \\

&
$\sigma$ &
$\Delta$T &
$\Delta$g &
$\Delta$Fe &
$\Delta$ &
$\sigma$ &
$\Delta$T &
$\Delta$g &
$\Delta$Fe &
$\Delta$ &
$\sigma$ &
$\Delta$T &
$\Delta$g &
$\Delta$Fe &
$\Delta$  \\

\hline

P008.5638+28.1855 & & & & & & 0.05 & -0.06 & -0.03 & -0.04 & 0.09 & 0.08 & -0.10 & -0.10 & -0.11 & 0.20 \\
P016.2907+28.3957 & & & & & & 0.07 & 0.28 & 0.31 & 0.29 & 0.51 & 0.07 & 0.11 & 0.12 & 0.12 & 0.21 \\
P021.6938+29.0039 & & & & & & 0.13 & -0.27 & -0.22 & -0.21 & 0.42 & 0.10 & -0.33 & -0.28 & -0.27 & 0.52 \\
P021.9576+32.4131 & & & & & & 0.07 & 0.12 & 0.12 & 0.08 & 0.19 & 0.06 & 0.02 & 0.03 & 0.07 & 0.10 \\
P031.9938+27.7363 & & & & & & 0.10 & -0.11 & -0.10 & -0.09 & 0.20 & 0.09 & -0.25 & -0.26 & -0.25 & 0.45 \\
P113.8240+45.1863 & & & & & & 0.26 & -0.02 & -0.01 & -0.03 & 0.26 & 0.11 & -0.12 & -0.11 & -0.12 & 0.23 \\
P116.9657+33.5337 & 0.24 & 0.13 & 0.08 & 0.14 & 0.32 & 0.09 & 0.02 & 0.03 & 0.04 & 0.11 & 0.07 & -0.01 & -0.01 & 0.00 & 0.07 \\
P133.0683+28.7219 & & & & & & 0.39 & 0.17 & 0.14 & 0.20 & 0.49 & 0.07 & 0.00 & 0.00 & 0.01 & 0.07 \\
P180.3206+02.5788 & & & & & & 0.10 & 0.07 & 0.07 & 0.07 & 0.15 & 0.08 & -0.04 & -0.03 & -0.03 & 0.10 \\
P182.5866+09.8940 & & & & & & 0.10 & 0.08 & 0.10 & 0.08 & 0.18 & 0.20 & -0.08 & -0.08 & -0.05 & 0.24 \\
P184.1783+01.0664 & 0.15 & 0.41 & 0.23 & 0.29 & 0.58 & 0.12 & 0.29 & 0.39 & 0.40 & 0.63 & & & & & \\
P184.2997+43.4721 & & & & & & 0.22 & -0.18 & -0.18 & -0.18 & 0.38 & & & & & \\
P188.0262+00.2055 & & & & & & 0.06 & 0.10 & 0.10 & 0.10 & 0.18 & 0.21 & -0.08 & -0.05 & -0.06 & 0.24 \\
P191.8535+12.0508 & & & & & & 0.05 & -0.36 & -0.30 & -0.33 & 0.58 & & & & & \\
P192.3242+13.3956 & & & & & & 0.09 & -0.14 & -0.09 & -0.09 & 0.20 & 0.06 & -0.24 & -0.18 & -0.18 & 0.35 \\
P194.9935+12.0585 & & & & & & 0.22 & 0.21 & 0.21 & 0.21 & 0.42 & 0.16 & -0.14 & -0.15 & -0.15 & 0.30 \\
P198.0851+08.9428 & 0.17 & 0.29 & 0.16 & 0.24 & 0.44 & 0.13 & 0.05 & 0.13 & 0.12 & 0.22 & 0.10 & -0.17 & -0.09 & -0.10 & 0.24 \\
P207.3454+14.1268 & 0.18 & 0.17 & 0.12 & 0.14 & 0.31 & 0.08 & 0.09 & 0.10 & 0.11 & 0.20 & & & & & \\
P207.9290+03.2767 & & & & & & 0.11 & -0.12 & -0.09 & -0.11 & 0.21 & 0.14 & -0.25 & -0.21 & -0.23 & 0.42 \\
P209.0986+09.8244 & & & & & & 0.09 & -0.10 & -0.05 & -0.02 & 0.14 & 0.19 & -0.34 & -0.24 & -0.25 & 0.52 \\
P224.8444+02.3043 & 0.31 & -0.22 & -0.22 & -0.32 & 0.54 & 0.09 & 0.07 & 0.14 & 0.09 & 0.20 & 0.11 & -0.06 & -0.01 & -0.06 & 0.14 \\
P236.9604+11.6155 & & & & & & 0.33 & 0.16 & 0.14 & 0.14 & 0.42 & 0.19 & -0.03 & -0.03 & -0.02 & 0.20 \\
P237.8589+12.5660 & & & & & & 0.13 & -0.40 & -0.33 & -0.36 & 0.65 & & & & & \\
P244.8986+10.9310 & & & & & & 0.16 & -0.23 & -0.16 & -0.18 & 0.37 & 0.14 & -0.30 & -0.25 & -0.21 & 0.47 \\
P246.9682+08.5360 & & & & & & 0.12 & -0.22 & -0.09 & -0.09 & 0.28 & & & & & \\
P247.2115+06.6348 & & & & & & 0.08 & 0.03 & 0.03 & 0.06 & 0.11 & 0.09 & -0.05 & -0.02 & -0.02 & 0.11 \\
P257.3131+12.8939 & & & & & & 0.08 & 0.01 & 0.03 & 0.03 & 0.09 & 0.07 & 0.02 & 0.05 & 0.07 & 0.11 \\
P258.1048+40.5405 & & & & & & 0.15 & -0.03 & -0.03 & -0.03 & 0.16 & & & & & \\
P339.1417+25.5503 & 0.16 & 0.22 & 0.17 & 0.18 & 0.37 & 0.15 & 0.14 & 0.14 & 0.15 & 0.29 & 0.08 & -0.10 & -0.06 & -0.09 & 0.17 \\
P339.3203+25.8764 & & & & & & 0.16 & 0.21 & 0.18 & 0.20 & 0.38 & 0.06 & -0.04 & -0.04 & -0.04 & 0.09 \\
HD 122563 & & & & & & 0.06 & -0.05 & 0.00 & 0.01 & 0.08 & 0.05 & -0.04 & 0.00 & 0.01 & 0.07 \\

\hline
\end{tabular}
\label{tab:graces_alpha_sys_1}

\end{table*}

%% file: Tables/graces_alpha_systematics_2.txt
\begin{table*}

\caption{Systematic errors for the $\alpha$-elements (pt. 2).  See the caption for Table~\ref{tab:graces_alpha_sys_1}.
}
\begin{tabular}{lccccccccccccccc}

\hline

ID &
\multicolumn{5}{c}{[Ca II/H]} & 
\multicolumn{5}{c}{[Ti I/H]} &
\multicolumn{5}{c}{[Ti II/H]}  \\

&
$\sigma$ &
$\Delta$T &
$\Delta$g &
$\Delta$Fe &
$\Delta$ &
$\sigma$ &
$\Delta$T &
$\Delta$g &
$\Delta$Fe &
$\Delta$ &
$\sigma$ &
$\Delta$T &
$\Delta$g &
$\Delta$Fe &
$\Delta$  \\

\hline

P008.5638+28.1855 & 0.10 & -0.03 & -0.00 & 0.01 & 0.11 & & & & & & & & & & \\
P016.2907+28.3957 & 0.11 & 0.18 & 0.18 & 0.21 & 0.34 & 0.07 & 0.04 & 0.05 & 0.07 & 0.12 & 0.07 & 0.02 & -0.01 & 0.02 & 0.07 \\
P021.6938+29.0039 & 0.15 & -0.25 & -0.29 & -0.21 & 0.46 & & & & & & & & & & \\
P021.9576+32.4131 & 0.10 & -0.08 & -0.09 & -0.06 & 0.17 & & & & & & 0.18 & -0.07 & -0.12 & -0.09 & 0.25 \\
P031.9938+27.7363 & 0.10 & -0.18 & -0.20 & -0.19 & 0.34 & & & & & & & & & & \\
P113.8240+45.1863 & 0.16 & -0.17 & -0.12 & -0.12 & 0.29 & & & & & & & & & & \\
P116.9657+33.5337 & 0.19 & -0.17 & -0.20 & -0.18 & 0.37 & 0.08 & -0.10 & -0.08 & -0.07 & 0.17 & 0.10 & -0.14 & -0.15 & -0.14 & 0.27 \\
P133.0683+28.7219 & 0.15 & -0.17 & -0.18 & -0.15 & 0.33 & 0.15 & -0.10 & -0.05 & -0.07 & 0.20 & 0.11 & -0.10 & -0.10 & -0.07 & 0.19 \\
P180.3206+02.5788 & 0.13 & -0.09 & -0.11 & -0.06 & 0.21 & 0.18 & -0.12 & -0.09 & -0.09 & 0.25 & 0.21 & -0.14 & -0.14 & -0.14 & 0.32 \\
P182.5866+09.8940 & 0.11 & 0.07 & 0.07 & 0.05 & 0.15 & & & & & & & & & & \\
P184.1783+01.0664 & 0.08 & -0.00 & -0.05 & 0.01 & 0.10 & & & & & & & & & & \\
P184.2997+43.4721 & 0.09 & -0.04 & -0.02 & -0.04 & 0.11 & & & & & & & & & & \\
P188.0262+00.2055 & 0.12 & 0.06 & 0.06 & 0.05 & 0.16 & & & & & & & & & & \\
P191.8535+12.0508 & 0.10 & 0.04 & 0.01 & 0.03 & 0.11 & & & & & & & & & & \\
P192.3242+13.3956 & 0.24 & -0.21 & -0.21 & -0.23 & 0.44 & & & & & & 0.18 & -0.25 & -0.25 & -0.20 & 0.45 \\
P194.9935+12.0585 & 0.19 & 0.71 & 0.71 & 0.65 & 1.21 & & & & & & & & & & \\
P198.0851+08.9428 & 0.08 & -0.09 & -0.13 & -0.03 & 0.18 & & & & & & 0.12 & -0.14 & -0.17 & -0.14 & 0.29 \\
P207.3454+14.1268 & 0.10 & 0.22 & 0.22 & 0.24 & 0.41 & & & & & & & & & & \\
P207.9290+03.2767 & 0.09 & -0.16 & -0.17 & -0.14 & 0.29 & & & & & & & & & & \\
P209.0986+09.8244 & 0.10 & -0.20 & -0.21 & -0.16 & 0.34 & & & & & & & & & & \\
P224.8444+02.3043 & 0.17 & -0.12 & -0.09 & -0.09 & 0.25 & & & & & & & & & & \\
P236.9604+11.6155 & 0.13 & -0.14 & -0.15 & -0.12 & 0.27 & & & & & & & & & & \\
P237.8589+12.5660 & 0.16 & 0.28 & 0.25 & 0.32 & 0.52 & & & & & & & & & & \\
P244.8986+10.9310 & 0.08 & -0.18 & -0.19 & -0.16 & 0.31 & & & & & & & & & & \\
P246.9682+08.5360 & 0.10 & -0.10 & -0.07 & -0.06 & 0.17 & & & & & & & & & & \\
P247.2115+06.6348 & 0.33 & -0.25 & -0.29 & -0.27 & 0.57 & 0.09 & -0.08 & -0.09 & -0.05 & 0.16 & 0.18 & -0.14 & -0.15 & -0.12 & 0.30 \\
P257.3131+12.8939 & 0.11 & -0.14 & -0.16 & -0.14 & 0.28 & 0.08 & -0.08 & -0.01 & -0.04 & 0.12 & 0.08 & -0.13 & -0.13 & -0.11 & 0.23 \\
P258.1048+40.5405 & 0.15 & -0.03 & -0.03 & -0.03 & 0.16 & & & & & & & & & & \\
P339.1417+25.5503 & 0.11 & -0.04 & -0.04 & -0.03 & 0.13 & 0.18 & -0.16 & -0.16 & -0.15 & 0.33 & & & & & \\
P339.3203+25.8764 & 0.09 & -0.09 & -0.10 & -0.08 & 0.18 & & & & & & & & & & \\
HD 122563 & & & & & & 0.11 & -0.10 & 0.00 & -0.04 & 0.16 & 0.06 & -0.01 & -0.02 & 0.01 & 0.06 \\
\hline
\end{tabular}
\label{tab:graces_alpha_sys_2}

\end{table*}

%% file: Tables/graces_light_systematics.txt
\begin{table*}

\caption{Systematic errors for [Na I/H] and [K I/H].   See the caption for Table~\ref{tab:graces_alpha_sys_1}.
}

\begin{tabular}{lcccccccccc}

\hline

ID &
\multicolumn{5}{c}{[Na I/H]} & 
\multicolumn{5}{c}{[K I/H]} \\

&
$\sigma$ &
$\Delta$T &
$\Delta$g &
$\Delta$Fe &
$\Delta$ &
$\sigma$ &
$\Delta$T &
$\Delta$g &
$\Delta$Fe &
$\Delta$  \\

\hline

P008.5638+28.1855 & 0.06 & -0.05 & 0.00 & -0.01 & 0.08 & & & & & \\
P016.2907+28.3957 & 0.16 & 0.58 & 0.58 & 0.56 & 1.00 & & & & & \\
P021.6938+29.0039 & & & & & & & & & & \\
P021.9576+32.4131 & 0.09 & 0.11 & 0.16 & 0.14 & 0.26 & 0.15 & 0.03 & 0.08 & 0.06 & 0.19 \\
P031.9938+27.7363 & 0.06 & -0.32 & -0.27 & -0.28 & 0.51 & & & & & \\
P113.8240+45.1863 & 0.21 & 0.05 & 0.09 & 0.09 & 0.25 & & & & & \\
P116.9657+33.5337 & 0.15 & 0.10 & 0.15 & 0.11 & 0.26 & 0.17 & -0.05 & -0.05 & -0.04 & 0.19 \\
P133.0683+28.7219 & 0.11 & 0.10 & 0.10 & 0.13 & 0.23 & 0.15 & 0.00 & 0.00 & -0.02 & 0.16 \\
P180.3206+02.5788 & 0.13 & 0.03 & 0.06 & 0.06 & 0.16 & 0.17 & -0.04 & 0.01 & 0.01 & 0.18 \\
P182.5866+09.8940 & 0.13 & -0.10 & -0.08 & -0.10 & 0.21 & & & & & \\
P184.1783+01.0664 & 0.15 & 0.44 & 0.54 & 0.55 & 0.89 & & & & & \\
P184.2997+43.4721 & 0.10 & -0.10 & -0.10 & -0.09 & 0.19 & & & & & \\
P188.0262+00.2055 & 0.10 & -0.03 & -0.03 & -0.04 & 0.12 & & & & & \\
P191.8535+12.0508 & & & & & & & & & & \\
P192.3242+13.3956 & 0.17 & -0.27 & -0.19 & -0.19 & 0.42 & & & & & \\
P194.9935+12.0585 & & & & & & & & & & \\
P198.0851+08.9428 & 0.14 & -0.11 & -0.06 & -0.07 & 0.20 & & & & & \\
P207.3454+14.1268 & 0.12 & 0.10 & 0.15 & 0.12 & 0.25 & & & & & \\
P207.9290+03.2767 & 0.14 & -0.28 & -0.25 & -0.25 & 0.47 & & & & & \\
P209.0986+09.8244 & 0.08 & -0.36 & -0.26 & -0.27 & 0.52 & & & & & \\
P224.8444+02.3043 & 0.06 & 0.00 & 0.09 & 0.04 & 0.12 & & & & & \\
P236.9604+11.6155 & & & & & & & & & & \\
P237.8589+12.5660 & & & & & & & & & & \\
P244.8986+10.9310 & 0.10 & -0.37 & -0.28 & -0.30 & 0.56 & & & & & \\
P246.9682+08.5360 & & & & & & 0.17 & -0.18 & -0.18 & -0.23 & 0.39 \\
P247.2115+06.6348 & 0.16 & 0.09 & 0.09 & 0.12 & 0.24 & & & & & \\
P257.3131+12.8939 & 0.36 & 0.14 & 0.18 & 0.16 & 0.45 & 0.13 & -0.03 & -0.03 & -0.01 & 0.14 \\
P258.1048+40.5405 & 0.16 & -0.13 & -0.08 & -0.08 & 0.23 & & & & & \\
P339.1417+25.5503 & & & & & & & & & & \\
P339.3203+25.8764 & 0.10 & 0.00 & 0.00 & 0.02 & 0.10 & & & & & \\
HD 122563 & 0.08 & -0.05 & -0.04 & 0.01 & 0.10 &  &  &  &  &  \\
\hline
\end{tabular}
\label{tab:graces_light_sys_1}

\end{table*}

%% file: Tables/graces_fe_peak_systematics.txt
\begin{table}
\centering
\caption{Systematic errors for Fe and Fe-peak elements. 
 See the caption for Table~\ref{tab:graces_alpha_sys_1}.
Since only upper limits were determined for [Mn I/H], [Cu I/H], and [Zn I/H], for all stars in the sample, the systematic errors are not shown here.}

\begin{tabular}{lcccccccccccccccccccc}

\hline

ID &
\multicolumn{5}{c}{[Cr I/H]} & 
\multicolumn{5}{c}{[Fe I/H]} &
\multicolumn{5}{c}{[Fe II/H]} &
\multicolumn{5}{c}{[Ni I/H]}

\\

&
$\sigma$ &
$\Delta$T &
$\Delta$g &
$\Delta$Fe &
$\Delta$ &
$\sigma$ &
$\Delta$T &
$\Delta$g &
$\Delta$Fe &
$\Delta$ &
$\sigma$ &
$\Delta$T &
$\Delta$g &
$\Delta$Fe &
$\Delta$ &
$\sigma$ &
$\Delta$T &
$\Delta$g &
$\Delta$Fe &
$\Delta$  \\

\hline

P008.5638+28.1855 & 0.11 & -0.15 & -0.15 & -0.11 & 0.26 & 0.05 & -0.15 & -0.12 & -0.11 & 0.23 & 0.12 & 0.01 & 0.01 & 0.02 & 0.12 & & & & & \\
P016.2907+28.3957 & 0.11 & 0.06 & 0.09 & 0.08 & 0.17 & 0.04 & 0.13 & 0.16 & 0.16 & 0.26 & 0.09 & 0.16 & 0.13 & 0.16 & 0.28 & 0.11 & 0.11 & 0.11 & 0.09 & 0.21 \\
P021.6938+29.0039 & 0.18 & -0.62 & -0.52 & -0.51 & 0.97 & 0.06 & -0.62 & -0.51 & -0.50 & 0.95 & 0.17 & -0.23 & -0.28 & -0.20 & 0.45 & 0.23 & -0.57 & -0.47 & -0.46 & 0.90 \\
P021.9576+32.4131 & 0.07 & 0.05 & 0.07 & 0.05 & 0.12 & 0.05 & 0.05 & 0.07 & 0.07 & 0.13 & 0.07 & -0.06 & -0.07 & -0.04 & 0.13 & 0.11 & 0.03 & 0.08 & 0.06 & 0.15 \\
P031.9938+27.7363 & 0.18 & -0.47 & -0.42 & -0.43 & 0.78 & 0.06 & -0.48 & -0.43 & -0.42 & 0.78 & 0.09 & -0.15 & -0.15 & -0.13 & 0.27 & & & & & \\
P113.8240+45.1863 & 0.15 & -0.22 & -0.20 & -0.20 & 0.38 & 0.06 & -0.16 & -0.14 & -0.15 & 0.27 & 0.17 & 0.04 & 0.04 & 0.04 & 0.18 & & & & & \\
P116.9657+33.5337 & 0.09 & -0.11 & -0.09 & -0.08 & 0.18 & 0.04 & -0.07 & -0.05 & -0.05 & 0.11 & 0.08 & -0.06 & -0.07 & -0.05 & 0.13 & 0.11 & -0.10 & -0.05 & -0.09 & 0.18 \\
P133.0683+28.7219 & 0.08 & -0.07 & -0.04 & -0.04 & 0.12 & 0.04 & -0.06 & -0.03 & -0.05 & 0.09 & 0.07 & -0.04 & -0.08 & -0.05 & 0.13 & 0.27 & -0.09 & -0.05 & -0.06 & 0.30 \\
P180.3206+02.5788 & 0.16 & -0.09 & -0.04 & -0.04 & 0.19 & 0.06 & 0.13 & 0.16 & 0.15 & 0.26 & 0.17 & 0.16 & 0.13 & 0.16 & 0.31 & & & & & \\
P182.5866+09.8940 & & & & & & 0.06 & -0.12 & -0.10 & -0.09 & 0.19 & 0.17 & -0.08 & -0.08 & -0.06 & 0.21 & & & & & \\
P184.1783+01.0664 & & & & & & 0.08 & -0.36 & -0.16 & -0.17 & 0.43 & 0.16 & -0.01 & 0.03 & 0.00 & 0.16 & & & & & \\
P184.2997+43.4721 & 0.20 & -0.21 & -0.16 & -0.18 & 0.38 & 0.07 & -0.18 & -0.13 & -0.15 & 0.28 & & & & & & & & & & \\
P188.0262+00.2055 & & & & & & 0.05 & -0.07 & -0.04 & -0.04 & 0.10 & 0.10 & 0.00 & 0.00 & -0.01 & 0.10 & & & & & \\
P191.8535+12.0508 & & & & & & 0.11 & -0.61 & -0.56 & -0.57 & 1.01 & 0.13 & -0.22 & -0.22 & -0.20 & 0.39 & & & & & \\
P192.3242+13.3956 & 0.17 & -0.41 & -0.33 & -0.33 & 0.64 & 0.05 & -0.38 & -0.29 & -0.29 & 0.56 & 0.11 & -0.11 & -0.12 & -0.08 & 0.21 & 0.11 & -0.40 & -0.35 & -0.35 & 0.65 \\
P194.9935+12.0585 & & & & & & 0.07 & -0.24 & -0.22 & -0.22 & 0.40 & 0.16 & 0.05 & 0.05 & 0.05 & 0.18 & & & & & \\
P198.0851+08.9428 & 0.14 & -0.25 & -0.15 & -0.19 & 0.37 & 0.05 & -0.21 & -0.13 & -0.14 & 0.29 & 0.17 & 0.01 & -0.06 & -0.03 & 0.19 & & & & & \\
P207.3454+14.1268 & & & & & & 0.10 & 0.01 & 0.04 & 0.03 & 0.11 & 0.25 & 0.02 & 0.02 & 0.01 & 0.25 & & & & & \\
P207.9290+03.2767 & 0.14 & -0.39 & -0.37 & -0.36 & 0.66 & 0.05 & -0.41 & -0.39 & -0.38 & 0.68 & 0.11 & -0.10 & -0.13 & -0.10 & 0.22 & & & & & \\
P209.0986+09.8244 & & & & & & 0.05 & -0.53 & -0.40 & -0.40 & 0.78 & 0.09 & -0.16 & -0.16 & -0.14 & 0.28 & 0.25 & -0.49 & -0.39 & -0.35 & 0.76 \\
P224.8444+02.3043 & 0.16 & -0.11 & -0.01 & -0.06 & 0.20 & 0.05 & -0.10 & 0.02 & -0.06 & 0.13 & 0.11 & -0.04 & -0.09 & -0.06 & 0.16 & & & & & \\
P236.9604+11.6155 & & & & & & 0.09 & -0.05 & -0.02 & -0.02 & 0.11 & 0.08 & -0.11 & -0.12 & -0.11 & 0.21 & & & & & \\
P237.8589+12.5660 & & & & & & & & & & & & & & & & & & & & \\
P244.8986+10.9310 & & & & & & 0.05 & -0.54 & -0.45 & -0.42 & 0.82 & 0.11 & -0.17 & -0.17 & -0.15 & 0.31 & & & & & \\
P246.9682+08.5360 & & & & & & & & & & & & & & & & & & & & \\
P247.2115+06.6348 & 0.12 & -0.07 & -0.05 & -0.05 & 0.16 & 0.05 & -0.07 & -0.06 & -0.06 & 0.12 & 0.10 & -0.07 & -0.09 & -0.08 & 0.18 & 0.11 & -0.10 & -0.05 & -0.07 & 0.17 \\
P257.3131+12.8939 & 0.15 & -0.00 & 0.03 & 0.06 & 0.16 & 0.04 & -0.05 & 0.00 & -0.00 & 0.07 & 0.12 & -0.00 & -0.00 & 0.02 & 0.13 & 0.22 & -0.05 & 0.00 & -0.01 & 0.23 \\
P258.1048+40.5405 & & & & & & & & & & & & & & & & & & & & \\
P339.1417+25.5503 & 0.10 & -0.16 & -0.16 & -0.15 & 0.29 & 0.05 & -0.16 & -0.12 & -0.14 & 0.25 & 0.08 & -0.04 & -0.05 & -0.04 & 0.11 & & & & & \\
P339.3203+25.8764 & 0.16 & -0.12 & -0.12 & -0.10 & 0.25 & 0.06 & -0.09 & -0.07 & -0.05 & 0.14 & 0.09 & -0.07 & -0.10 & -0.08 & 0.17 & & & & & \\
HD 122563 & 0.08 & -0.08 & 0.02 & 0.01 & 0.11 & 0.04 & -0.09 & -0.00 & -0.00 & 0.10 & 0.07 & -0.00 & -0.05 & -0.01 & 0.09 & 0.06 & -0.10 & 0.00 & 0.01 & 0.12 \\
\hline
\end{tabular}
\label{tab:graces_fe_peak_sys}

\end{table}

%% file: Tables/graces_Ba_systematics.txt
\begin{table}

\caption{Systematic errors for [Ba II/H].
 See the caption for Table~\ref{tab:graces_alpha_sys_1}.
}

\begin{tabular}{lccccc}

\hline

ID &
\multicolumn{5}{c}{[Ba II/H]}

\\

&
$\sigma$ &
$\Delta$T &
$\Delta$g &
$\Delta$Fe &
$\Delta$  \\

\hline

P008.5638+28.1855 & & & & & \\
P016.2907+28.3957 & 0.12 & 0.14 & 0.11 & 0.14 & 0.25 \\
P021.6938+29.0039 & & & & & \\
P021.9576+32.4131 & & & & & \\
P031.9938+27.7363 & & & & & \\
P113.8240+45.1863 & 0.12 & -0.14 & -0.14 & -0.12 & 0.27 \\
P116.9657+33.5337 & & & & & \\
P133.0683+28.7219 & 0.11 & -0.20 & -0.20 & -0.17 & 0.35 \\
P180.3206+02.5788 & & & & & \\
P182.5866+09.8940 & & & & & \\
P184.1783+01.0664 & 0.12 & -0.16 & -0.06 & -0.02 & 0.21 \\
P184.2997+43.4721 & & & & & \\
P188.0262+00.2055 & & & & & \\
P191.8535+12.0508 & & & & & \\
P192.3242+13.3956 & & & & & \\
P194.9935+12.0585 & & & & & \\
P198.0851+08.9428 & & & & & \\
P207.3454+14.1268 & & & & & \\
P207.9290+03.2767 & & & & & \\
P209.0986+09.8244 & & & & & \\
P224.8444+02.3043 & & & & & \\
P236.9604+11.6155 & & & & & \\
P237.8589+12.5660 & & & & & \\
P244.8986+10.9310 & 0.21 & -0.43 & -0.38 & -0.33 & 0.69 \\
P246.9682+08.5360 & & & & & \\
P247.2115+06.6348 & 0.13 & -0.22 & -0.22 & -0.22 & 0.40 \\
P257.3131+12.8939 & 0.12 & -0.05 & -0.02 & -0.03 & 0.14 \\
P258.1048+40.5405 & & & & & \\
P339.1417+25.5503 & & & & & \\
P339.3203+25.8764 & & & & & \\
HD 122563 & 0.12 & -0.05 & -0.05 & -0.01 & 0.14 \\
\hline
\end{tabular}
\label{tab:graces_Ba_sys}

\end{table}

%% file: Tables/graces_P016_sys.txt
\begin{table*}

\caption{Systematic errors for [Sc II/H], [La II/H] and [Nd II/H] for P016.2907+28.3957.}

\begin{tabular}{lccccccccccccccc}

\hline

ID &
\multicolumn{5}{c}{[Sc II/H]} & 
\multicolumn{5}{c}{[La II/H]} &
\multicolumn{5}{c}{[Nd II/H]}  \\

&
$\sigma$ &
$\Delta$T &
$\Delta$g &
$\Delta$Fe &
$\Delta$ &
$\sigma$ &
$\Delta$T &
$\Delta$g &
$\Delta$Fe &
$\Delta$ &
$\sigma$ &
$\Delta$T &
$\Delta$g &
$\Delta$Fe &
$\Delta$  \\

\hline

P016.2907+28.3957 & 0.30 & 0.21 & 0.16 & 0.19 & 0.44 & 0.23 & 0.21 & 0.21 & 0.19 & 0.42 & 0.23 & -0.06 & -0.06 & -0.03 & 0.25 \\

\hline
\end{tabular}
\label{tab:P016_sys}

\end{table*}

%% file: Tables/graces_nlte_corrections.txt
\begin{table*}

\caption{NLTE Corrections. NLTE deviations were calculated on a line-by-line basis for each star given its line list and stellar parameters; the averaged NLTE correction is given below. Corrections in oxygen are from \citet{Sitnova2013}, Na from \citet{Lind2011}, Mg from \citet{Bergemann2017}, Ca from \citet{Mashonkina2007}, Ti from \citet{Bergemann2011}, and Cr from \citet{Bergemann2010b}. }

\begin{tabular}{lccccccc}

\hline

ID &
$\Delta$[O I/H] &
$\Delta$[Na I/H] &
$\Delta$[Mg I/H] &
$\Delta$[Ca I/H] &
$\Delta$[Ti I/H] &
$\Delta$[Ti II/H] &
$\Delta$[Cr I/H]  \\

\hline

P008.5638+28.1855 &  & -0.17 (1) & 0.05 (4) & 0.13 (3) & 0.35 (2) & 0.04 (2) & 0.45 (1) \\
P016.2907+28.3957 &  & -0.40 (2) & 0.16 (4) & 0.26 (6) & 0.62 (3) & 0.04 (6) & 0.62 (4) \\
P021.6938+29.0039 &  & -0.06 (2) & 0.14 (2) & 0.21 (3) &  & 0.04 (2) & 0.61 (1) \\
P021.9576+32.4131 &  & -0.18 (2) & 0.20 (3) & 0.31 (6) & 0.59 (2) & 0.06 (1) & 0.69 (3) \\
P031.9938+27.7363 &  & -0.13 (1) & 0.12 (3) & 0.25 (3) &  & 0.05 (2) & 0.57 (1) \\
P113.8240+45.1863 & -0.05 (2) & -0.42 (2) & 0.03 (2) & 0.08 (4) & 0.34 (2) & 0.04 (2) & 0.36 (2) \\
P116.9657+33.5337 &  & -0.30 (2) & 0.15 (3) & 0.26 (5) & 0.62 (3) & 0.05 (2) & 0.61 (3) \\
P133.0683+28.7219 &  & -0.31 (2) & 0.15 (2) & 0.27 (6) & 0.62 (2) & 0.04 (1) & 0.60 (3) \\
P180.3206+02.5788 &  & -0.13 (2) & 0.16 (3) & 0.26 (4) & 0.60 (1) & 0.01 (1) & 0.66 (1) \\
P182.5866+09.8940 &  & -0.07 (2) & 0.07 (2) &  &  & 0.04 (2) & 0.47 (2) \\
P184.1783+01.0664 &  & -0.27 (2) & 0.17 (2) & 0.35 (1) &  & 0.04 (1) & 0.63 (2) \\
P184.2997+43.4721 &  & -0.09 (2) & 0.00 (2) & 0.12 (5) & 0.21 (2) & 0.03 (2) & 0.40 (1) \\
P188.0262+0.2055 &  & -0.08 (2) & 0.07 (3) & 0.16 (2) &  & 0.04 (2) & 0.47 (2) \\
P191.8535+12.0508 &  & -0.05 (2) & 0.20 (2) &  &  &  &  \\
P192.3242+13.3956 & -0.07 (2) & -0.10 (2) & 0.09 (3) & 0.18 (4) & 0.49 (2) & 0.05 (1) & 0.53 (2) \\
P194.9935+12.0585 &  & -0.05 (2) & 0.04 (2) & 0.13 (2) & 0.34 (2) & 0.03 (2) & 0.49 (3) \\
P198.0851+08.9428 & -0.08 (2) & -0.21 (2) & 0.09 (2) & 0.17 (3) & 0.49 (2) & 0.05 (2) & 0.53 (2) \\
P207.3454+14.1268 &  & -0.20 (2) & 0.09 (3) & 0.22 (4) &  & 0.04 (1) & 0.48 (2) \\
P207.9290+03.2767 &  & -0.07 (2) & 0.14 (2) & 0.27 (2) &  & 0.05 (2) & 0.58 (2) \\
P209.0986+09.8244 &  & -0.09 (2) & 0.14 (4) & 0.29 (2) &  & 0.04 (2) & 0.61 (2) \\
P224.8444+02.3043 &  & -0.10 (2) & 0.20 (3) & 0.26 (1) &  & -0.00 (2) & 0.72 (1) \\
P236.9604+11.6155 &  & -0.31 (2) & 0.26 (2) & 0.30 (2) &  & 0.05 (2) & 0.76 (2) \\
P237.8589+12.5660 &  & -0.04 (1) & 0.23 (2) &  &  &  &  \\
P244.8986+10.9310 &  & -0.06 (2) & 0.15 (4) & 0.31 (1) &  & 0.03 (1) & 0.62 (2) \\
P246.9682+08.5360 &  & -0.06 (2) & 0.19 (2) &  &  &  & 0.65 (1) \\
P247.2115+06.6348 &  & -0.17 (2) & 0.12 (3) & 0.30 (4) & 0.59 (3) & 0.10 (4) & 0.59 (4) \\
P257.3131+12.8939 &  & -0.32 (2) & 0.15 (4) & 0.24 (6) & 0.61 (3) & 0.07 (5) & 0.59 (3) \\
P258.1048+40.5405 &  & -0.07 (1) & 0.14 (2) &  &  &  &  \\
P339.1417+25.5503 & -0.12 (2) & -0.10 (2) & 0.07 (2) & 0.18 (3) &  & 0.05 (2) & 0.47 (2) \\
P339.3203+25.8764 &  & -0.13 (2) & 0.16 (2) & 0.20 (2) &  & 0.04 (2) & 0.66 (1) \\
HD 122563 &  & -0.33 (2) & 0.08 (3) & 0.22 (7) & 0.48 (4) & 0.08 (7) & 0.39 (5) \\
\hline
\end{tabular}
\label{tab:graces_nlte}

\end{table*}

%% file: Tables/sample_linelist.txt
\begin{table*}

\caption{Sample line list of atomic lines used in this paper. The best fit abundance per line in each star is listed as $A(X)$ = log[$N(X)$/$N(H)$] + 12. Shortened target names are listed by their RA and DEC. The full table is available online. }

\begin{tabular}{lccccccc}

\hline

Elem & 
$\lambda$ & 
$\chi$ & 
$\log(gf)$ & 
P016 & P237 & P247 & HD122563
\\

 & 
(\AA) & 
(eV) & 
 & 
+28 & +12 & +06 & 
\\

\hline

Fe I & 5269.537 & 0.858 & -1.33 & 4.70 & <3.22 & 3.95 & 4.51 \\
Fe I & 5328.039 & 0.914 & -1.47 & 4.70 & <3.37 & 4.00 & 4.61 \\
Fe II & 5169.028 & 2.891 & -0.87 & 4.75 & <3.27 & 4.20 &  \\
Li I & 6707.924 & 0.0 & -0.299 & <0.82 & 1.49 & <0.50 & <-0.18 \\
O I & 7774.166 & 9.139 & 0.17 & 7.38 & <6.80 & <7.06 &  \\
Na I & 5889.951 & 0.0 & 0.12 & 3.71 & <1.98 & 2.94 & 3.81 \\
Mg I & 5172.684 & 2.71 & -0.4 & 5.22 & 3.39 & 4.70 & 5.32 \\
Mg I & 5183.604 & 2.715 & -0.18 & 5.37 & 3.59 & 4.65 & 5.52 \\
K I & 7698.974 & 0.0 & -0.17 & <2.65 & <2.77 & <2.18 & \\
Ca I & 6122.217 & 1.884 & -0.41 & 3.86 & <3.48 & 3.34 & 3.91 \\
Ca I & 6162.173 & 1.897 & 0.1 & 3.51 & <3.03 & 3.14 & 3.61 \\
Sc I & 4670.4 & 1.356 & -1.324 & 1.12 & <1.79 & <1.25 &  \\
Ti I & 4981.731 & 0.848 & 0.57 & 2.42 & <2.09 & 1.90 & 2.22 \\
Ti I & 4991.066 & 0.835 & 0.45 & 2.47 & <2.09 & 1.95 & 2.32 \\
Ti II & 5188.68 & 1.581 & -1.21 & 2.72 & <2.09 & 2.00 & 2.42 \\
Ti II & 5226.543 & 1.565 & -1.3 & 2.67 & <2.09 & 1.80 & 2.27 \\
Cr I & 5206.023 & 0.941 & 0.02 & 2.51 & <2.28 & 2.04 & 2.36 \\
Cr I & 5208.409 & 0.941 & 0.17 & 2.56 & <2.18 & 1.94 & 2.51 \\
Mn I & 4823.514 & 2.317 & -0.466 & <2.65 & <3.02 & <2.48 & 2.35 \\
Ni I & 5476.904 & 1.825 & -0.78 & 3.24 & <2.96 & 2.87 & 3.34 \\
Zn I & 4810.528 & 4.075 & -0.14 & <2.33 & <2.75 & <2.36 & 1.98 \\
Y II & 4900.11 & 1.032 & -0.09 & <-0.12 & <0.30 & <-0.69 & -0.92 \\
Ba II & 6141.73 & 0.704 & -0.077 & 1.35 & <-0.83 & -1.92 & -1.70 \\
Ba II & 6496.91 & 0.604 & -0.38 & 1.40 & <-0.68 & -1.92 & -1.60 \\
La II & 4920.798 & 0.59 & -0.27 & -0.20 & <0.19 & <-1.00 & <-1.33 \\
Nd II & 4811.342 & 0.064 & -1.14 & 0.69 & <1.56 & <0.22 & <-1.06 \\
Nd II & 4825.48 & 0.182 & -0.42 & -0.06 & <0.81 & <-0.53 & <-1.41 \\
Eu II & 6645.072 & 1.379 & -0.517 & <-0.66 & <0.01 & <-1.33 & <-1.76 \\

\hline
\end{tabular}
\label{tab:sample_lines}
\end{table*}

%% file: Tables/inv_par_orbits.txt
\begin{table}

\centering

\caption{Gaia DR2 proper motions, parallaxes, orbital parameters, and action vectors for stars in this paper, based on orbits calculated using the 1/$\varpi$ distances. 
Actions and energies have been normalized to the Sun for brevity ($J_{\phi\odot} = 2009.92$ kpc km s$^{-1}$, 
$J_{z\odot} = 0.35$ kpc km s$^{-1}$, 
$E_{\odot} = -64943.61$ km$^{2}$ s$^{-2}$; \citealt{Sestito19UMP})}
\begin{tabular}{ccccccccccc}
\hline
ID &
$\mu_{\alpha}$ &
$\mu_{\delta}$ &
$\varpi$ &
$r_{peri}$ &
$r_{apo}$ &
Eccentricity &
|Z|$_{max}$ &
$J_{\phi}/J_{\phi\odot}$ &
$J_{z}/J_{z\odot}$ &
$E/E_{\odot}$ \\

&
($\mu$as yr$^{-1}$) &
($\mu$as yr$^{-1}$) &
($\mu$as) &
(kpc) &
(kpc) &
 &
(kpc) &
& & \\
\hline

P008.5638+28.1855 & 44.54 $\pm$ 0.07 & 12.19 $\pm$ 0.07 & 0.8800 $\pm$ 0.0436 & 6.0 $\pm$ 0.1 & 11.7 $\pm$ 0.7 & 0.32 $\pm$ 0.02 & 10.9 $\pm$ 0.6 & -0.27 $\pm$ 0.02 & 3113.98 $\pm$ 11.60 & 0.803 $\pm$ -0.028 \\
P016.2907+28.3957 & 3.66 $\pm$ 0.06 & -1.95 $\pm$ 0.05 & 0.0840 $\pm$ 0.0365 & 5.7 $\pm$ 2.1 & 29.3 $\pm$ 7.0 & 0.68 $\pm$ 0.03 & 20.3 $\pm$ 1.4 & -0.70 $\pm$ 0.40 & 2061.04 $\pm$ 572.95 & 0.391 $\pm$ -0.117 \\
P021.6938+29.0039 & 13.42 $\pm$ 0.09 & 4.54 $\pm$ 0.07 & 0.3643 $\pm$ 0.0452 & 4.8 $\pm$ 0.1 & 11.6 $\pm$ 0.9 & 0.42 $\pm$ 0.03 & 8.4 $\pm$ 1.6 & 0.45 $\pm$ 0.05 & 1532.21 $\pm$ 403.68 & 0.849 $\pm$ -0.038 \\
P021.9576+32.4131 & 6.83 $\pm$ 0.09 & -4.16 $\pm$ 0.08 & 0.1301 $\pm$ 0.0519 & 3.3 $\pm$ 3.3 & 13.2 $\pm$ 3.1 & 0.67 $\pm$ 0.21 & 4.0 $\pm$ 1.2 & -0.39 $\pm$ 0.48 & 254.75 $\pm$ 112.52 & 0.860 $\pm$ -0.159 \\
P031.9938+27.7363 & 12.91 $\pm$ 0.09 & -26.81 $\pm$ 0.10 & 0.4654 $\pm$ 0.0553 & 2.5 $\pm$ 1.1 & 11.8 $\pm$ 0.6 & 0.66 $\pm$ 0.11 & 3.5 $\pm$ 1.4 & -0.42 $\pm$ 0.12 & 307.35 $\pm$ 178.99 & 0.922 $\pm$ -0.062 \\
P113.8240+45.1863 & -33.61 $\pm$ 0.07 & -23.27 $\pm$ 0.05 & 0.7620 $\pm$ 0.0506 & 9.0 $\pm$ 0.1 & 20.9 $\pm$ 2.1 & 0.40 $\pm$ 0.04 & 17.9 $\pm$ 2.3 & 0.64 $\pm$ 0.02 & 3755.69 $\pm$ 437.76 & 0.496 $\pm$ -0.046 \\
P116.9657+33.5337 & 0.94 $\pm$ 0.07 & -11.58 $\pm$ 0.04 & 0.1269 $\pm$ 0.0452 & 5.7 $\pm$ 3.9 & 45.4 $\pm$ 80.6 & 0.76 $\pm$ 0.09 & 24.0 $\pm$ 30.6 & -0.77 $\pm$ 0.79 & 1440.86 $\pm$ 297.30 & 0.325 $\pm$ -0.292 \\
P133.0683+28.7219 & -0.73 $\pm$ 0.08 & -11.92 $\pm$ 0.05 & 0.1443 $\pm$ 0.0514 & 5.7 $\pm$ 4.9 & 16.4 $\pm$ 14.8 & 0.55 $\pm$ 0.25 & 6.9 $\pm$ 7.0 & -0.61 $\pm$ 0.66 & 635.43 $\pm$ 383.09 & 0.748 $\pm$ -0.252 \\
P180.3206+02.5788 & -0.05 $\pm$ 0.10 & -11.36 $\pm$ 0.06 & 0.1616 $\pm$ 0.0608 & 4.4 $\pm$ 1.1 & 59.9 $\pm$ 78.7 & 0.82 $\pm$ 0.10 & 51.2 $\pm$ 58.1 & 0.25 $\pm$ 0.37 & 3365.47 $\pm$ 795.44 & 0.136 $\pm$ -0.228 \\
P182.5866+09.8940 & -76.32 $\pm$ 0.09 & -46.52 $\pm$ 0.06 & 0.7807 $\pm$ 0.0476 & 4.5 $\pm$ 0.4 & 34.2 $\pm$ 12.0 & 0.75 $\pm$ 0.05 & 10.5 $\pm$ 4.6 & -0.87 $\pm$ 0.10 & 372.56 $\pm$ 89.22 & 0.339 $\pm$ -0.162 \\
P184.1783+01.0664 & -1.92 $\pm$ 0.12 & -6.88 $\pm$ 0.07 & 0.2203 $\pm$ 0.0587 & 8.3 $\pm$ 0.1 & 27.3 $\pm$ 0.8 & 0.53 $\pm$ 0.01 & 26.2 $\pm$ 0.8 & -0.30 $\pm$ 0.12 & 5894.07 $\pm$ 637.49 & 0.383 $\pm$ -0.014 \\
P184.2997+43.4721 & -19.30 $\pm$ 0.05 & -37.97 $\pm$ 0.06 & 1.4142 $\pm$ 0.0506 & 1.8 $\pm$ 0.1 & 8.5 $\pm$ 0.0 & 0.64 $\pm$ 0.02 & 2.5 $\pm$ 0.2 & 0.31 $\pm$ 0.02 & 245.74 $\pm$ 4.87 & 1.140 $\pm$ -0.005 \\
P188.0262+00.2055 & 25.18 $\pm$ 0.10 & -35.02 $\pm$ 0.07 & 0.7269 $\pm$ 0.0441 & 1.6 $\pm$ 0.2 & 21.4 $\pm$ 1.9 & 0.86 $\pm$ 0.02 & 4.1 $\pm$ 0.4 & 0.34 $\pm$ 0.03 & 145.12 $\pm$ 5.97 & 0.600 $\pm$ -0.051 \\
P191.8535+12.0508 & -11.78 $\pm$ 0.08 & -0.00 $\pm$ 0.05 & 0.4436 $\pm$ 0.0362 & 4.7 $\pm$ 0.2 & 9.5 $\pm$ 0.2 & 0.34 $\pm$ 0.03 & 2.5 $\pm$ 0.2 & 0.64 $\pm$ 0.02 & 233.92 $\pm$ 21.55 & 0.958 $\pm$ -0.003 \\
P192.3242+13.3956 & -12.93 $\pm$ 0.09 & -1.36 $\pm$ 0.06 & 0.2900 $\pm$ 0.0435 & 2.6 $\pm$ 0.4 & 11.1 $\pm$ 0.8 & 0.62 $\pm$ 0.07 & 4.2 $\pm$ 0.7 & 0.42 $\pm$ 0.05 & 392.90 $\pm$ 52.80 & 0.946 $\pm$ -0.031 \\
P194.9935+12.0585 & -32.89 $\pm$ 0.07 & -27.50 $\pm$ 0.05 & 0.8620 $\pm$ 0.0335 & 0.6 $\pm$ 0.2 & 8.2 $\pm$ 0.4 & 0.86 $\pm$ 0.04 & 3.8 $\pm$ 0.4 & 0.10 $\pm$ 0.03 & 349.95 $\pm$ 37.17 & 1.183 $\pm$ -0.004 \\
P198.0851+8.9428 & -0.22 $\pm$ 0.10 & -0.72 $\pm$ 0.07 & -0.0585 $\pm$ 0.0555 & 23.9 $\pm$ 28.5 & 86.0 $\pm$ 247.6 & 0.33 $\pm$ 0.20 & 55.6 $\pm$ 85.7 & -0.06 $\pm$ 0.56 & 15444.69 $\pm$ 23639.44 & 0.204 $\pm$ -0.445 \\
P207.3454+14.1268 & -6.03 $\pm$ 0.14 & -16.95 $\pm$ 0.10 & 0.5402 $\pm$ 0.0814 & 2.6 $\pm$ 0.6 & 8.0 $\pm$ 0.1 & 0.52 $\pm$ 0.09 & 3.6 $\pm$ 0.4 & 0.35 $\pm$ 0.08 & 485.66 $\pm$ 64.42 & 1.124 $\pm$ -0.021 \\
P207.9290+03.2767 & -27.35 $\pm$ 0.07 & -9.93 $\pm$ 0.05 & 0.4321 $\pm$ 0.0411 & 0.9 $\pm$ 0.4 & 11.6 $\pm$ 65.9 & 0.83 $\pm$ 0.07 & 8.4 $\pm$ 59.7 & -0.12 $\pm$ 0.07 & 813.71 $\pm$ 58.63 & 1.065 $\pm$ -0.043 \\
P209.0986+9.8244 & -20.90 $\pm$ 0.08 & -1.71 $\pm$ 0.06 & 0.3218 $\pm$ 0.0435 & 1.5 $\pm$ 0.4 & 14.1 $\pm$ 2.0 & 0.80 $\pm$ 0.07 & 10.4 $\pm$ 3.3 & 0.19 $\pm$ 0.08 & 1086.27 $\pm$ 316.95 & 0.825 $\pm$ -0.073 \\
P224.8444+02.3043 & -2.31 $\pm$ 0.06 & -0.10 $\pm$ 0.06 & 0.0194 $\pm$ 0.0397 & 13.4 $\pm$ 13.6 & 331.6 $\pm$ 662.9 & 0.76 $\pm$ 0.15 & 239.9 $\pm$ 476.0 & 0.57 $\pm$ 1.10 & 11107.29 $\pm$ 13163.12 & -0.105 $\pm$ -0.728 \\
P236.9604+11.6155 & -0.86 $\pm$ 0.09 & -3.12 $\pm$ 0.14 & 0.1705 $\pm$ 0.0599 & 4.1 $\pm$ 0.6 & 6.3 $\pm$ 0.3 & 0.21 $\pm$ 0.06 & 3.9 $\pm$ 1.0 & 0.38 $\pm$ 0.11 & 883.84 $\pm$ 371.10 & 1.147 $\pm$ -0.035 \\
P237.8589+12.5660 & -5.94 $\pm$ 0.06 & -7.86 $\pm$ 0.05 & 0.2971 $\pm$ 0.0410 & 2.1 $\pm$ 0.4 & 6.6 $\pm$ 0.1 & 0.52 $\pm$ 0.07 & 2.5 $\pm$ 0.4 & 0.30 $\pm$ 0.06 & 328.14 $\pm$ 71.82 & 1.268 $\pm$ -0.026 \\
P244.8986+10.9310 & -16.05 $\pm$ 0.06 & -12.30 $\pm$ 0.04 & 0.3805 $\pm$ 0.0588 & 1.1 $\pm$ 0.6 & 7.9 $\pm$ 14.5 & 0.74 $\pm$ 0.14 & 2.3 $\pm$ 6.0 & -0.17 $\pm$ 0.09 & 209.31 $\pm$ 17.05 & 1.291 $\pm$ -0.013 \\
P246.9682+08.5360 & -6.41 $\pm$ 0.05 & 12.37 $\pm$ 0.04 & 0.2747 $\pm$ 0.0392 & 3.7 $\pm$ 0.4 & 28.7 $\pm$ 6.8 & 0.76 $\pm$ 0.07 & 9.5 $\pm$ 3.5 & 0.69 $\pm$ 0.04 & 398.12 $\pm$ 84.91 & 0.424 $\pm$ -0.118 \\
P247.2115+06.6348 & -1.27 $\pm$ 0.06 & -2.96 $\pm$ 0.03 & 0.0830 $\pm$ 0.0410 & 2.6 $\pm$ 0.9 & 7.9 $\pm$ 2.6 & 0.47 $\pm$ 0.21 & 7.2 $\pm$ 1.9 & -0.02 $\pm$ 0.15 & 1900.73 $\pm$ 456.17 & 1.101 $\pm$ -0.116 \\
P257.3131+12.8939 & 2.14 $\pm$ 0.07 & -1.62 $\pm$ 0.06 & 0.0565 $\pm$ 0.0370 & 8.4 $\pm$ 3.5 & 23.5 $\pm$ 32.4 & 0.37 $\pm$ 0.12 & 22.7 $\pm$ 32.3 & -0.06 $\pm$ 0.26 & 5781.40 $\pm$ 3366.22 & 0.610 $\pm$ -0.363 \\
P258.1048+40.5405 & 0.83 $\pm$ 0.06 & 1.07 $\pm$ 0.08 & 0.3890 $\pm$ 0.0364 & 3.0 $\pm$ 0.0 & 8.3 $\pm$ 0.0 & 0.47 $\pm$ 0.00 & 4.7 $\pm$ 0.2 & 0.37 $\pm$ 0.01 & 736.29 $\pm$ 55.41 & 1.076 $\pm$ -0.005 \\
P339.1417+25.5503 & 33.41 $\pm$ 0.06 & 11.41 $\pm$ 0.05 & 0.9142 $\pm$ 0.0342 & 2.6 $\pm$ 0.1 & 12.8 $\pm$ 0.4 & 0.66 $\pm$ 0.02 & 0.8 $\pm$ 0.0 & 0.50 $\pm$ 0.01 & 22.68 $\pm$ 0.72 & 0.883 $\pm$ -0.014 \\
P339.3203+25.8764 & 20.30 $\pm$ 0.08 & 2.16 $\pm$ 0.06 & 0.2368 $\pm$ 0.0438 & 2.0 $\pm$ 1.1 & 46.0 $\pm$ 76.7 & 0.90 $\pm$ 0.02 & 9.4 $\pm$ 19.7 & -0.45 $\pm$ 0.22 & 157.15 $\pm$ 57.13 & 0.337 $\pm$ -0.308 \\

\hline
\end{tabular}
\label{tab:inv_par_orbits}
\end{table}

%% file: pristine_hr.bbl
\begin{thebibliography}{}
\makeatletter
\relax
\def\mn@urlcharsother{\let\do\@makeother \do\$\do\&\do\#\do\^\do\_\do\%\do\~}
\def\mn@doi{\begingroup\mn@urlcharsother \@ifnextchar [ {\mn@doi@}
  {\mn@doi@[]}}
\def\mn@doi@[#1]#2{\def\@tempa{#1}\ifx\@tempa\@empty \href
  {http://dx.doi.org/#2} {doi:#2}\else \href {http://dx.doi.org/#2} {#1}\fi
  \endgroup}
\def\mn@eprint#1#2{\mn@eprint@#1:#2::\@nil}
\def\mn@eprint@arXiv#1{\href {http://arxiv.org/abs/#1} {{\tt arXiv:#1}}}
\def\mn@eprint@dblp#1{\href {http://dblp.uni-trier.de/rec/bibtex/#1.xml}
  {dblp:#1}}
\def\mn@eprint@#1:#2:#3:#4\@nil{\def\@tempa {#1}\def\@tempb {#2}\def\@tempc
  {#3}\ifx \@tempc \@empty \let \@tempc \@tempb \let \@tempb \@tempa \fi \ifx
  \@tempb \@empty \def\@tempb {arXiv}\fi \@ifundefined
  {mn@eprint@\@tempb}{\@tempb:\@tempc}{\expandafter \expandafter \csname
  mn@eprint@\@tempb\endcsname \expandafter{\@tempc}}}

\bibitem[\protect\citeauthoryear{{Abate}, {Pols}, {Izzard}, {Mohamed}  \& {de
  Mink}}{{Abate} et~al.}{2013}]{Abate13}
{Abate} C.,  {Pols} O.~R.,  {Izzard} R.~G.,  {Mohamed} S.~S.,   {de Mink}
  S.~E.,  2013, \mn@doi [\aap] {10.1051/0004-6361/201220007}, \href
  {http://adsabs.harvard.edu/abs/2013A%26A...552A..26A} {552, A26}

\bibitem[\protect\citeauthoryear{{Abel}, {Bryan}  \& {Norman}}{{Abel}
  et~al.}{2002}]{Abel02}
{Abel} T.,  {Bryan} G.~L.,   {Norman} M.~L.,  2002, \mn@doi [Science]
  {10.1126/science.295.5552.93}, \href
  {https://ui.adsabs.harvard.edu/abs/2002Sci...295...93A} {295, 93}

\bibitem[\protect\citeauthoryear{{Aguado}, {Gonz{\'a}lez Hern{\'a}ndez},
  {Allende Prieto}  \& {Rebolo}}{{Aguado} et~al.}{2017}]{Aguado2017WHT}
{Aguado} D.~S.,  {Gonz{\'a}lez Hern{\'a}ndez} J.~I.,  {Allende Prieto} C.,
  {Rebolo} R.,  2017, \mn@doi [\aap] {10.1051/0004-6361/201730654}, \href
  {https://ui.adsabs.harvard.edu/abs/2017A&A...605A..40A} {605, A40}

\bibitem[\protect\citeauthoryear{{Aguado}, {Gonz{\'a}lez Hern{\'a}ndez},
  {Allende Prieto}  \& {Rebolo}}{{Aguado} et~al.}{2018a}]{Aguado18UMPb}
{Aguado} D.~S.,  {Gonz{\'a}lez Hern{\'a}ndez} J.~I.,  {Allende Prieto} C.,
  {Rebolo} R.,  2018a, \mn@doi [\apjl] {10.3847/2041-8213/aaa23a}, \href
  {https://ui.adsabs.harvard.edu/abs/2018ApJ...852L..20A} {852, L20}

\bibitem[\protect\citeauthoryear{{Aguado}, {Allende Prieto}, {Gonz{\'a}lez
  Hern{\'a}ndez}  \& {Rebolo}}{{Aguado} et~al.}{2018b}]{Aguado18UMPa}
{Aguado} D.~S.,  {Allende Prieto} C.,  {Gonz{\'a}lez Hern{\'a}ndez} J.~I.,
  {Rebolo} R.,  2018b, \mn@doi [\apjl] {10.3847/2041-8213/aaadb8}, \href
  {https://ui.adsabs.harvard.edu/abs/2018ApJ...854L..34A} {854, L34}

\bibitem[\protect\citeauthoryear{{Aguado} et~al.,}{{Aguado}
  et~al.}{2019}]{Aguado19Pristine}
{Aguado} D.~S.,  et~al., 2019, \mn@doi [\mnras] {10.1093/mnras/stz2643}, \href
  {https://ui.adsabs.harvard.edu/abs/2019MNRAS.490.2241A} {490, 2241}

\bibitem[\protect\citeauthoryear{{Andrae} et~al.,}{{Andrae}
  et~al.}{2018}]{Andrae2018}
{Andrae} R.,  et~al., 2018, \mn@doi [\aap] {10.1051/0004-6361/201732516}, \href
  {https://ui.adsabs.harvard.edu/abs/2018A&A...616A...8A} {616, A8}

\bibitem[\protect\citeauthoryear{{Aoki} et~al.,}{{Aoki} et~al.}{2009}]{Aoki09}
{Aoki} W.,  et~al., 2009, \mn@doi [\aap] {10.1051/0004-6361/200911959}, \href
  {http://adsabs.harvard.edu/abs/2009A%26A...502..569A} {502, 569}

\bibitem[\protect\citeauthoryear{{Aoki} et~al.,}{{Aoki} et~al.}{2013}]{Aoki13}
{Aoki} W.,  et~al., 2013, \mn@doi [\aj] {10.1088/0004-6256/145/1/13}, \href
  {http://adsabs.harvard.edu/abs/2013AJ....145...13A} {145, 13}

\bibitem[\protect\citeauthoryear{{Arentsen} et~al.,}{{Arentsen}
  et~al.}{2020a}]{Arentsen20}
{Arentsen} A.,  et~al., 2020a, \mn@doi [\mnras] {10.1093/mnrasl/slz156}, \href
  {https://ui.adsabs.harvard.edu/abs/2020MNRAS.491L..11A} {491, L11}

\bibitem[\protect\citeauthoryear{{Arentsen} et~al.,}{{Arentsen}
  et~al.}{2020b}]{Arentsen20b}
{Arentsen} A.,  et~al., 2020b, \mn@doi [\mnras] {10.1093/mnras/staa1661}, \href
  {https://ui.adsabs.harvard.edu/abs/2020MNRAS.496.4964A} {496, 4964}

\bibitem[\protect\citeauthoryear{{Asplund}, {Grevesse}, {Sauval}  \&
  {Scott}}{{Asplund} et~al.}{2009}]{Asplund09}
{Asplund} M.,  {Grevesse} N.,  {Sauval} A.~J.,   {Scott} P.,  2009, \mn@doi
  [\araa] {10.1146/annurev.astro.46.060407.145222}, \href
  {http://adsabs.harvard.edu/abs/2009ARA%26A..47..481A} {47, 481}

\bibitem[\protect\citeauthoryear{{Astropy Collaboration} et~al.,}{{Astropy
  Collaboration} et~al.}{2013}]{Astropy1}
{Astropy Collaboration} et~al., 2013, \mn@doi [\aap]
  {10.1051/0004-6361/201322068}, \href
  {http://adsabs.harvard.edu/abs/2013A%26A...558A..33A} {558, A33}

\bibitem[\protect\citeauthoryear{{Astropy Collaboration} et~al.,}{{Astropy
  Collaboration} et~al.}{2018}]{Astropy2}
{Astropy Collaboration} et~al., 2018, \mn@doi [aj] {10.3847/1538-3881/aabc4f},
  \href {https://ui.adsabs.harvard.edu/abs/2018AJ....156..123A} {156, 123}

\bibitem[\protect\citeauthoryear{{Barb{\'a}}, {Minniti}, {Geisler},
  {Alonso-Garc{\'\i}a}, {Hempel}, {Monachesi}, {Arias}  \&
  {G{\'o}mez}}{{Barb{\'a}} et~al.}{2019}]{Barba19}
{Barb{\'a}} R.~H.,  {Minniti} D.,  {Geisler} D.,  {Alonso-Garc{\'\i}a} J.,
  {Hempel} M.,  {Monachesi} A.,  {Arias} J.~I.,   {G{\'o}mez} F.~A.,  2019,
  \mn@doi [\apjl] {10.3847/2041-8213/aaf811}, \href
  {https://ui.adsabs.harvard.edu/abs/2019ApJ...870L..24B} {870, L24}

\bibitem[\protect\citeauthoryear{{Beers} \& {Christlieb}}{{Beers} \&
  {Christlieb}}{2005}]{BC05}
{Beers} T.~C.,  {Christlieb} N.,  2005, \mn@doi [\araa]
  {10.1146/annurev.astro.42.053102.134057}, \href
  {http://adsabs.harvard.edu/abs/2005ARA%26A..43..531B} {43, 531}

\bibitem[\protect\citeauthoryear{{Beers}, {Preston}  \& {Shectman}}{{Beers}
  et~al.}{1985}]{Beers85}
{Beers} T.~C.,  {Preston} G.~W.,   {Shectman} S.~A.,  1985, \mn@doi [\aj]
  {10.1086/113917}, \href {http://adsabs.harvard.edu/abs/1985AJ.....90.2089B}
  {90, 2089}

\bibitem[\protect\citeauthoryear{{Beers}, {Preston}  \& {Shectman}}{{Beers}
  et~al.}{1992}]{Beers92}
{Beers} T.~C.,  {Preston} G.~W.,   {Shectman} S.~A.,  1992, \mn@doi [\aj]
  {10.1086/116207}, \href
  {https://ui.adsabs.harvard.edu/abs/1992AJ....103.1987B} {103, 1987}

\bibitem[\protect\citeauthoryear{{Beers}, {Rossi}, {Norris}, {Ryan}  \&
  {Shefler}}{{Beers} et~al.}{1999}]{Beers99}
{Beers} T.~C.,  {Rossi} S.,  {Norris} J.~E.,  {Ryan} S.~G.,   {Shefler} T.,
  1999, \mn@doi [\aj] {10.1086/300727}, \href
  {http://adsabs.harvard.edu/abs/1999AJ....117..981B} {117, 981}

\bibitem[\protect\citeauthoryear{{Belokurov}, {Erkal}, {Evans}, {Koposov}  \&
  {Deason}}{{Belokurov} et~al.}{2018}]{Belokurov18}
{Belokurov} V.,  {Erkal} D.,  {Evans} N.~W.,  {Koposov} S.~E.,   {Deason}
  A.~J.,  2018, \mn@doi [\mnras] {10.1093/mnras/sty982}, \href
  {https://ui.adsabs.harvard.edu/abs/2018MNRAS.478..611B} {478, 611}

\bibitem[\protect\citeauthoryear{{Bergemann}}{{Bergemann}}{2011}]{Bergemann2011}
{Bergemann} M.,  2011, \mn@doi [\mnras] {10.1111/j.1365-2966.2011.18295.x},
  \href {https://ui.adsabs.harvard.edu/abs/2011MNRAS.413.2184B} {413, 2184}

\bibitem[\protect\citeauthoryear{{Bergemann} \& {Cescutti}}{{Bergemann} \&
  {Cescutti}}{2010}]{Bergemann2010b}
{Bergemann} M.,  {Cescutti} G.,  2010, \mn@doi [\aap]
  {10.1051/0004-6361/201014250}, \href
  {https://ui.adsabs.harvard.edu/abs/2010A&A...522A...9B} {522, A9}

\bibitem[\protect\citeauthoryear{{Bergemann}, {Lind}, {Collet}, {Magic}  \&
  {Asplund}}{{Bergemann} et~al.}{2012}]{Bergemann2012}
{Bergemann} M.,  {Lind} K.,  {Collet} R.,  {Magic} Z.,   {Asplund} M.,  2012,
  \mn@doi [\mnras] {10.1111/j.1365-2966.2012.21687.x}, \href
  {https://ui.adsabs.harvard.edu/abs/2012MNRAS.427...27B} {427, 27}

\bibitem[\protect\citeauthoryear{{Bergemann}, {Collet}, {Amarsi}, {Kovalev},
  {Ruchti}  \& {Magic}}{{Bergemann} et~al.}{2017}]{Bergemann2017}
{Bergemann} M.,  {Collet} R.,  {Amarsi} A.~M.,  {Kovalev} M.,  {Ruchti} G.,
  {Magic} Z.,  2017, \mn@doi [\apj] {10.3847/1538-4357/aa88cb}, \href
  {https://ui.adsabs.harvard.edu/abs/2017ApJ...847...15B} {847, 15}

\bibitem[\protect\citeauthoryear{{Bond}}{{Bond}}{1980}]{bond1980}
{Bond} H.~E.,  1980, \mn@doi [\apjs] {10.1086/190703}, \href
  {https://ui.adsabs.harvard.edu/abs/1980ApJS...44..517B} {44, 517}

\bibitem[\protect\citeauthoryear{{Bonifacio} et~al.,}{{Bonifacio}
  et~al.}{2007}]{Bonifacio2007a}
{Bonifacio} P.,  et~al., 2007, \mn@doi [\aap] {10.1051/0004-6361:20064834},
  \href {https://ui.adsabs.harvard.edu/abs/2007A&A...462..851B} {462, 851}

\bibitem[\protect\citeauthoryear{{Bonifacio}, {Sbordone}, {Caffau}, {Ludwig},
  {Spite}, {Gonz{\'a}lez Hern{\'a}ndez}  \& {Behara}}{{Bonifacio}
  et~al.}{2012}]{Bonifacio2012}
{Bonifacio} P.,  {Sbordone} L.,  {Caffau} E.,  {Ludwig} H.~G.,  {Spite} M.,
  {Gonz{\'a}lez Hern{\'a}ndez} J.~I.,   {Behara} N.~T.,  2012, \mn@doi [\aap]
  {10.1051/0004-6361/201219004}, \href
  {https://ui.adsabs.harvard.edu/abs/2012A&A...542A..87B} {542, A87}

\bibitem[\protect\citeauthoryear{{Bonifacio} et~al.,}{{Bonifacio}
  et~al.}{2018}]{Bonifacio18}
{Bonifacio} P.,  et~al., 2018, \mn@doi [\aap] {10.1051/0004-6361/201732320},
  \href {https://ui.adsabs.harvard.edu/abs/2018A&A...612A..65B} {612, A65}

\bibitem[\protect\citeauthoryear{{Bonifacio}, {Caffau}, {Spite}  \&
  {Spite}}{{Bonifacio} et~al.}{2019a}]{Bonifacio19}
{Bonifacio} P.,  {Caffau} E.,  {Spite} M.,   {Spite} F.,  2019a, \mn@doi
  [Research Notes of the American Astronomical Society]
  {10.3847/2515-5172/ab1b4d}, \href
  {https://ui.adsabs.harvard.edu/abs/2019RNAAS...3...64B} {3, 64}

\bibitem[\protect\citeauthoryear{{Bonifacio} et~al.,}{{Bonifacio}
  et~al.}{2019b}]{Bonifacio2019}
{Bonifacio} P.,  et~al., 2019b, \mn@doi [\mnras] {10.1093/mnras/stz1378}, \href
  {https://ui.adsabs.harvard.edu/abs/2019MNRAS.487.3797B} {487, 3797}

\bibitem[\protect\citeauthoryear{{Bovy}}{{Bovy}}{2015}]{Galpy}
{Bovy} J.,  2015, \mn@doi [\apjs] {10.1088/0067-0049/216/2/29}, \href
  {https://ui.adsabs.harvard.edu/abs/2015ApJS..216...29B} {216, 29}

\bibitem[\protect\citeauthoryear{{Bovy}, {Rix}, {Schlafly}, {Nidever},
  {Holtzman}, {Shetrone}  \& {Beers}}{{Bovy} et~al.}{2016}]{Bovy16b}
{Bovy} J.,  {Rix} H.-W.,  {Schlafly} E.~F.,  {Nidever} D.~L.,  {Holtzman}
  J.~A.,  {Shetrone} M.,   {Beers} T.~C.,  2016, \mn@doi [\apj]
  {10.3847/0004-637X/823/1/30}, \href
  {http://adsabs.harvard.edu/abs/2016ApJ...823...30B} {823, 30}

\bibitem[\protect\citeauthoryear{{Bromm}, {Coppi}  \& {Larson}}{{Bromm}
  et~al.}{1999}]{Bromm1999}
{Bromm} V.,  {Coppi} P.~S.,   {Larson} R.~B.,  1999, \mn@doi [\apjl]
  {10.1086/312385}, \href
  {https://ui.adsabs.harvard.edu/abs/1999ApJ...527L...5B} {527, L5}

\bibitem[\protect\citeauthoryear{{Buck}}{{Buck}}{2020}]{Buck2020}
{Buck} T.,  2020, \mn@doi [\mnras] {10.1093/mnras/stz3289}, \href
  {https://ui.adsabs.harvard.edu/abs/2020MNRAS.491.5435B} {491, 5435}

\bibitem[\protect\citeauthoryear{{Buzzoni} et~al.,}{{Buzzoni}
  et~al.}{1984}]{Buzzoni1984}
{Buzzoni} B.,  et~al., 1984, The Messenger, \href
  {https://ui.adsabs.harvard.edu/abs/1984Msngr..38....9B} {38, 9}

\bibitem[\protect\citeauthoryear{{Caffau} et~al.,}{{Caffau}
  et~al.}{2012}]{Caffau12}
{Caffau} E.,  et~al., 2012, \mn@doi [\aap] {10.1051/0004-6361/201118744}, \href
  {https://ui.adsabs.harvard.edu/abs/2012A&A...542A..51C} {542, A51}

\bibitem[\protect\citeauthoryear{{Caffau} et~al.,}{{Caffau}
  et~al.}{2017}]{Caffau2017}
{Caffau} E.,  et~al., 2017, \mn@doi [Astronomische Nachrichten]
  {10.1002/asna.201713368}, \href
  {https://ui.adsabs.harvard.edu/abs/2017AN....338..686C} {338, 686}

\bibitem[\protect\citeauthoryear{{Carney} \& {Peterson}}{{Carney} \&
  {Peterson}}{1981}]{carney1981}
{Carney} B.~W.,  {Peterson} R.~C.,  1981, \mn@doi [\apj] {10.1086/158804},
  \href {https://ui.adsabs.harvard.edu/abs/1981ApJ...245..238C} {245, 238}

\bibitem[\protect\citeauthoryear{{Carney}, {Latham}, {Stefanik}, {Laird}  \&
  {Morse}}{{Carney} et~al.}{2003}]{Carney03}
{Carney} B.~W.,  {Latham} D.~W.,  {Stefanik} R.~P.,  {Laird} J.~B.,   {Morse}
  J.~A.,  2003, \mn@doi [\aj] {10.1086/345386}, \href
  {http://adsabs.harvard.edu/abs/2003AJ....125..293C} {125, 293}

\bibitem[\protect\citeauthoryear{{Casagrande} et~al.,}{{Casagrande}
  et~al.}{2014}]{Casagrande2014}
{Casagrande} L.,  et~al., 2014, \mn@doi [\mnras] {10.1093/mnras/stu089}, \href
  {https://ui.adsabs.harvard.edu/abs/2014MNRAS.439.2060C} {439, 2060}

\bibitem[\protect\citeauthoryear{{Casagrande} et~al.,}{{Casagrande}
  et~al.}{2016}]{Casagrande16}
{Casagrande} L.,  et~al., 2016, \mn@doi [\mnras] {10.1093/mnras/stv2320}, \href
  {https://ui.adsabs.harvard.edu/abs/2016MNRAS.455..987C} {455, 987}

\bibitem[\protect\citeauthoryear{{Casagrande} et~al.,}{{Casagrande}
  et~al.}{2020}]{Casagrande2020}
{Casagrande} L.,  et~al., 2020, arXiv e-prints, \href
  {https://ui.adsabs.harvard.edu/abs/2020arXiv201102517C} {p. arXiv:2011.02517}

\bibitem[\protect\citeauthoryear{{Chene} et~al.,}{{Chene}
  et~al.}{2014}]{Chene2014}
{Chene} A.-N.,  et~al., 2014, {GRACES: Gemini remote access to CFHT ESPaDOnS
  spectrograph through the longest astronomical fiber ever made: experimental
  phase completed}.
p. 915147, \mn@doi{10.1117/12.2057417}

\bibitem[\protect\citeauthoryear{Choi, Dotter, Conroy, Cantiello, Paxton  \&
  Johnson}{Choi et~al.}{2016}]{choi2016mesa}
Choi J.,  Dotter A.,  Conroy C.,  Cantiello M.,  Paxton B.,   Johnson B.~D.,
  2016, The Astrophysical Journal, 823, 102

\bibitem[\protect\citeauthoryear{{Christlieb}, {Wisotzki}  \&
  {Gra{\ss}hoff}}{{Christlieb} et~al.}{2002}]{Christlieb02}
{Christlieb} N.,  {Wisotzki} L.,   {Gra{\ss}hoff} G.,  2002, \mn@doi [\aap]
  {10.1051/0004-6361:20020830}, \href
  {http://adsabs.harvard.edu/abs/2002A%26A...391..397C} {391, 397}

\bibitem[\protect\citeauthoryear{{Clark}, {Glover}, {Klessen}  \&
  {Bromm}}{{Clark} et~al.}{2011}]{Clark11}
{Clark} P.~C.,  {Glover} S. C.~O.,  {Klessen} R.~S.,   {Bromm} V.,  2011,
  \mn@doi [\apj] {10.1088/0004-637X/727/2/110}, \href
  {https://ui.adsabs.harvard.edu/abs/2011ApJ...727..110C} {727, 110}

\bibitem[\protect\citeauthoryear{{Clarkson}, {Herwig}  \&
  {Pignatari}}{{Clarkson} et~al.}{2018}]{Clarkson18}
{Clarkson} O.,  {Herwig} F.,   {Pignatari} M.,  2018, \mn@doi [\mnras]
  {10.1093/mnrasl/slx190}, \href
  {https://ui.adsabs.harvard.edu/abs/2018MNRAS.474L..37C} {474, L37}

\bibitem[\protect\citeauthoryear{{Coc} \& {Vangioni}}{{Coc} \&
  {Vangioni}}{2017}]{Coc2017}
{Coc} A.,  {Vangioni} E.,  2017, \mn@doi [International Journal of Modern
  Physics E] {10.1142/S0218301317410026}, \href
  {https://ui.adsabs.harvard.edu/abs/2017IJMPE..2641002C} {26, 1741002}

\bibitem[\protect\citeauthoryear{{Cohen} \& {Kirby}}{{Cohen} \&
  {Kirby}}{2012}]{Cohen2012}
{Cohen} J.~G.,  {Kirby} E.~N.,  2012, \mn@doi [\apj]
  {10.1088/0004-637X/760/1/86}, \href
  {https://ui.adsabs.harvard.edu/abs/2012ApJ...760...86C} {760, 86}

\bibitem[\protect\citeauthoryear{{Cohen}, {McWilliam}, {Christlieb},
  {Shectman}, {Thompson}, {Melendez}, {Wisotzki}  \& {Reimers}}{{Cohen}
  et~al.}{2007}]{Cohen2007}
{Cohen} J.~G.,  {McWilliam} A.,  {Christlieb} N.,  {Shectman} S.,  {Thompson}
  I.,  {Melendez} J.,  {Wisotzki} L.,   {Reimers} D.,  2007, \mn@doi [\apjl]
  {10.1086/518031}, \href
  {https://ui.adsabs.harvard.edu/abs/2007ApJ...659L.161C} {659, L161}

\bibitem[\protect\citeauthoryear{{Cohen}, {Huang}  \& {Kirby}}{{Cohen}
  et~al.}{2011}]{Cohen2011}
{Cohen} J.~G.,  {Huang} W.,   {Kirby} E.~N.,  2011, \mn@doi [\apj]
  {10.1088/0004-637X/740/2/60}, \href
  {https://ui.adsabs.harvard.edu/abs/2011ApJ...740...60C} {740, 60}

\bibitem[\protect\citeauthoryear{{Cohen}, {Christlieb}, {Thompson},
  {McWilliam}, {Shectman}, {Reimers}, {Wisotzki}  \& {Kirby}}{{Cohen}
  et~al.}{2013}]{Cohen13}
{Cohen} J.~G.,  {Christlieb} N.,  {Thompson} I.,  {McWilliam} A.,  {Shectman}
  S.,  {Reimers} D.,  {Wisotzki} L.,   {Kirby} E.,  2013, \mn@doi [\apj]
  {10.1088/0004-637X/778/1/56}, \href
  {http://adsabs.harvard.edu/abs/2013ApJ...778...56C} {778, 56}

\bibitem[\protect\citeauthoryear{{Collet}, {Nordlund}, {Asplund}, {Hayek}  \&
  {Trampedach}}{{Collet} et~al.}{2018}]{Collet2018}
{Collet} R.,  {Nordlund} {\r{A}}.,  {Asplund} M.,  {Hayek} W.,   {Trampedach}
  R.,  2018, \mn@doi [\mnras] {10.1093/mnras/sty002}, \href
  {https://ui.adsabs.harvard.edu/abs/2018MNRAS.475.3369C} {475, 3369}

\bibitem[\protect\citeauthoryear{{Cooke} \& {Madau}}{{Cooke} \&
  {Madau}}{2014}]{Cooke14}
{Cooke} R.~J.,  {Madau} P.,  2014, \mn@doi [\apj]
  {10.1088/0004-637X/791/2/116}, \href
  {https://ui.adsabs.harvard.edu/abs/2014ApJ...791..116C} {791, 116}

\bibitem[\protect\citeauthoryear{{Cordoni} et~al.,}{{Cordoni}
  et~al.}{2021}]{Cordoni2020}
{Cordoni} G.,  et~al., 2021, \mn@doi [\mnras] {10.1093/mnras/staa3417}, \href
  {https://ui.adsabs.harvard.edu/abs/2021MNRAS.503.2539C} {503, 2539}

\bibitem[\protect\citeauthoryear{{C{\^o}t{\'e}}, {West}, {Heger}, {Ritter},
  {O'Shea}, {Herwig}, {Travaglio}  \& {Bisterzo}}{{C{\^o}t{\'e}}
  et~al.}{2016}]{Cote16}
{C{\^o}t{\'e}} B.,  {West} C.,  {Heger} A.,  {Ritter} C.,  {O'Shea} B.~W.,
  {Herwig} F.,  {Travaglio} C.,   {Bisterzo} S.,  2016, \mn@doi [\mnras]
  {10.1093/mnras/stw2244}, \href
  {http://adsabs.harvard.edu/abs/2016MNRAS.463.3755C} {463, 3755}

\bibitem[\protect\citeauthoryear{{Cristallo}, {Straniero}, {Piersanti}  \&
  {Gobrecht}}{{Cristallo} et~al.}{2015}]{Cristallo2015}
{Cristallo} S.,  {Straniero} O.,  {Piersanti} L.,   {Gobrecht} D.,  2015,
  \mn@doi [\apjs] {10.1088/0067-0049/219/2/40}, \href
  {https://ui.adsabs.harvard.edu/abs/2015ApJS..219...40C} {219, 40}

\bibitem[\protect\citeauthoryear{{Cui} et~al.,}{{Cui} et~al.}{2012}]{Cui12}
{Cui} X.-Q.,  et~al., 2012, \mn@doi [Research in Astronomy and Astrophysics]
  {10.1088/1674-4527/12/9/003}, \href
  {https://ui.adsabs.harvard.edu/abs/2012RAA....12.1197C} {12, 1197}

\bibitem[\protect\citeauthoryear{{Cyburt}, {Fields}, {Olive}  \&
  {Yeh}}{{Cyburt} et~al.}{2016}]{Cyburt16}
{Cyburt} R.~H.,  {Fields} B.~D.,  {Olive} K.~A.,   {Yeh} T.-H.,  2016, \mn@doi
  [Reviews of Modern Physics] {10.1103/RevModPhys.88.015004}, \href
  {https://ui.adsabs.harvard.edu/abs/2016RvMP...88a5004C} {88, 015004}

\bibitem[\protect\citeauthoryear{{Da Costa} et~al.,}{{Da Costa}
  et~al.}{2019}]{DaCosta2019}
{Da Costa} G.~S.,  et~al., 2019, \mn@doi [\mnras] {10.1093/mnras/stz2550},
  \href {https://ui.adsabs.harvard.edu/abs/2019MNRAS.489.5900D} {489, 5900}

\bibitem[\protect\citeauthoryear{{Dalton} et~al.,}{{Dalton}
  et~al.}{2014}]{Dalton14}
{Dalton} G.,  et~al., 2014, in Ground-based and Airborne Instrumentation for
  Astronomy V. p. 91470L (\mn@eprint {arXiv} {1412.0843}),
  \mn@doi{10.1117/12.2055132}

\bibitem[\protect\citeauthoryear{{Dalton} et~al.,}{{Dalton}
  et~al.}{2018}]{Dalton2018}
{Dalton} G.,  et~al., 2018, in {Evans} C.~J.,  {Simard} L.,   {Takami} H.,
  eds,  Society of Photo-Optical Instrumentation Engineers (SPIE) Conference
  Series Vol. 10702, Ground-based and Airborne Instrumentation for Astronomy
  VII. p. 107021B, \mn@doi{10.1117/12.2312031}

\bibitem[\protect\citeauthoryear{{Demarque}, {Woo}, {Kim}  \& {Yi}}{{Demarque}
  et~al.}{2004}]{YY1}
{Demarque} P.,  {Woo} J.-H.,  {Kim} Y.-C.,   {Yi} S.~K.,  2004, \mn@doi [\apjs]
  {10.1086/424966}, \href
  {https://ui.adsabs.harvard.edu/abs/2004ApJS..155..667D} {155, 667}

\bibitem[\protect\citeauthoryear{{Di Matteo}, {Spite}, {Haywood}, {Bonifacio},
  {G{\'o}mez}, {Spite}  \& {Caffau}}{{Di Matteo} et~al.}{2020}]{dimatteo2020}
{Di Matteo} P.,  {Spite} M.,  {Haywood} M.,  {Bonifacio} P.,  {G{\'o}mez} A.,
  {Spite} F.,   {Caffau} E.,  2020, \mn@doi [\aap]
  {10.1051/0004-6361/201937016}, \href
  {https://ui.adsabs.harvard.edu/abs/2020A&A...636A.115D} {636, A115}

\bibitem[\protect\citeauthoryear{{Donati}, {Catala}, {Landstreet}  \&
  {Petit}}{{Donati} et~al.}{2006}]{Donati2006}
{Donati} J.~F.,  {Catala} C.,  {Landstreet} J.~D.,   {Petit} P.,  2006, in
  {Casini} R.,  {Lites} B.~W.,  eds,  Astronomical Society of the Pacific
  Conference Series Vol. 358, Solar Polarization 4. p.~362

\bibitem[\protect\citeauthoryear{{Dotter}}{{Dotter}}{2016}]{Dotter2016}
{Dotter} A.,  2016, \mn@doi [\apjs] {10.3847/0067-0049/222/1/8}, \href
  {https://ui.adsabs.harvard.edu/abs/2016ApJS..222....8D} {222, 8}

\bibitem[\protect\citeauthoryear{{Eisenstein} et~al.,}{{Eisenstein}
  et~al.}{2011}]{Eisenstein11}
{Eisenstein} D.~J.,  et~al., 2011, \mn@doi [\aj] {10.1088/0004-6256/142/3/72},
  \href {http://adsabs.harvard.edu/abs/2011AJ....142...72E} {142, 72}

\bibitem[\protect\citeauthoryear{{El-Badry} et~al.,}{{El-Badry}
  et~al.}{2018}]{ElBadry18EMP}
{El-Badry} K.,  et~al., 2018, \mn@doi [\mnras] {10.1093/mnras/sty1864}, \href
  {https://ui.adsabs.harvard.edu/abs/2018MNRAS.480..652E} {480, 652}

\bibitem[\protect\citeauthoryear{{Emerick}, {Bryan}, {Mac Low}, {C{\^o}t{\'e}},
  {Johnston}  \& {O'Shea}}{{Emerick} et~al.}{2018}]{Emerick2018}
{Emerick} A.,  {Bryan} G.~L.,  {Mac Low} M.-M.,  {C{\^o}t{\'e}} B.,  {Johnston}
  K.~V.,   {O'Shea} B.~W.,  2018, \mn@doi [\apj] {10.3847/1538-4357/aaec7d},
  \href {https://ui.adsabs.harvard.edu/abs/2018ApJ...869...94E} {869, 94}

\bibitem[\protect\citeauthoryear{{Frebel} \& {Norris}}{{Frebel} \&
  {Norris}}{2015}]{FN15}
{Frebel} A.,  {Norris} J.~E.,  2015, \mn@doi [\araa]
  {10.1146/annurev-astro-082214-122423}, \href
  {http://adsabs.harvard.edu/abs/2015ARA%26A..53..631F} {53, 631}

\bibitem[\protect\citeauthoryear{{Frebel}, {Simon}  \& {Kirby}}{{Frebel}
  et~al.}{2014}]{Frebel14}
{Frebel} A.,  {Simon} J.~D.,   {Kirby} E.~N.,  2014, \mn@doi [\apj]
  {10.1088/0004-637X/786/1/74}, \href
  {http://adsabs.harvard.edu/abs/2014ApJ...786...74F} {786, 74}

\bibitem[\protect\citeauthoryear{{Frebel}, {Ji}, {Ezzeddine}, {Hansen},
  {Chiti}, {Thompson}  \& {Merle}}{{Frebel} et~al.}{2019}]{Frebel19}
{Frebel} A.,  {Ji} A.~P.,  {Ezzeddine} R.,  {Hansen} T.~T.,  {Chiti} A.,
  {Thompson} I.~B.,   {Merle} T.,  2019, \mn@doi [\apj]
  {10.3847/1538-4357/aae848}, \href
  {https://ui.adsabs.harvard.edu/abs/2019ApJ...871..146F} {871, 146}

\bibitem[\protect\citeauthoryear{{Freeman} \& {Bland-Hawthorn}}{{Freeman} \&
  {Bland-Hawthorn}}{2002}]{Freeman02}
{Freeman} K.,  {Bland-Hawthorn} J.,  2002, \mn@doi [\araa]
  {10.1146/annurev.astro.40.060401.093840}, \href
  {http://adsabs.harvard.edu/abs/2002ARA%26A..40..487F} {40, 487}

\bibitem[\protect\citeauthoryear{{Fryer}, {Belczynski}, {Wiktorowicz},
  {Dominik}, {Kalogera}  \& {Holz}}{{Fryer} et~al.}{2012}]{Fryer12}
{Fryer} C.~L.,  {Belczynski} K.,  {Wiktorowicz} G.,  {Dominik} M.,  {Kalogera}
  V.,   {Holz} D.~E.,  2012, \mn@doi [\apj] {10.1088/0004-637X/749/1/91}, \href
  {http://adsabs.harvard.edu/abs/2012ApJ...749...91F} {749, 91}

\bibitem[\protect\citeauthoryear{{Gaia Collaboration} et~al.,}{{Gaia
  Collaboration} et~al.}{2018}]{GaiaDR2}
{Gaia Collaboration} et~al., 2018, \mn@doi [\aap]
  {10.1051/0004-6361/201833051}, \href
  {https://ui.adsabs.harvard.edu/abs/2018A&A...616A...1G} {616, A1}

\bibitem[\protect\citeauthoryear{{Garc{\'{\i}}a P{\'e}rez}
  et~al.,}{{Garc{\'{\i}}a P{\'e}rez} et~al.}{2015}]{GarciaPerez15}
{Garc{\'{\i}}a P{\'e}rez} A.~E.,  et~al., 2015, preprint, \href
  {http://adsabs.harvard.edu/abs/2015arXiv151007635G} {} (\mn@eprint {arXiv}
  {1510.07635})

\bibitem[\protect\citeauthoryear{{Gonz{\'a}lez Hern{\'a}ndez} \&
  {Bonifacio}}{{Gonz{\'a}lez Hern{\'a}ndez} \& {Bonifacio}}{2009}]{GHB2009}
{Gonz{\'a}lez Hern{\'a}ndez} J.~I.,  {Bonifacio} P.,  2009, \mn@doi [\aap]
  {10.1051/0004-6361/200810904}, \href
  {https://ui.adsabs.harvard.edu/abs/2009A&A...497..497G} {497, 497}

\bibitem[\protect\citeauthoryear{{Gonz{\'a}lez Hern{\'a}ndez}
  et~al.,}{{Gonz{\'a}lez Hern{\'a}ndez} et~al.}{2008}]{GonzalezHernandez2008}
{Gonz{\'a}lez Hern{\'a}ndez} J.~I.,  et~al., 2008, \mn@doi [\aap]
  {10.1051/0004-6361:20078847}, \href
  {https://ui.adsabs.harvard.edu/abs/2008A&A...480..233G} {480, 233}

\bibitem[\protect\citeauthoryear{{Gonz{\'a}lez Hern{\'a}ndez}, {Bonifacio},
  {Caffau}, {Ludwig}, {Steffen}, {Monaco}  \& {Cayrel}}{{Gonz{\'a}lez
  Hern{\'a}ndez} et~al.}{2019}]{GonzalezHernandez2019}
{Gonz{\'a}lez Hern{\'a}ndez} J.~I.,  {Bonifacio} P.,  {Caffau} E.,  {Ludwig}
  H.~G.,  {Steffen} M.,  {Monaco} L.,   {Cayrel} R.,  2019, \mn@doi [\aap]
  {10.1051/0004-6361/201936011}, \href
  {https://ui.adsabs.harvard.edu/abs/2019A&A...628A.111G} {628, A111}

\bibitem[\protect\citeauthoryear{{Greif}, {Bromm}, {Clark}, {Glover}, {Smith},
  {Klessen}, {Yoshida}  \& {Springel}}{{Greif} et~al.}{2012}]{Greif12}
{Greif} T.~H.,  {Bromm} V.,  {Clark} P.~C.,  {Glover} S. C.~O.,  {Smith} R.~J.,
   {Klessen} R.~S.,  {Yoshida} N.,   {Springel} V.,  2012, \mn@doi [\mnras]
  {10.1111/j.1365-2966.2012.21212.x}, \href
  {https://ui.adsabs.harvard.edu/abs/2012MNRAS.424..399G} {424, 399}

\bibitem[\protect\citeauthoryear{{Guo}, {White}, {Li}  \&
  {Boylan-Kolchin}}{{Guo} et~al.}{2010}]{guo2010}
{Guo} Q.,  {White} S.,  {Li} C.,   {Boylan-Kolchin} M.,  2010, \mn@doi [\mnras]
  {10.1111/j.1365-2966.2010.16341.x}, \href
  {https://ui.adsabs.harvard.edu/abs/2010MNRAS.404.1111G} {404, 1111}

\bibitem[\protect\citeauthoryear{{Gustafsson}, {Edvardsson}, {Eriksson},
  {J{\o}rgensen}, {Nordlund}  \& {Plez}}{{Gustafsson}
  et~al.}{2008}]{Gustafsson08}
{Gustafsson} B.,  {Edvardsson} B.,  {Eriksson} K.,  {J{\o}rgensen} U.~G.,
  {Nordlund} {\AA}.,   {Plez} B.,  2008, \mn@doi [\aap]
  {10.1051/0004-6361:200809724}, \href
  {http://adsabs.harvard.edu/abs/2008A%26A...486..951G} {486, 951}

\bibitem[\protect\citeauthoryear{{Hansen}, {Andersen}, {Nordstr{\"o}m},
  {Beers}, {Placco}, {Yoon}  \& {Buchhave}}{{Hansen} et~al.}{2016}]{Hansen16s}
{Hansen} T.~T.,  {Andersen} J.,  {Nordstr{\"o}m} B.,  {Beers} T.~C.,  {Placco}
  V.~M.,  {Yoon} J.,   {Buchhave} L.~A.,  2016, preprint, \href
  {http://adsabs.harvard.edu/abs/2016arXiv160103385H} {} (\mn@eprint {arXiv}
  {1601.03385})

\bibitem[\protect\citeauthoryear{{Hansen} et~al.,}{{Hansen}
  et~al.}{2017}]{hansen2017}
{Hansen} T.~T.,  et~al., 2017, \mn@doi [\apj] {10.3847/1538-4357/aa634a}, \href
  {https://ui.adsabs.harvard.edu/abs/2017ApJ...838...44H} {838, 44}

\bibitem[\protect\citeauthoryear{{Hartwig} et~al.,}{{Hartwig}
  et~al.}{2018}]{Hartwig18}
{Hartwig} T.,  et~al., 2018, \mn@doi [\mnras] {10.1093/mnras/sty1176}, \href
  {https://ui.adsabs.harvard.edu/abs/2018MNRAS.478.1795H} {478, 1795}

\bibitem[\protect\citeauthoryear{{Hasselquist} et~al.,}{{Hasselquist}
  et~al.}{2017}]{Hasselquist17}
{Hasselquist} S.,  et~al., 2017, \mn@doi [\apj] {10.3847/1538-4357/aa7ddc},
  \href {http://adsabs.harvard.edu/abs/2017ApJ...845..162H} {845, 162}

\bibitem[\protect\citeauthoryear{{Hayden}, {Bovy}  \& {Holtzman}}{{Hayden}
  et~al.}{2015}]{Hayden15}
{Hayden} M.~R.,  {Bovy} J.,   {Holtzman} J. A. e.~a.,  2015, \mn@doi [ApJ]
  {10.1088/0004-637X/808/2/132}, \href
  {https://ui.adsabs.harvard.edu/#abs/2015ApJ...808..132H} {808, 132}

\bibitem[\protect\citeauthoryear{{Hayes} et~al.,}{{Hayes}
  et~al.}{2018}]{Hayes18}
{Hayes} C.~R.,  et~al., 2018, \mn@doi [\apj] {10.3847/1538-4357/aa9cec}, \href
  {https://ui.adsabs.harvard.edu/abs/2018ApJ...852...49H} {852, 49}

\bibitem[\protect\citeauthoryear{{Heger} \& {Woosley}}{{Heger} \&
  {Woosley}}{2010}]{HegerWoosley10}
{Heger} A.,  {Woosley} S.~E.,  2010, \mn@doi [\apj]
  {10.1088/0004-637X/724/1/341}, \href
  {http://adsabs.harvard.edu/abs/2010ApJ...724..341H} {724, 341}

\bibitem[\protect\citeauthoryear{{Helmi}, {White}, {de Zeeuw}  \&
  {Zhao}}{{Helmi} et~al.}{1999}]{Helmi99}
{Helmi} A.,  {White} S. D.~M.,  {de Zeeuw} P.~T.,   {Zhao} H.,  1999, \mn@doi
  [\nat] {10.1038/46980}, \href
  {https://ui.adsabs.harvard.edu/abs/1999Natur.402...53H} {402, 53}

\bibitem[\protect\citeauthoryear{{Helmi}, {Veljanoski}, {Breddels}, {Tian}  \&
  {Sales}}{{Helmi} et~al.}{2017}]{Helmi2017}
{Helmi} A.,  {Veljanoski} J.,  {Breddels} M.~A.,  {Tian} H.,   {Sales} L.~V.,
  2017, \mn@doi [\aap] {10.1051/0004-6361/201629990}, \href
  {https://ui.adsabs.harvard.edu/abs/2017A&A...598A..58H} {598, A58}

\bibitem[\protect\citeauthoryear{{Helmi}, {Babusiaux}, {Koppelman}, {Massari},
  {Veljanoski}  \& {Brown}}{{Helmi} et~al.}{2018}]{Helmi18}
{Helmi} A.,  {Babusiaux} C.,  {Koppelman} H.~H.,  {Massari} D.,  {Veljanoski}
  J.,   {Brown} A. G.~A.,  2018, \mn@doi [\nat] {10.1038/s41586-018-0625-x},
  \href {https://ui.adsabs.harvard.edu/abs/2018Natur.563...85H} {563, 85}

\bibitem[\protect\citeauthoryear{{Howes} et~al.,}{{Howes}
  et~al.}{2015}]{Howes15}
{Howes} L.~M.,  et~al., 2015, \mn@doi [\nat] {10.1038/nature15747}, \href
  {http://adsabs.harvard.edu/abs/2015Natur.527..484H} {527, 484}

\bibitem[\protect\citeauthoryear{{Howes} et~al.,}{{Howes}
  et~al.}{2016}]{Howes16}
{Howes} L.~M.,  et~al., 2016, \mn@doi [\mnras] {10.1093/mnras/stw1004}, \href
  {https://ui.adsabs.harvard.edu/abs/2016MNRAS.460..884H} {460, 884}

\bibitem[\protect\citeauthoryear{{Ibata}, {Gilmore}  \& {Irwin}}{{Ibata}
  et~al.}{1994}]{Ibata94}
{Ibata} R.~A.,  {Gilmore} G.,   {Irwin} M.~J.,  1994, \mn@doi [\nat]
  {10.1038/370194a0}, \href
  {https://ui.adsabs.harvard.edu/abs/1994Natur.370..194I} {370, 194}

\bibitem[\protect\citeauthoryear{{Ishigaki}, {Tominaga}, {Kobayashi}  \&
  {Nomoto}}{{Ishigaki} et~al.}{2014}]{Ishigaki14}
{Ishigaki} M.~N.,  {Tominaga} N.,  {Kobayashi} C.,   {Nomoto} K.,  2014,
  \mn@doi [\apjl] {10.1088/2041-8205/792/2/L32}, \href
  {http://adsabs.harvard.edu/abs/2014ApJ...792L..32I} {792, L32}

\bibitem[\protect\citeauthoryear{{Iwamoto}, {Umeda}, {Tominaga}, {Nomoto}  \&
  {Maeda}}{{Iwamoto} et~al.}{2005}]{Iwamoto2005}
{Iwamoto} N.,  {Umeda} H.,  {Tominaga} N.,  {Nomoto} K.,   {Maeda} K.,  2005,
  \mn@doi [Science] {10.1126/science.1112997}, \href
  {https://ui.adsabs.harvard.edu/abs/2005Sci...309..451I} {309, 451}

\bibitem[\protect\citeauthoryear{{Ji}, {Frebel}, {Chiti}  \& {Simon}}{{Ji}
  et~al.}{2016}]{Ji16}
{Ji} A.~P.,  {Frebel} A.,  {Chiti} A.,   {Simon} J.~D.,  2016, \mn@doi [\nat]
  {10.1038/nature17425}, \href
  {https://ui.adsabs.harvard.edu/abs/2016Natur.531..610J} {531, 610}

\bibitem[\protect\citeauthoryear{{Johnston}, {Bullock}, {Sharma}, {Font},
  {Robertson}  \& {Leitner}}{{Johnston} et~al.}{2008}]{Johnston2008}
{Johnston} K.~V.,  {Bullock} J.~S.,  {Sharma} S.,  {Font} A.,  {Robertson}
  B.~E.,   {Leitner} S.~N.,  2008, \mn@doi [\apj] {10.1086/592228}, \href
  {https://ui.adsabs.harvard.edu/abs/2008ApJ...689..936J} {689, 936}

\bibitem[\protect\citeauthoryear{{Joyce} \& {Chaboyer}}{{Joyce} \&
  {Chaboyer}}{2015}]{Joyce2015}
{Joyce} M.,  {Chaboyer} B.,  2015, \mn@doi [\apj]
  {10.1088/0004-637X/814/2/142}, \href
  {https://ui.adsabs.harvard.edu/abs/2015ApJ...814..142J} {814, 142}

\bibitem[\protect\citeauthoryear{{Joyce} \& {Chaboyer}}{{Joyce} \&
  {Chaboyer}}{2018}]{Joyce2018}
{Joyce} M.,  {Chaboyer} B.,  2018, \mn@doi [\apj] {10.3847/1538-4357/aab200},
  \href {https://ui.adsabs.harvard.edu/abs/2018ApJ...856...10J} {856, 10}

\bibitem[\protect\citeauthoryear{{Keller} et~al.,}{{Keller}
  et~al.}{2007}]{Keller07}
{Keller} S.~C.,  et~al., 2007, \mn@doi [\pasa] {10.1071/AS07001}, \href
  {http://adsabs.harvard.edu/abs/2007PASA...24....1K} {24, 1}

\bibitem[\protect\citeauthoryear{{Keller} et~al.,}{{Keller}
  et~al.}{2014}]{Keller14}
{Keller} S.~C.,  et~al., 2014, \mn@doi [\nat] {10.1038/nature12990}, \href
  {https://ui.adsabs.harvard.edu/abs/2014Natur.506..463K} {506, 463}

\bibitem[\protect\citeauthoryear{{Kobayashi}, {Nomoto}  \&
  {Hachisu}}{{Kobayashi} et~al.}{2015}]{Kobayashi2015}
{Kobayashi} C.,  {Nomoto} K.,   {Hachisu} I.,  2015, \mn@doi [\apjl]
  {10.1088/2041-8205/804/1/L24}, \href
  {https://ui.adsabs.harvard.edu/abs/2015ApJ...804L..24K} {804, L24}

\bibitem[\protect\citeauthoryear{{Kobayashi}, {Karakas}  \&
  {Lugaro}}{{Kobayashi} et~al.}{2020}]{Kobayashi2020}
{Kobayashi} C.,  {Karakas} A.~I.,   {Lugaro} M.,  2020, \mn@doi [\apj]
  {10.3847/1538-4357/abae65}, \href
  {https://ui.adsabs.harvard.edu/abs/2020ApJ...900..179K} {900, 179}

\bibitem[\protect\citeauthoryear{{Koch}, {McWilliam}, {Grebel}, {Zucker}  \&
  {Belokurov}}{{Koch} et~al.}{2008}]{Koch08}
{Koch} A.,  {McWilliam} A.,  {Grebel} E.~K.,  {Zucker} D.~B.,   {Belokurov} V.,
   2008, \mn@doi [\apjl] {10.1086/595001}, \href
  {http://adsabs.harvard.edu/abs/2008ApJ...688L..13K} {688, L13}

\bibitem[\protect\citeauthoryear{{Koppelman}, {Helmi}  \&
  {Veljanoski}}{{Koppelman} et~al.}{2018}]{Koppelman2018}
{Koppelman} H.,  {Helmi} A.,   {Veljanoski} J.,  2018, \mn@doi [\apjl]
  {10.3847/2041-8213/aac882}, \href
  {https://ui.adsabs.harvard.edu/abs/2018ApJ...860L..11K} {860, L11}

\bibitem[\protect\citeauthoryear{{Korobkin}, {Rosswog}, {Arcones}  \&
  {Winteler}}{{Korobkin} et~al.}{2012}]{Korobkin2012}
{Korobkin} O.,  {Rosswog} S.,  {Arcones} A.,   {Winteler} C.,  2012, \mn@doi
  [\mnras] {10.1111/j.1365-2966.2012.21859.x}, \href
  {https://ui.adsabs.harvard.edu/abs/2012MNRAS.426.1940K} {426, 1940}

\bibitem[\protect\citeauthoryear{{Kovalev}, {Bergemann}, {Ting}  \&
  {Rix}}{{Kovalev} et~al.}{2019}]{Kovalev2019}
{Kovalev} M.,  {Bergemann} M.,  {Ting} Y.-S.,   {Rix} H.-W.,  2019, \mn@doi
  [\aap] {10.1051/0004-6361/201935861}, \href
  {https://ui.adsabs.harvard.edu/abs/2019A&A...628A..54K} {628, A54}

\bibitem[\protect\citeauthoryear{{Kraft} \& {Ivans}}{{Kraft} \&
  {Ivans}}{2003}]{KI2003}
{Kraft} R.~P.,  {Ivans} I.~I.,  2003, \mn@doi [\pasp] {10.1086/345914}, \href
  {https://ui.adsabs.harvard.edu/abs/2003PASP..115..143K} {115, 143}

\bibitem[\protect\citeauthoryear{{Kratz}, {Farouqi}, {Pfeiffer}, {Truran},
  {Sneden}  \& {Cowan}}{{Kratz} et~al.}{2007}]{Kratz2007}
{Kratz} K.-L.,  {Farouqi} K.,  {Pfeiffer} B.,  {Truran} J.~W.,  {Sneden} C.,
  {Cowan} J.~J.,  2007, \mn@doi [\apj] {10.1086/517495}, \href
  {https://ui.adsabs.harvard.edu/abs/2007ApJ...662...39K} {662, 39}

\bibitem[\protect\citeauthoryear{{Lamb} et~al.,}{{Lamb} et~al.}{2017}]{Lamb17}
{Lamb} M.,  et~al., 2017, \mn@doi [\mnras] {10.1093/mnras/stw2865}, \href
  {http://adsabs.harvard.edu/abs/2017MNRAS.465.3536L} {465, 3536}

\bibitem[\protect\citeauthoryear{{Lee} et~al.,}{{Lee} et~al.}{2013}]{Lee13}
{Lee} Y.~S.,  et~al., 2013, \mn@doi [\aj] {10.1088/0004-6256/146/5/132}, \href
  {http://adsabs.harvard.edu/abs/2013AJ....146..132L} {146, 132}

\bibitem[\protect\citeauthoryear{{Lejeune}, {Cuisinier}  \& {Buser}}{{Lejeune}
  et~al.}{1998}]{YYL}
{Lejeune} T.,  {Cuisinier} F.,   {Buser} R.,  1998, VizieR Online Data Catalog,
  \href {https://ui.adsabs.harvard.edu/abs/1998yCat..41300065L} {pp
  J/A+AS/130/65}

\bibitem[\protect\citeauthoryear{{Limberg} et~al.,}{{Limberg}
  et~al.}{2020}]{Limberg2020}
{Limberg} G.,  et~al., 2020, arXiv e-prints, \href
  {https://ui.adsabs.harvard.edu/abs/2020arXiv201108305L} {p. arXiv:2011.08305}

\bibitem[\protect\citeauthoryear{{Lind}, {Asplund}, {Barklem}  \&
  {Belyaev}}{{Lind} et~al.}{2011}]{Lind2011}
{Lind} K.,  {Asplund} M.,  {Barklem} P.~S.,   {Belyaev} A.~K.,  2011, \mn@doi
  [\aap] {10.1051/0004-6361/201016095}, \href
  {https://ui.adsabs.harvard.edu/abs/2011A&A...528A.103L} {528, A103}

\bibitem[\protect\citeauthoryear{{Lind}, {Bergemann}  \& {Asplund}}{{Lind}
  et~al.}{2012}]{Lind2012}
{Lind} K.,  {Bergemann} M.,   {Asplund} M.,  2012, \mn@doi [\mnras]
  {10.1111/j.1365-2966.2012.21686.x}, \href
  {https://ui.adsabs.harvard.edu/abs/2012MNRAS.427...50L} {427, 50}

\bibitem[\protect\citeauthoryear{{Lindegren} et~al.,}{{Lindegren}
  et~al.}{2018}]{Lindegren2018}
{Lindegren} L.,  et~al., 2018, \mn@doi [\aap] {10.1051/0004-6361/201832727},
  \href {https://ui.adsabs.harvard.edu/abs/2018A&A...616A...2L} {616, A2}

\bibitem[\protect\citeauthoryear{{Lucatello}, {Tsangarides}, {Beers},
  {Carretta}, {Gratton}  \& {Ryan}}{{Lucatello} et~al.}{2005}]{Lucatello05}
{Lucatello} S.,  {Tsangarides} S.,  {Beers} T.~C.,  {Carretta} E.,  {Gratton}
  R.~G.,   {Ryan} S.~G.,  2005, \mn@doi [\apj] {10.1086/428104}, \href
  {http://adsabs.harvard.edu/abs/2005ApJ...625..825L} {625, 825}

\bibitem[\protect\citeauthoryear{{Lucchesi} et~al.,}{{Lucchesi}
  et~al.}{2020}]{Lucchesi20}
{Lucchesi} R.,  et~al., 2020, arXiv e-prints, \href
  {https://ui.adsabs.harvard.edu/abs/2020arXiv200111033L} {p. arXiv:2001.11033}

\bibitem[\protect\citeauthoryear{{Lucey} et~al.,}{{Lucey}
  et~al.}{2019}]{Lucey2019}
{Lucey} M.,  et~al., 2019, \mn@doi [\mnras] {10.1093/mnras/stz1847}, \href
  {https://ui.adsabs.harvard.edu/abs/2019MNRAS.488.2283L} {488, 2283}

\bibitem[\protect\citeauthoryear{{Lucey} et~al.,}{{Lucey}
  et~al.}{2020}]{Lucey2020}
{Lucey} M.,  et~al., 2020, arXiv e-prints, \href
  {https://ui.adsabs.harvard.edu/abs/2020arXiv200903886L} {p. arXiv:2009.03886}

\bibitem[\protect\citeauthoryear{{Lugaro}, {Karakas}, {Stancliffe}  \&
  {Rijs}}{{Lugaro} et~al.}{2012}]{Lugaro2012}
{Lugaro} M.,  {Karakas} A.~I.,  {Stancliffe} R.~J.,   {Rijs} C.,  2012, \mn@doi
  [\apj] {10.1088/0004-637X/747/1/2}, \href
  {https://ui.adsabs.harvard.edu/abs/2012ApJ...747....2L} {747, 2}

\bibitem[\protect\citeauthoryear{{Lupton} et~al.,}{{Lupton}
  et~al.}{2005}]{Lupton2005}
{Lupton} R.~H.,  et~al., 2005, in American Astronomical Society Meeting
  Abstracts. p. 133.08

\bibitem[\protect\citeauthoryear{{Maeder} \& {Meynet}}{{Maeder} \&
  {Meynet}}{2015}]{Meynet2015}
{Maeder} A.,  {Meynet} G.,  2015, \mn@doi [\aap] {10.1051/0004-6361/201526234},
  \href {https://ui.adsabs.harvard.edu/abs/2015A&A...580A..32M} {580, A32}

\bibitem[\protect\citeauthoryear{Majewski, Schiavon, Frinchaboy  \& et~al.\
  2015}{Majewski et~al.}{2015}]{Majewski2015}
Majewski S.~R.,  Schiavon R.~P.,  Frinchaboy P.~M.,   et~al.\ 2015 2015, arXiv
  preprint arXiv:1509.05420

\bibitem[\protect\citeauthoryear{{Martioli}, {Teeple}, {Manset}, {Devost},
  {Withington}, {Venne}  \& {Tannock}}{{Martioli} et~al.}{2012}]{OPERA}
{Martioli} E.,  {Teeple} D.,  {Manset} N.,  {Devost} D.,  {Withington} K.,
  {Venne} A.,   {Tannock} M.,  2012, {Open source pipeline for ESPaDOnS
  reduction and analysis}.
p. 84512B, \mn@doi{10.1117/12.926627}

\bibitem[\protect\citeauthoryear{{Mashonkina} \& {Belyaev}}{{Mashonkina} \&
  {Belyaev}}{2019}]{Mashonkina2019}
{Mashonkina} L.~I.,  {Belyaev} A.~K.,  2019, \mn@doi [Astronomy Letters]
  {10.1134/S1063773719060033}, \href
  {https://ui.adsabs.harvard.edu/abs/2019AstL...45..341M} {45, 341}

\bibitem[\protect\citeauthoryear{{Mashonkina}, {Gehren}  \&
  {Bikmaev}}{{Mashonkina} et~al.}{1999}]{Mashonkina1999}
{Mashonkina} L.,  {Gehren} T.,   {Bikmaev} I.,  1999, \aap, \href
  {https://ui.adsabs.harvard.edu/abs/1999A&A...343..519M} {343, 519}

\bibitem[\protect\citeauthoryear{{Mashonkina}, {Korn}  \&
  {Przybilla}}{{Mashonkina} et~al.}{2007}]{Mashonkina2007}
{Mashonkina} L.,  {Korn} A.~J.,   {Przybilla} N.,  2007, \mn@doi [\aap]
  {10.1051/0004-6361:20065999}, \href
  {https://ui.adsabs.harvard.edu/abs/2007A&A...461..261M} {461, 261}

\bibitem[\protect\citeauthoryear{{Mashonkina}, {Jablonka}, {Pakhomov},
  {Sitnova}  \& {North}}{{Mashonkina} et~al.}{2017}]{Mashonkina17}
{Mashonkina} L.,  {Jablonka} P.,  {Pakhomov} Y.,  {Sitnova} T.,   {North} P.,
  2017, \mn@doi [\aap] {10.1051/0004-6361/201730779}, \href
  {https://ui.adsabs.harvard.edu/abs/2017A&A...604A.129M} {604, A129}

\bibitem[\protect\citeauthoryear{{Matsuno}, {Aoki}, {Beers}, {Lee}  \&
  {Honda}}{{Matsuno} et~al.}{2017}]{Matsuno17}
{Matsuno} T.,  {Aoki} W.,  {Beers} T.~C.,  {Lee} Y.~S.,   {Honda} S.,  2017,
  \mn@doi [\aj] {10.3847/1538-3881/aa7a08}, \href
  {http://adsabs.harvard.edu/abs/2017AJ....154...52M} {154, 52}

\bibitem[\protect\citeauthoryear{{Matsuno}, {Aoki}  \& {Suda}}{{Matsuno}
  et~al.}{2019}]{Matsuno2019}
{Matsuno} T.,  {Aoki} W.,   {Suda} T.,  2019, \mn@doi [\apjl]
  {10.3847/2041-8213/ab0ec0}, \href
  {https://ui.adsabs.harvard.edu/abs/2019ApJ...874L..35M} {874, L35}

\bibitem[\protect\citeauthoryear{{McWilliam}, {Wallerstein}  \&
  {Mottini}}{{McWilliam} et~al.}{2013}]{McWilliam13}
{McWilliam} A.,  {Wallerstein} G.,   {Mottini} M.,  2013, \mn@doi [\apj]
  {10.1088/0004-637X/778/2/149}, \href
  {https://ui.adsabs.harvard.edu/abs/2013ApJ...778..149M} {778, 149}

\bibitem[\protect\citeauthoryear{{Meynet}, {Ekstr{\"o}m}  \& {Maeder}}{{Meynet}
  et~al.}{2006}]{Meynet2006}
{Meynet} G.,  {Ekstr{\"o}m} S.,   {Maeder} A.,  2006, \mn@doi [\aap]
  {10.1051/0004-6361:20053070}, \href
  {https://ui.adsabs.harvard.edu/abs/2006A&A...447..623M} {447, 623}

\bibitem[\protect\citeauthoryear{{Meza}, {Navarro}, {Abadi}  \&
  {Steinmetz}}{{Meza} et~al.}{2005}]{Meza2005}
{Meza} A.,  {Navarro} J.~F.,  {Abadi} M.~G.,   {Steinmetz} M.,  2005, \mn@doi
  [\mnras] {10.1111/j.1365-2966.2005.08869.x}, \href
  {https://ui.adsabs.harvard.edu/abs/2005MNRAS.359...93M} {359, 93}

\bibitem[\protect\citeauthoryear{{Monty}, {Venn}, {Lane}, {Lokhorst}  \&
  {Yong}}{{Monty} et~al.}{2020}]{Monty2020}
{Monty} S.,  {Venn} K.~A.,  {Lane} J. M.~M.,  {Lokhorst} D.,   {Yong} D.,
  2020, \mn@doi [\mnras] {10.1093/mnras/staa1995}, \href
  {https://ui.adsabs.harvard.edu/abs/2020MNRAS.497.1236M} {497, 1236}

\bibitem[\protect\citeauthoryear{{Mucciarelli} \& {Bellazzini}}{{Mucciarelli}
  \& {Bellazzini}}{2020}]{MB2020}
{Mucciarelli} A.,  {Bellazzini} M.,  2020, \mn@doi [Research Notes of the
  American Astronomical Society] {10.3847/2515-5172/ab8820}, \href
  {https://ui.adsabs.harvard.edu/abs/2020RNAAS...4...52M} {4, 52}

\bibitem[\protect\citeauthoryear{{Myeong}, {Evans}, {Belokurov}, {Sand ers}  \&
  {Koposov}}{{Myeong} et~al.}{2018}]{Myeong18}
{Myeong} G.~C.,  {Evans} N.~W.,  {Belokurov} V.,  {Sand ers} J.~L.,   {Koposov}
  S.~E.,  2018, \mn@doi [\apjl] {10.3847/2041-8213/aad7f7}, \href
  {https://ui.adsabs.harvard.edu/abs/2018ApJ...863L..28M} {863, L28}

\bibitem[\protect\citeauthoryear{{Myeong}, {Vasiliev}, {Iorio}, {Evans}  \&
  {Belokurov}}{{Myeong} et~al.}{2019}]{Myeong19}
{Myeong} G.~C.,  {Vasiliev} E.,  {Iorio} G.,  {Evans} N.~W.,   {Belokurov} V.,
  2019, \mn@doi [\mnras] {10.1093/mnras/stz1770}, \href
  {https://ui.adsabs.harvard.edu/abs/2019MNRAS.488.1235M} {488, 1235}

\bibitem[\protect\citeauthoryear{{Nakamura} \& {Umemura}}{{Nakamura} \&
  {Umemura}}{2001}]{Nakamura2001}
{Nakamura} F.,  {Umemura} M.,  2001, \mn@doi [\apj] {10.1086/318663}, \href
  {https://ui.adsabs.harvard.edu/abs/2001ApJ...548...19N} {548, 19}

\bibitem[\protect\citeauthoryear{{Nishimura}, {Takiwaki}  \&
  {Thielemann}}{{Nishimura} et~al.}{2015}]{Nishimura2015}
{Nishimura} N.,  {Takiwaki} T.,   {Thielemann} F.-K.,  2015, \mn@doi [\apj]
  {10.1088/0004-637X/810/2/109}, \href
  {https://ui.adsabs.harvard.edu/abs/2015ApJ...810..109N} {810, 109}

\bibitem[\protect\citeauthoryear{{Nissen} \& {Schuster}}{{Nissen} \&
  {Schuster}}{2010}]{Nissen10}
{Nissen} P.~E.,  {Schuster} W.~J.,  2010, \mn@doi [\aap]
  {10.1051/0004-6361/200913877}, \href
  {https://ui.adsabs.harvard.edu/abs/2010A&A...511L..10N} {511, L10}

\bibitem[\protect\citeauthoryear{{Nordlander} et~al.,}{{Nordlander}
  et~al.}{2019}]{Nordlander19}
{Nordlander} T.,  et~al., 2019, \mn@doi [\mnras] {10.1093/mnrasl/slz109}, \href
  {https://ui.adsabs.harvard.edu/abs/2019MNRAS.488L.109N} {488, L109}

\bibitem[\protect\citeauthoryear{{Norris} \& {Yong}}{{Norris} \&
  {Yong}}{2019}]{Norris2019}
{Norris} J.~E.,  {Yong} D.,  2019, \mn@doi [\apj] {10.3847/1538-4357/ab1f84},
  \href {https://ui.adsabs.harvard.edu/abs/2019ApJ...879...37N} {879, 37}

\bibitem[\protect\citeauthoryear{{Norris}, {Ryan}  \& {Beers}}{{Norris}
  et~al.}{1997}]{Norris1997}
{Norris} J.~E.,  {Ryan} S.~G.,   {Beers} T.~C.,  1997, \mn@doi [\apj]
  {10.1086/304695}, \href
  {https://ui.adsabs.harvard.edu/abs/1997ApJ...488..350N} {488, 350}

\bibitem[\protect\citeauthoryear{{Norris} et~al.,}{{Norris}
  et~al.}{2013}]{Norris2013}
{Norris} J.~E.,  et~al., 2013, \mn@doi [\apj] {10.1088/0004-637X/762/1/28},
  \href {https://ui.adsabs.harvard.edu/abs/2013ApJ...762...28N} {762, 28}

\bibitem[\protect\citeauthoryear{{Norris}, {Yong}, {Venn}, {Gilmore},
  {Casagrande}  \& {Dotter}}{{Norris} et~al.}{2017}]{Norris2017}
{Norris} J.~E.,  {Yong} D.,  {Venn} K.~A.,  {Gilmore} G.,  {Casagrande} L.,
  {Dotter} A.,  2017, \mn@doi [\apjs] {10.3847/1538-4365/aa755e}, \href
  {https://ui.adsabs.harvard.edu/abs/2017ApJS..230...28N} {230, 28}

\bibitem[\protect\citeauthoryear{{Onken} et~al.,}{{Onken}
  et~al.}{2020}]{Onken2020}
{Onken} C.~A.,  et~al., 2020, arXiv e-prints, \href
  {https://ui.adsabs.harvard.edu/abs/2020arXiv200810359O} {p. arXiv:2008.10359}

\bibitem[\protect\citeauthoryear{{Paxton}, {Bildsten}, {Dotter}, {Herwig},
  {Lesaffre}  \& {Timmes}}{{Paxton} et~al.}{2011}]{Paxton2011}
{Paxton} B.,  {Bildsten} L.,  {Dotter} A.,  {Herwig} F.,  {Lesaffre} P.,
  {Timmes} F.,  2011, \mn@doi [\apjs] {10.1088/0067-0049/192/1/3}, \href
  {https://ui.adsabs.harvard.edu/abs/2011ApJS..192....3P} {192, 3}

\bibitem[\protect\citeauthoryear{{Pazder}, {Fournier}, {Pawluczyk}  \& {van
  Kooten}}{{Pazder} et~al.}{2014}]{Pazder2014}
{Pazder} J.,  {Fournier} P.,  {Pawluczyk} R.,   {van Kooten} M.,  2014, {The
  FRD and transmission of the 270-m GRACES optical fiber link and a high
  numerical aperture fiber for astronomy}.
p. 915124, \mn@doi{10.1117/12.2057327}

\bibitem[\protect\citeauthoryear{{Pignatari} et~al.,}{{Pignatari}
  et~al.}{2016}]{Pignatari2016}
{Pignatari} M.,  et~al., 2016, \mn@doi [\apjs] {10.3847/0067-0049/225/2/24},
  \href {https://ui.adsabs.harvard.edu/abs/2016ApJS..225...24P} {225, 24}

\bibitem[\protect\citeauthoryear{{Pilachowski}, {Sneden}  \&
  {Kraft}}{{Pilachowski} et~al.}{1996}]{Pilachowski1996}
{Pilachowski} C.~A.,  {Sneden} C.,   {Kraft} R.~P.,  1996, \mn@doi [\aj]
  {10.1086/117909}, \href
  {https://ui.adsabs.harvard.edu/abs/1996AJ....111.1689P} {111, 1689}

\bibitem[\protect\citeauthoryear{{Placco}, {Frebel}, {Beers}  \&
  {Stancliffe}}{{Placco} et~al.}{2014}]{Placco14}
{Placco} V.~M.,  {Frebel} A.,  {Beers} T.~C.,   {Stancliffe} R.~J.,  2014,
  \mn@doi [\apj] {10.1088/0004-637X/797/1/21}, \href
  {http://adsabs.harvard.edu/abs/2014ApJ...797...21P} {797, 21}

\bibitem[\protect\citeauthoryear{Prieto, Beers, Wilhelm, Newberg, Rockosi,
  Yanny  \& Lee}{Prieto et~al.}{2006}]{AllendePrieto2006}
Prieto C.~A.,  Beers T.~C.,  Wilhelm R.,  Newberg H.~J.,  Rockosi C.~M.,  Yanny
  B.,   Lee Y.~S.,  2006, \mn@doi [The Astrophysical Journal] {10.1086/498131},
  636, 804

\bibitem[\protect\citeauthoryear{{Roederer}, {Preston}, {Thompson}, {Shectman},
  {Sneden}, {Burley}  \& {Kelson}}{{Roederer} et~al.}{2014}]{Roederer14}
{Roederer} I.~U.,  {Preston} G.~W.,  {Thompson} I.~B.,  {Shectman} S.~A.,
  {Sneden} C.,  {Burley} G.~S.,   {Kelson} D.~D.,  2014, \mn@doi [\aj]
  {10.1088/0004-6256/147/6/136}, \href
  {http://adsabs.harvard.edu/abs/2014AJ....147..136R} {147, 136}

\bibitem[\protect\citeauthoryear{{Roederer} et~al.,}{{Roederer}
  et~al.}{2016}]{RoedererReticulum16}
{Roederer} I.~U.,  et~al., 2016, \aj, 151, 82

\bibitem[\protect\citeauthoryear{{Salvadori}, {Bonifacio}, {Caffau}, {Korotin},
  {Andreevsky}, {Spite}  \& {Sk{\'u}lad{\'o}ttir}}{{Salvadori}
  et~al.}{2019}]{Salvadori19}
{Salvadori} S.,  {Bonifacio} P.,  {Caffau} E.,  {Korotin} S.,  {Andreevsky} S.,
   {Spite} M.,   {Sk{\'u}lad{\'o}ttir} {\'A}.,  2019, \mn@doi [\mnras]
  {10.1093/mnras/stz1464}, \href
  {https://ui.adsabs.harvard.edu/abs/2019MNRAS.487.4261S} {487, 4261}

\bibitem[\protect\citeauthoryear{{Sbordone} et~al.,}{{Sbordone}
  et~al.}{2010}]{Sbordone2010}
{Sbordone} L.,  et~al., 2010, \mn@doi [\aap] {10.1051/0004-6361/200913282},
  \href {https://ui.adsabs.harvard.edu/abs/2010A&A...522A..26S} {522, A26}

\bibitem[\protect\citeauthoryear{{Schlafly} \& {Finkbeiner}}{{Schlafly} \&
  {Finkbeiner}}{2011}]{SF2011}
{Schlafly} E.~F.,  {Finkbeiner} D.~P.,  2011, \mn@doi [\apj]
  {10.1088/0004-637X/737/2/103}, \href
  {https://ui.adsabs.harvard.edu/abs/2011ApJ...737..103S} {737, 103}

\bibitem[\protect\citeauthoryear{{Schlaufman}, {Thompson}  \&
  {Casey}}{{Schlaufman} et~al.}{2018}]{Schlaufman18}
{Schlaufman} K.~C.,  {Thompson} I.~B.,   {Casey} A.~R.,  2018, \mn@doi [\apj]
  {10.3847/1538-4357/aadd97}, \href
  {https://ui.adsabs.harvard.edu/abs/2018ApJ...867...98S} {867, 98}

\bibitem[\protect\citeauthoryear{{Schlegel}, {Finkbeiner}  \&
  {Davis}}{{Schlegel} et~al.}{1998}]{Schlegel1998}
{Schlegel} D.~J.,  {Finkbeiner} D.~P.,   {Davis} M.,  1998, \mn@doi [\apj]
  {10.1086/305772}, \href
  {https://ui.adsabs.harvard.edu/abs/1998ApJ...500..525S} {500, 525}

\bibitem[\protect\citeauthoryear{{Schneider}, {Omukai}, {Bianchi}  \&
  {Valiante}}{{Schneider} et~al.}{2012}]{Schneider12}
{Schneider} R.,  {Omukai} K.,  {Bianchi} S.,   {Valiante} R.,  2012, \mn@doi
  [\mnras] {10.1111/j.1365-2966.2011.19818.x}, \href
  {https://ui.adsabs.harvard.edu/abs/2012MNRAS.419.1566S} {419, 1566}

\bibitem[\protect\citeauthoryear{{Sestito} et~al.,}{{Sestito}
  et~al.}{2019}]{Sestito19UMP}
{Sestito} F.,  et~al., 2019, \mn@doi [\mnras] {10.1093/mnras/stz043}, \href
  {https://ui.adsabs.harvard.edu/abs/2019MNRAS.484.2166S} {484, 2166}

\bibitem[\protect\citeauthoryear{{Sestito} et~al.,}{{Sestito}
  et~al.}{2020}]{Sestito19Pristine}
{Sestito} F.,  et~al., 2020, \mn@doi [\mnras] {10.1093/mnrasl/slaa022}, \href
  {https://ui.adsabs.harvard.edu/abs/2020MNRAS.497L...7S} {497, L7}

\bibitem[\protect\citeauthoryear{{Sestito} et~al.,}{{Sestito}
  et~al.}{2021}]{Sestito2020}
{Sestito} F.,  et~al., 2021, \mn@doi [\mnras] {10.1093/mnras/staa3479}, \href
  {https://ui.adsabs.harvard.edu/abs/2021MNRAS.500.3750S} {500, 3750}

\bibitem[\protect\citeauthoryear{{Silk}}{{Silk}}{1983}]{Silk83}
{Silk} J.,  1983, \mn@doi [\mnras] {10.1093/mnras/205.3.705}, \href
  {https://ui.adsabs.harvard.edu/abs/1983MNRAS.205..705S} {205, 705}

\bibitem[\protect\citeauthoryear{{Sitnova}, {Mashonkina}  \&
  {Ryabchikova}}{{Sitnova} et~al.}{2013}]{Sitnova2013}
{Sitnova} T.~M.,  {Mashonkina} L.~I.,   {Ryabchikova} T.~A.,  2013, \mn@doi
  [Astronomy Letters] {10.1134/S1063773713020084}, \href
  {https://ui.adsabs.harvard.edu/abs/2013AstL...39..126S} {39, 126}

\bibitem[\protect\citeauthoryear{{Sitnova} et~al.,}{{Sitnova}
  et~al.}{2015}]{Sitnova2015}
{Sitnova} T.,  et~al., 2015, \mn@doi [\apj] {10.1088/0004-637X/808/2/148},
  \href {https://ui.adsabs.harvard.edu/abs/2015ApJ...808..148S} {808, 148}

\bibitem[\protect\citeauthoryear{{Sitnova}, {Mashonkina}, {Ezzeddine}  \&
  {Frebel}}{{Sitnova} et~al.}{2019}]{Sitnova2019}
{Sitnova} T.~M.,  {Mashonkina} L.~I.,  {Ezzeddine} R.,   {Frebel} A.,  2019,
  \mn@doi [\mnras] {10.1093/mnras/stz626}, \href
  {https://ui.adsabs.harvard.edu/abs/2019MNRAS.485.3527S} {485, 3527}

\bibitem[\protect\citeauthoryear{{Sneden}, {Cowan}  \& {Gallino}}{{Sneden}
  et~al.}{2008}]{Sneden08}
{Sneden} C.,  {Cowan} J.~J.,   {Gallino} R.,  2008, \mn@doi [\araa]
  {10.1146/annurev.astro.46.060407.145207}, \href
  {http://adsabs.harvard.edu/abs/2008ARA%26A..46..241S} {46, 241}

\bibitem[\protect\citeauthoryear{{Spite} \& {Spite}}{{Spite} \&
  {Spite}}{1982}]{Spite82}
{Spite} M.,  {Spite} F.,  1982, \mn@doi [\nat] {10.1038/297483a0}, \href
  {https://ui.adsabs.harvard.edu/abs/1982Natur.297..483S} {297, 483}

\bibitem[\protect\citeauthoryear{{Starkenburg}, {Shetrone}, {McConnachie}  \&
  {Venn}}{{Starkenburg} et~al.}{2014}]{Starkenburg14}
{Starkenburg} E.,  {Shetrone} M.~D.,  {McConnachie} A.~W.,   {Venn} K.~A.,
  2014, \mn@doi [\mnras] {10.1093/mnras/stu623}, \href
  {http://adsabs.harvard.edu/abs/2014MNRAS.441.1217S} {441, 1217}

\bibitem[\protect\citeauthoryear{{Starkenburg}, {Oman}, {Navarro}, {Crain},
  {Fattahi}, {Frenk}, {Sawala}  \& {Schaye}}{{Starkenburg}
  et~al.}{2017a}]{Starkenburg17b}
{Starkenburg} E.,  {Oman} K.~A.,  {Navarro} J.~F.,  {Crain} R.~A.,  {Fattahi}
  A.,  {Frenk} C.~S.,  {Sawala} T.,   {Schaye} J.,  2017a, \mn@doi [\mnras]
  {10.1093/mnras/stw2873}, \href
  {http://adsabs.harvard.edu/abs/2017MNRAS.465.2212S} {465, 2212}

\bibitem[\protect\citeauthoryear{{Starkenburg} et~al.,}{{Starkenburg}
  et~al.}{2017b}]{Starkenburg17a}
{Starkenburg} E.,  et~al., 2017b, \mn@doi [\mnras] {10.1093/mnras/stx1068},
  \href {https://ui.adsabs.harvard.edu/abs/2017MNRAS.471.2587S} {471, 2587}

\bibitem[\protect\citeauthoryear{{Steigman}}{{Steigman}}{2007}]{Steigman07}
{Steigman} G.,  2007, \mn@doi [Annual Review of Nuclear and Particle Science]
  {10.1146/annurev.nucl.56.080805.140437}, \href
  {http://adsabs.harvard.edu/abs/2007ARNPS..57..463S} {57, 463}

\bibitem[\protect\citeauthoryear{{Susa}, {Hasegawa}  \& {Tominaga}}{{Susa}
  et~al.}{2014}]{Susa14}
{Susa} H.,  {Hasegawa} K.,   {Tominaga} N.,  2014, \mn@doi [\apj]
  {10.1088/0004-637X/792/1/32}, \href
  {http://adsabs.harvard.edu/abs/2014ApJ...792...32S} {792, 32}

\bibitem[\protect\citeauthoryear{{Tegmark}, {Silk}, {Rees}, {Blanchard}, {Abel}
   \& {Palla}}{{Tegmark} et~al.}{1997}]{Tegmark97}
{Tegmark} M.,  {Silk} J.,  {Rees} M.~J.,  {Blanchard} A.,  {Abel} T.,   {Palla}
  F.,  1997, \mn@doi [\apj] {10.1086/303434}, \href
  {https://ui.adsabs.harvard.edu/abs/1997ApJ...474....1T} {474, 1}

\bibitem[\protect\citeauthoryear{{Theler} et~al.,}{{Theler}
  et~al.}{2020}]{Theler2020}
{Theler} R.,  et~al., 2020, \mn@doi [\aap] {10.1051/0004-6361/201937146}, \href
  {https://ui.adsabs.harvard.edu/abs/2020A&A...642A.176T} {642, A176}

\bibitem[\protect\citeauthoryear{{Tolstoy}, {Hill}  \& {Tosi}}{{Tolstoy}
  et~al.}{2009}]{Tolstoy09}
{Tolstoy} E.,  {Hill} V.,   {Tosi} M.,  2009, \mn@doi [\araa]
  {10.1146/annurev-astro-082708-101650}, \href
  {http://adsabs.harvard.edu/abs/2009ARA%26A..47..371T} {47, 371}

\bibitem[\protect\citeauthoryear{{Tominaga}, {Iwamoto}  \& {Nomoto}}{{Tominaga}
  et~al.}{2014}]{Tominaga14}
{Tominaga} N.,  {Iwamoto} N.,   {Nomoto} K.,  2014, \mn@doi [\apj]
  {10.1088/0004-637X/785/2/98}, \href
  {http://adsabs.harvard.edu/abs/2014ApJ...785...98T} {785, 98}

\bibitem[\protect\citeauthoryear{{Tsujimoto} \& {Nishimura}}{{Tsujimoto} \&
  {Nishimura}}{2015}]{Tsujimoto2015}
{Tsujimoto} T.,  {Nishimura} N.,  2015, \mn@doi [\apjl]
  {10.1088/2041-8205/811/1/L10}, \href
  {https://ui.adsabs.harvard.edu/abs/2015ApJ...811L..10T} {811, L10}

\bibitem[\protect\citeauthoryear{{VandenBerg}, {Bergbusch}, {Dotter},
  {Ferguson}, {Michaud}, {Richer}  \& {Proffitt}}{{VandenBerg}
  et~al.}{2012}]{Vandenberg2012}
{VandenBerg} D.~A.,  {Bergbusch} P.~A.,  {Dotter} A.,  {Ferguson} J.~W.,
  {Michaud} G.,  {Richer} J.,   {Proffitt} C.~R.,  2012, \mn@doi [\apj]
  {10.1088/0004-637X/755/1/15}, \href
  {https://ui.adsabs.harvard.edu/abs/2012ApJ...755...15V} {755, 15}

\bibitem[\protect\citeauthoryear{{Venn}, {Irwin}, {Shetrone}, {Tout}, {Hill}
  \& {Tolstoy}}{{Venn} et~al.}{2004}]{Venn04}
{Venn} K.~A.,  {Irwin} M.,  {Shetrone} M.~D.,  {Tout} C.~A.,  {Hill} V.,
  {Tolstoy} E.,  2004, \mn@doi [\aj] {10.1086/422734}, \href
  {http://adsabs.harvard.edu/abs/2004AJ....128.1177V} {128, 1177}

\bibitem[\protect\citeauthoryear{{Venn}, {Starkenburg}, {Malo}, {Martin}  \&
  {Laevens}}{{Venn} et~al.}{2017}]{Venn17}
{Venn} K.~A.,  {Starkenburg} E.,  {Malo} L.,  {Martin} N.,   {Laevens}
  B.~P.~M.,  2017, \mn@doi [\mnras] {10.1093/mnras/stw3198}, \href
  {https://ui.adsabs.harvard.edu/abs/2017MNRAS.466.3741V} {466, 3741}

\bibitem[\protect\citeauthoryear{{Venn} et~al.,}{{Venn}
  et~al.}{2020}]{Venn2020}
{Venn} K.~A.,  et~al., 2020, \mn@doi [\mnras] {10.1093/mnras/stz3546}, \href
  {https://ui.adsabs.harvard.edu/abs/2020MNRAS.492.3241V} {492, 3241}

\bibitem[\protect\citeauthoryear{{Wanajo}, {M{\"u}ller}, {Janka}  \&
  {Heger}}{{Wanajo} et~al.}{2018}]{Wanajo18}
{Wanajo} S.,  {M{\"u}ller} B.,  {Janka} H.-T.,   {Heger} A.,  2018, \mn@doi
  [\apj] {10.3847/1538-4357/aa9d97}, \href
  {https://ui.adsabs.harvard.edu/abs/2018ApJ...852...40W} {852, 40}

\bibitem[\protect\citeauthoryear{{Wang}, {Dutton}, {Stinson}, {Macci{\`o}},
  {Penzo}, {Kang}, {Keller}  \& {Wadsley}}{{Wang} et~al.}{2015}]{Wang2015}
{Wang} L.,  {Dutton} A.~A.,  {Stinson} G.~S.,  {Macci{\`o}} A.~V.,  {Penzo} C.,
   {Kang} X.,  {Keller} B.~W.,   {Wadsley} J.,  2015, \mn@doi [\mnras]
  {10.1093/mnras/stv1937}, \href
  {https://ui.adsabs.harvard.edu/abs/2015MNRAS.454...83W} {454, 83}

\bibitem[\protect\citeauthoryear{{White} \& {Springel}}{{White} \&
  {Springel}}{2000}]{White00}
{White} S. D.~M.,  {Springel} V.,  2000, in {Weiss} A.,  {Abel} T.~G.,   {Hill}
  V.,  eds, The First Stars. p.~327 (\mn@eprint {arXiv} {astro-ph/9911378}),
  \mn@doi{10.1007/10719504_62}

\bibitem[\protect\citeauthoryear{{Wise}, {Turk}, {Norman}  \& {Abel}}{{Wise}
  et~al.}{2012}]{Wise12}
{Wise} J.~H.,  {Turk} M.~J.,  {Norman} M.~L.,   {Abel} T.,  2012, \mn@doi
  [\apj] {10.1088/0004-637X/745/1/50}, \href
  {https://ui.adsabs.harvard.edu/abs/2012ApJ...745...50W} {745, 50}

\bibitem[\protect\citeauthoryear{{Woosley} \& {Weaver}}{{Woosley} \&
  {Weaver}}{1995}]{Woosley1995}
{Woosley} S.~E.,  {Weaver} T.~A.,  1995, \mn@doi [\apjs] {10.1086/192237},
  \href {https://ui.adsabs.harvard.edu/abs/1995ApJS..101..181W} {101, 181}

\bibitem[\protect\citeauthoryear{{Woosley}, {Heger}  \& {Weaver}}{{Woosley}
  et~al.}{2002}]{Woosley2002}
{Woosley} S.~E.,  {Heger} A.,   {Weaver} T.~A.,  2002, \mn@doi [Reviews of
  Modern Physics] {10.1103/RevModPhys.74.1015}, \href
  {https://ui.adsabs.harvard.edu/abs/2002RvMP...74.1015W} {74, 1015}

\bibitem[\protect\citeauthoryear{{Yanny} et~al.,}{{Yanny}
  et~al.}{2009}]{Yanny09a}
{Yanny} B.,  et~al., 2009, \mn@doi [\aj] {10.1088/0004-6256/137/5/4377}, \href
  {http://adsabs.harvard.edu/abs/2009AJ....137.4377Y} {137, 4377}

\bibitem[\protect\citeauthoryear{{Yong} et~al.,}{{Yong} et~al.}{2013a}]{Yong13}
{Yong} D.,  et~al., 2013a, \mn@doi [\apj] {10.1088/0004-637X/762/1/26}, \href
  {http://adsabs.harvard.edu/abs/2013ApJ...762...26Y} {762, 26}

\bibitem[\protect\citeauthoryear{{Yong} et~al.,}{{Yong}
  et~al.}{2013b}]{yong2013CEMP}
{Yong} D.,  et~al., 2013b, \apj, 762, 27

\bibitem[\protect\citeauthoryear{{Yoon} et~al.,}{{Yoon} et~al.}{2016}]{Yoon16}
{Yoon} J.,  et~al., 2016, preprint, \href
  {http://adsabs.harvard.edu/abs/2016arXiv160706336Y} {} (\mn@eprint {arXiv}
  {1607.06336})

\bibitem[\protect\citeauthoryear{{York} et~al.,}{{York} et~al.}{2000}]{York00}
{York} D.~G.,  et~al., 2000, \mn@doi [\aj] {10.1086/301513}, \href
  {http://adsabs.harvard.edu/abs/2000AJ....120.1579Y} {120, 1579}

\bibitem[\protect\citeauthoryear{{Yoshida}, {Omukai}, {Hernquist}  \&
  {Abel}}{{Yoshida} et~al.}{2006}]{Yoshida06}
{Yoshida} N.,  {Omukai} K.,  {Hernquist} L.,   {Abel} T.,  2006, \mn@doi [\apj]
  {10.1086/507978}, \href
  {https://ui.adsabs.harvard.edu/abs/2006ApJ...652....6Y} {652, 6}

\bibitem[\protect\citeauthoryear{{Youakim} et~al.,}{{Youakim}
  et~al.}{2017}]{Youakim17}
{Youakim} K.,  et~al., 2017, preprint, \href
  {http://adsabs.harvard.edu/abs/2017arXiv170801264Y} {} (\mn@eprint {arXiv}
  {1708.01264})

\bibitem[\protect\citeauthoryear{{Yuan} et~al.,}{{Yuan}
  et~al.}{2020}]{Yuan2020}
{Yuan} Z.,  et~al., 2020, \mn@doi [\apj] {10.3847/1538-4357/ab6ef7}, \href
  {https://ui.adsabs.harvard.edu/abs/2020ApJ...891...39Y} {891, 39}

\bibitem[\protect\citeauthoryear{{de Boer}, {Tolstoy}, {Lemasle}, {Saha},
  {Olszewski}, {Mateo}, {Irwin}  \& {Battaglia}}{{de Boer}
  et~al.}{2014}]{deBoer2014}
{de Boer} T.~J.~L.,  {Tolstoy} E.,  {Lemasle} B.,  {Saha} A.,  {Olszewski}
  E.~W.,  {Mateo} M.,  {Irwin} M.~J.,   {Battaglia} G.,  2014, \mn@doi [\aap]
  {10.1051/0004-6361/201424119}, \href
  {https://ui.adsabs.harvard.edu/abs/2014A&A...572A..10D} {572, A10}

\makeatother
\end{thebibliography}
